\documentclass[goettingen, gauss, print, additionalreferee]{thesis}

\usepackage{graphicx}	
\usepackage{amsmath}	
\usepackage{amssymb}	
\usepackage{todonotes}
\usepackage{multirow}
\usepackage{natbib}
\usepackage{newtxtext,newtxmath}
\usepackage{color,soul}
\definecolor{mygreen}{rgb}{0,0.47,0.06}
\usepackage{hyperref}
\usepackage{textcomp}
\usepackage[utf8]{inputenc}
\usepackage{bm}
\usepackage{float}
\usepackage{listings}
\usepackage{silence}
\title{Layerings in the nucleus of \\ comet 67P/Churyumov-Gerasimenko}
\author{Birko-Katarina Ruzicka}
\town{M\"unchen}
\thesisadvisorycommitteea{Dr. Hermann B\"ohnhardt}
\instthesisadvisorycommitteea{Max-Planck-Institut f\"ur Sonnensystemforschung, G\"ottingen, Germany}
\thesisadvisorycommitteeb{Prof. Dr. Andreas Pack}
\instthesisadvisorycommitteeb{Georg-August-Universit\"at G\"ottingen, Germany}
\thesisadvisorycommitteec{Prof. Dr. Ulrich Christensen}
\instthesisadvisorycommitteec{Max-Planck-Institut f\"ur Sonnensystemforschung, G\"ottingen, Germany}

\refereea{Prof. Dr. Andreas Pack} 
\instrefereea{Georg-August-Universit\"at G\"ottingen, Germany}
\refereeb{Dr. Hermann B\"ohnhardt}
\instrefereeb{Max-Planck-Institut f\"ur Sonnensystemforschung, G\"ottingen, Germany}

\commissiona{Prof. Dr. Ulrich Christensen}
\instcommissiona{Max-Planck-Institut f\"ur Sonnensystemforschung, G\"ottingen, Germany}
\commissionb{Dr. Matthias Schr\"oter}
\instcommissionb{Max-Planck Institut für Dynamik und Selbstorganisation, G\"ottingen, Germany}
\commissionc{Dr. Luca Penasa}
\instcommissionc{Center of Studies and Activities for Space, University of Padova, Italy}
\commissiond{Prof. Dr. Jonas Kley}
\instcommissiond{Georg-August-Universit\"at G\"ottingen, Germany}

\submittedyear{2019}
\publicationyear{2019}
\isbn{978-3-947208-20-3}

\newcommand{\comment}[1]{}
\usepackage[german,english]{babel}
\usepackage{natbib}
\usepackage{printlen}
\usepackage{color}
\begin{document}
\selectlanguage{english}
\maketitle
\tableofcontents \label{contents}
\listoffigures
\listoftables

\chapter*{Summary\markboth{Summary}{Summary}}
\addcontentsline{toc}{chapter}{Summary}

During the past decades, spacecraft missions to cometary nuclei revealed that their surfaces are complex and diverse. Most recently, the 'Rosetta' mission to comet 67P/Churyumov-Gerasimenko ('comet 67P') delivered spectacular images of its nucleus at unprecedented resolution, showing smooth plains, flat terraces, steep cliffs, circular pits, and indications of global layerings. In the years since Rosetta arrived at the comet, it has been a matter of intense study how and when those apparent layerings were formed in the cometary nucleus.
By merging techniques of structural geology, statistical image processing, and solar system science, this thesis aims to contribute to the understanding of the formation of the layerings, and consequently the formation of the nuclei as a whole.

Following an introduction that gives an overview of cometary science in general, the Rosetta mission in particular, as well as geological aspects of layerings, I describe the two distinctive approaches I used to study the layerings' orientation on both lobes of comet 67P's nucleus.  

In the first approach, I mapped layering-related linear features on the nucleus surface, meaning the edge-lines of morphological terraces as well as the traces of internal strata where they intersect with the surface along hill slopes. I mapped these lineaments on a three-dimensional shape model of the nucleus, onto which I projected high-resolution images for clearer spatial orientation. This method locally improved the spatial resolution by more than an order of magnitude. By mapping only lineaments of substantial curvature, I was then able to fit planes through the nodes that make up the lineaments. I compared those planes' normal vectors to normals determined by other authors in similar or identical locations, albeit using different methods. In this way, I confirmed their results, including that the layering systems on the comet's two lobes are geometrically independent from each other. My results rule out the proposal that 67P's lobes represent collisional fragments of a much larger, layered body.

In the second approach, I developed a Fourier-based image analysis algorithm to detect lineament structures at pixel-precision. I used this algorithm to analyse the Hathor cliff on the Small Lobe of comet 67P, where layering-related, sub-parallel linear features are freshly exposed. I found it to be a broadly applicable, powerful tool for automating the detection of layerings in images where conventional edge-detection algorithms are not effective. When correctly configured to the target conditions, I found the algorithm to have a higher signal-to-noise detection sensitivity than a human researcher while also reducing over-interpretation due to human biases.

In summary, I studied the layerings in the nucleus of comet 67P using several unconventional approaches and constrained their lateral extent, curvature, and to a degree also their thickness. Ultimately, I nominated two mechanisms that could have formed these layerings in cometary nuclei.

\newpage

\chapter{Introduction} \label{ch_1}

\section{Concerning comets} \label{ch_1_1}

Comets are planetesimals that formed within a few million years of the Sun's ignition, in the outer reaches of the primordial disk \citep{davidsson_primordial_2016}, at a distance to the Sun of approximately 15 to 30 AU (one 'astronomical unit' meaning the average distance between the Earth and the Sun). A widely accepted scenario is that after their formation in the young Solar System, most planetesimals were removed from their orbits by encounters with other bodies, as well as by gravitational interactions with the accreting terrestrial and giant planets \citep{morbidelli_2000_source}. As a result, they either got incorporated into the growing planets or fell into the Sun, but most of them were ejected outward and left the Solar System. A small portion of comets was diverted towards so-called dynamical reservoirs, which provided an environment that protected and preserved the primordial comets from thermal and collisional processing \citep{davidsson_primordial_2016}. Thus, comets are not only among the oldest objects, but likely also the least altered solid bodies surviving from the origin of the solar system who may provide a unique record of the physical processes involved in their formation.

I studied the proposed layerings on the nucleus of comet 67P/Churyumov-Gerasimenko ('comet 67P') and will provide an analysis of their geometry. In particular, I focused on deducing the comet's internal structure from a detailed study of the orientation of these layerings on the nucleus surface.

In this chapter, I will briefly introduce comets as well as some of their properties that are relevant for studying layerings in cometary nuclei. I will give an overview of the instruments onboard the Rosetta spacecraft and summarise the insights and data that my analysis is based on. Finally, I will briefly familiarise the non-geologist reader with the geological processes and conditions that create layerings on Earth and other bodies in the Solar System.

\subsection{Orbits and reservoirs} \label{ch_1_1_1}

Most comets are on elliptical orbits such that they spend the majority of a revolution far from the Sun and only a short section of a revolution is spent in closer proximity to the Sun. They can be classified by the length of their orbital periods: \textbf{Short period comets} (also called 'periodic comets') have orbital periods of less than 200 years, low eccentricity, and generally orbit close to in the plane of the ecliptic \citep{duncan_1988_shortperiod}. Their shorter orbital periods take them close to the giant planets, whose gravity can affect their orbits. \textbf{Long period comets} either have approximately circular orbits with a semi-major axis that lies far outside of the planetary system; or they have highly eccentric orbits that it can take them hundreds of thousands of years to complete. They should not be confused with comets on near-parabolic or hyperbolic orbits, who will most likely not ever return to the proximity of the Sun. When long period comets cross the orbit of one of the giant planets, they can be gravitationally captured into shorter orbits whose aphelia are all near that planet's orbit. This group of comets is then called that its 'family' \citep{wilson_1909_comet}. 67P/Churyumov-Gerasimenko is a member of the Jupiter family comets (JFC).

Which dynamical reservoir a comet ends up in depends on its orbital parameters at the time of formation, as well as which objects in the solar system it happened to interact with during its lifetime. There are three known reservoirs of comets in the solar system: The Oort cloud, the Kuiper belt, and the asteroid belt.


The \textbf{Oort cloud} is a presumed, roughly spherical envelope of predominantly icy planetesimals surrounding the Solar System at approximately 20,000 AU to 150,000 AU \citep{brasser_2013_oort}. For comparison, Pluto orbits at a distance to the Sun of roughly 40 AU \citep{stern_2018_pluto}. Due to the great distance and low object density, the Oort cloud has not yet been observed directly, but it is believed to be the source of most comets with long orbital periods \citep{kaib_2009_oort}.

The flattened \textbf{Kuiper belt} (also frequently referred to as 'Edgeworth-Kuiper belt') is oriented close to the plane in which the planets orbit the Sun. It extends from the orbit of Neptune (at 30 AU) to approximately 50 AU from the Sun \citep{stern_1997_kuiper}, and is believed to be the source of most short-period comets. Its population is part of the so-called 'trans-Neptunian objects' \citep{mcfadden_2006_encyclopedia}.

The \textbf{asteroid belt}, while eponymously consisting mostly of asteroids, also hosts a small number of 'main-belt comets' \citep{levison_2009_mainbelt}. It remains a matter of debate whether these comets formed within the asteroid belt, or were dynamically injected into the belt by Jupiter \citep{hsieh_2006_population}. Main-belt comets are observed to have orbits that are close to circular. 

The current understanding is that comets remain within these reservoirs until they are gravitationally disturbed. Comets in the Oort cloud are predominantly disturbed by gravitational interaction with massive objects in the solar neighbourhood (i.e. stars, nebulae and galactic structures), while comets in the Kuiper- and asteroid belts are affected most strongly by the gravity of the planets. Once a comet is expelled from its reservoir, its median dynamical lifetime is only about 300,000 years \citep{mcfadden_2006_encyclopedia}. This term describes the span of time until the comet is either ejected from the Solar System, or reaches an orbit that leads it close enough to the Sun that its nucleus begins to sublimate and we can see its brightly glowing coma in the night sky.  

\subsection{Composition and structure} \label{ch_1_1_2} 

\begin{figure} [h] 
\centering
\includegraphics[width=0.9\linewidth]{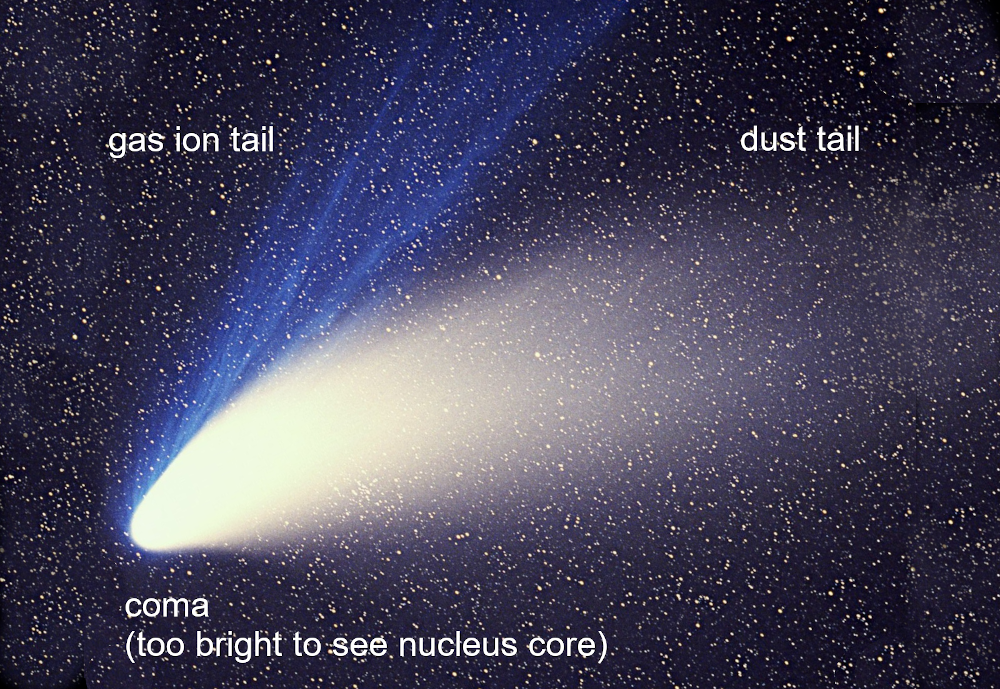}
\caption[Photograph of comet Hale-Bopp, cometary components labelled]{A photograph of comet Hale-Bopp, on which the components of a typical comet are labelled (modified from \citet{kolmhofer_1997_hale}, available under CC BY-SA 3.0 license)}
\label{fig:anatomy}
\end{figure}

The object we commonly refer to as "a comet" in fact consists of several components (\autoref{fig:anatomy}): At the core lies a solid body called the \textbf{nucleus}. When the nucleus is warmed above a critical temperature, it becomes 'active', i.e. it releases gases that form an extremely tenuous atmosphere. This envelope of dust particles and gases is called the \textbf{coma}. As long as the comet is far from the Sun, its coma is too faint to be seen by the unassisted eye. As the comet approaches perihelion, i.e. the point of its orbit where it is closest to the Sun, its coma grows rapidly. It is pushed outward by the Sun's radiation and particle wind to form the long, characteristic \textbf{tail}. Typically, the tail is divided into a curved stream of dust particles pointing in the direction of the comet's orbit, and a straight stream of gas particles along the direction of the Sun's magnetic field lines (\autoref{fig:anatomy}).

Comets have been studied using ground-based telescopes for centuries, and space telescopes for decades. Spectroscopic studies have revealed the molecular composition of their coma to be mostly \textbf{volatiles} (ices) with minor amounts of \textbf{refractories} (dust). By mass, the coma's volatiles consist mainly of water molecules \citep{bockelee_2011_overview}.

Our knowledge about the nuclei on the other hand has only been accumulated within the last decades, mainly from the six comets that were visited by spacecraft. Chronologically, those are: 1P/Halley, 19P/Borrelly, 81P/Wild 2, 9P/Tempel 1, 103P/Hartley 2, and 67P/Churyumov-Gerasimenko (\autoref{fig:comets_visited}). All of those encounters except one were fly-by missions with high relative velocities, meaning that a space probe was sent onto a trajectory that carried it close enough to a comet to record scientific data, but that the time interval for data collection was limited during those missions. The Rosetta mission was the first mission to place a spacecraft in orbit around a comet, where it had time to collect data for almost two years (cf. \autoref{ch_1_2}). From all data that was cumulatively collected by the space probes, but especially from data collected by Rosetta, the following is known about the composition and structure of cometary nuclei:

\begin{figure} [h] 
\centering
\includegraphics[width=\linewidth]{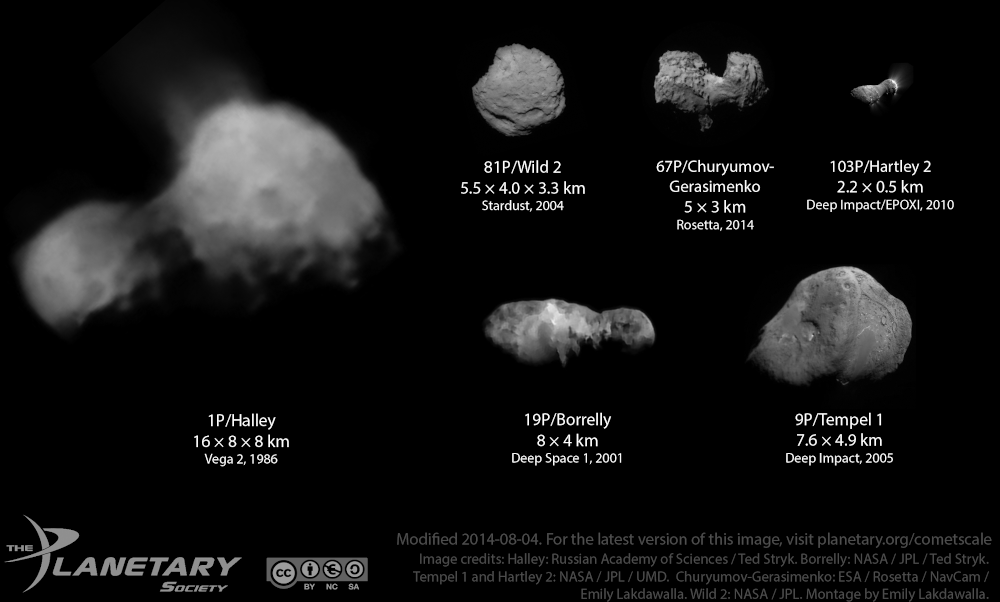}
\caption[Comets visited by spacecraft]{Comets visited by spacecraft, pictured at true relative scale (\citet{lakdawalla_2014_comets}, available under CC BY-NC-SA 3.0 license)}
\label{fig:comets_visited}
\end{figure}

\textbf{Size.} Nucleus radii range from 0.2 to 37 km. Comets whose orbits lie within the ecliptic tend to have a smaller nucleus (0.2--15 km) than comets whose orbits lie outside of the ecliptic (1.6--37 km). Several nuclei with sub-kilometre radii were observed, and only a few of the well-measured nuclei have radii larger than 5 km \citep{lamy_2004_sizes}.

\textbf{Composition.} Cometary nuclei are composed of volatiles and refractory materials at a ratio of about 1, although the ratio depends on several model assumptions and could vary between 0.1 and 10 \citep{ahearn_1995_comets}. Silicates and organic refractories each make up roughly 25\% of their mass, while the other half consists of a H$_2$O dominated mixture containing a few percent each of CO, CO$_2$, CH$_3$OH and other simple molecules \citep{greenberg_1998_making,bockelee_2011_overview}. 

\textbf{Albedo.} Albedo is a measure for the ratio of incoming light that is reflected back from a surface. A typical albedo for cometary nuclei is only 0.02--0.06, i.e. only 2--6\% of light is reflected. The rest of the light is absorbed by the surface, meaning that the nuclei are about as dark as charcoal and thus among the darkest known objects in the solar system \citep{lamy_2004_sizes} and photographs showing nuclei need to be exposed longer to compensate for this. \citet{cruikshank_1989_dark} proposed that complex organic compounds contribute to the astoundingly low albedo value. An additional factor might be their surface texture, which is in large parts lumpy and uneven and thus not well suited to reflect light. The nucleus is furthermore largely covered in 'airfall' material, a term that describes fine-grained dust that is expelled from the comet's surface and interior during outgassing events, but does not reach sufficient velocity to escape the comet's gravity \citep{thomas_2015_redistribution}. In a process not dissimilar to the deposition of ash after a volcanic eruption on Earth, the airfall material then gently settles back onto the comet's surface. As little light is reflected and much of it is therefore absorbed, the low albedo might explain the high observed surface temperature of cometary nuclei \citep[cf.][]{groussin_2013_temperature}.

\textbf{Density and porosity.} Density and porosity are fundamental physical properties that convey much about the internal structure and composition of a particular body \citep{weissman_2008_density}. The density can be calculated from the nucleus volume (derived from spacecraft images) and its mass, which could only be assessed indirectly for the fly-by missions before Rosetta. Most relevantly, mass estimates were derived from the non-gravitational accelerations on the nucleus due to outgassing \citep{rickman_1986_masses,skorov_1999_gasflow}. Rosetta determined its target's mass more directly through the gravitational effects of the nucleus on Rosetta's orbit. The bulk density of cometary nuclei was found to be below 1.0 g/$cm^3$, with most values around 0.6 g/$cm^3$. I compiled some examples in \autoref{tab:comet_properties}. Density therefore stays well below the theoretical bulk density of 1.65 g/$cm^3$ \citep{greenberg_1998_making} for a fully packed cometary nucleus, and therefore requires considerable micro- and macro-porosity. Depending on the assumed dust-to-ice ratio in the nucleus, the porosity lies upwards of 60\%  \citep{weissman_2008_density}. This means that they must be made up of loosely packed, fluffy particles.  

\vspace{0.1cm}

\begin{table} [h] 
    \centering
    {\renewcommand{\arraystretch}{1.2}
    \begin{tabular}{lrr}
    \hline
    \textbf{Name} & \textbf{Density [g/cm$^3$]} & \textbf{Bulk nucleus porosity}\\
    \hline
    1P/Halley    & 0.6 (+0.9,-0.4) $^1$ & > 80 \% $^6$  \\
    9P/Tempel 1  & 0.45 $\pm$ 0.25 $^2$ & 50--88 \% $^2$ \\
    19P/Borrelly & 0.49 (+0.34,-0.20)$^3$ &  \\
    81P/Wild 2   & 0.60-0.80  $^4$ & 30--60 \% $^4$ \\
    67P/C.-G.    & 0.5378 $\pm$ 0.0006 $^5$ &  72--74 \% $^7$\\
    \hline
    \end{tabular} }
    \caption[Density and bulk nucleus porosity for some cometary nuclei]{Density and bulk nucleus porosity for some cometary nuclei. 
    $^1$ \citet{sagdeev_1988_nucleus}, 
    $^2$ \citet{davidsson_2007_tempel}, 
    $^3$ \citet{farnham_2002_borrelly}, 
    $^4$ \citet{davidsson_2006_wild}, 
    $^5$ \citet{preusker_global_2017}, 
    $^6$ \citet{mekler_1990_porous}, 
    $^7$ \citet{paetzold_2018_nucleus}. }
    \label{tab:comet_properties}
\end{table}

\newpage

\section{The Rosetta mission to comet 67P} \label{ch_1_2}

Rosetta was an international space mission led by the European Space Agency ESA, with contributions from its member states and NASA. The Rosetta spacecraft, carrying 11 instruments and a lander module named 'Philae', left Earth in 2004, performed several gravitational manoeuvres around Earth and Mars, and passed by the two asteroids (2867) Steins in 2008 and (21) Lutetia in 2010 \citep{glassmeier_2007_rosetta}. It arrived at comet 67P in August of 2014, and stayed in orbit around the comet for more than two years. The mission ended in September of 2016 with an intentional impact onto the nucleus.

\subsection{Instrumentation onboard Rosetta} \label{ch_1_2_1}

\begin{figure} [h] 
\centering
\includegraphics[width=\linewidth]{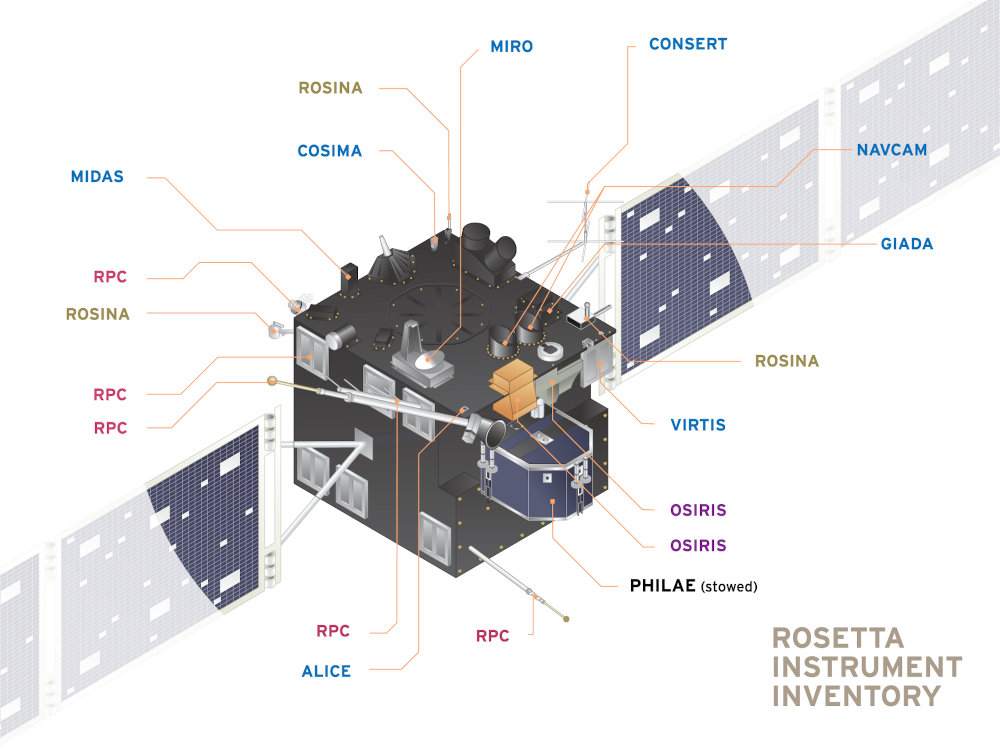}
\caption[Rosetta's inventory of instruments]{Rosetta's inventory of instruments. Some instruments are divided into sub-instruments mounted in several locations (modified from \citet{lakdawalla_2019_rosetta}, available under CC BY 3.0 license) \vspace{0.4cm}}
\label{fig:rosetta}
\end{figure}

The Rosetta mission's suite of instruments was designed with the goal to shed light on the origins of cometary formation and evolution, and thus to learn more about the early Solar System \citep{glassmeier_2007_rosetta}. According to \citet{schwehm_1999_rosetta}, the mission's suite of instruments was designed to be able to globally characterise the comet's dynamic properties, surface morphology and composition, study the origins and development of cometary activity, and determine the compositions of volatiles and refractories in the nucleus. The inventory of instruments onboard the Rosetta spacecraft is shown in \autoref{fig:rosetta}, their full names and purpose \citep{glassmeier_2007_rosetta} are as follows:

\begin{itemize}
    \item Alice (ultraviolet imaging spectrometer): Composition of the nucleus and coma
    \item CONSERT (COmet Nucleus Sounding Experiment by Radio-wave Transmission): Study the internal structure of the comet with Philae
    \item COSIMA (COmetary Secondary Ion Mass Analyser): Composition of dust in coma
    \item GIADA (Grain Impact Analyser and Dust Accumulator): Measure the number, mass, momentum, and velocity distribution of dust grains in the near-comet environment
    \item MIDAS (Micro Imaging Dust Analysis System): Dust environment of the comet
    \item MIRO (Microwave Instrument for the Rosetta Orbiter): Investigate outgassing from the nucleus and development of the coma
    \item NAVCAM (NAVigation CAMera): Locate spacecraft relative to background stars and nucleus
    \item OSIRIS (Optical, Spectroscopic and Infrared Remote Imaging System): Scientific camera system. This instrument is of particular relevance to this thesis and will be explained in more detail below.
    \item ROSINA (Rosetta Orbiter Spectrometer for Ion and Neutral Analysis): Determine composition of the comet's atmosphere and ionosphere
    \item RPC (Rosetta Plasma Consortium): Study the comet's plasma environment
    \item VIRTIS (Visible and InfraRed Thermal Imaging Spectrometer): Study the comet nucleus and the gases in the coma
\end{itemize}{}

The Philae lander carried ten additional instruments, including several spectrometers, a camera system, the counterpart instrument for CONSERT, and a magnetometer.

\citet{keller_osiris_2007} give a comprehensive description of the OSIRIS camera system, and I will summarise the relevant content here. OSIRIS consisted of two complementary sub-systems, a wide-angle camera (WAC, also sometimes referred to as OSIWAC) and a \textbf{narrow-angle camera} (NAC / OSINAC). The wide-angle camera had a lower spatial resolution but a much wider field of view (11.35\textdegree{} $\times$ 12.11\textdegree), as its purpose was to observe the nucleus activity as well as the 3-dimensional flow field of dust and gas around the nucleus. The narrow-angle camera, on the other hand, had a narrower \textbf{field of view} (2.20\textdegree{} $\times$ 2.22\textdegree) but high spatial resolution and was intended to take detailed images of the nucleus surface. The NAC images served as basis for my analyses. 

The NAC used a 2048 $\times$ 2048 back-side illuminated CCD detector with a UV optimised anti-reflection coating \citep{tubiana_scientific_2015}. The camera has an angular resolution of 1.88\,$\times$ 10$^{-5}$ degrees and a focal length of 0.7173 metres. It was equipped with a total of 12 discrete \textbf{filters}, mounted onto two filter wheels that could be rotated independently. Nominal operation was the filter combination 'F22', which combined an anti-reflection-coated far-focusing plate with an orange filter (centred at 645 nm with a bandwidth of 94 nm), but many images were taken with other filter combinations. I found the F22-filtered images most useful for visual analysis. The specific images I used for each type of analysis are identified in the respective chapter. 

Although OSIRIS images of 67P's nucleus appear to be greyscale images, it is worth noting that they are actually 32-bit colour images of a greyish comet \citep{tubiana_scientific_2015}. The native file format is IMG, which is conveniently opened in the Fairwood PDS Image Viewer \citep{hviid_2009_pds}, a software developed for viewing images in the NASA Planetary Data System (PDS). The full file name of an OSIRIS image is composed according to the key  CYYYYMMDDTHHMMSSUUUFFLIFAB.IMG, whose elements are described in \autoref{tab:img_key}.

\vspace{0.4cm}

\begin{table}[h]  
    \centering
    {\renewcommand{\arraystretch}{1.3}
    \begin{tabular}{ll}
    \hline
    \textbf{Field(s)}  &  \textbf{Description }\\
    \hline
    C           &  Camera: either N (for NAC) or W (for WAC) \\
    YYYYMMDD    &  year, month and day of image acquisition \\
    T           &  separator T (for 'time follows') \\
    HHMMSS      &  hour, minute and second of image acquisition \\
    UUU         &  milli-second of image acquisition \\
    FF          &  image file type, several options exist but only the option \\
                &  ID (for Image Data) is relevant for this thesis \\
    L           &  CODMAC$^1$ processing level of the image \\
    I           &  Instance (values > 0 indicate multiple transmissions of image) \\
    F           &  separator F (for 'filters follow') \\
    A           &  position index for filter wheel \#1 \\
    B           &  position index for filter wheel \#2 \\
    .IMG        &  file extension\\
    \hline
    \end{tabular} }
    \caption[Elements of the OSIRIS image file names]{Elements of the OSIRIS image file names, according to the instrument handbook \citep{tubiana_2018_handbook}. $^1$Committee on Data Management, Archiving, and Computing}
    \label{tab:img_key}
\end{table}{}

Each image file begins with a header that has extensive information about the image, its calibration status, as well as photometric data such as the camera's position relative to the Sun and the comet centre at the time of image acquisition. In addition, images of calibration level $\geq$5 contain layers with pixel-by-pixel geospatial information such as the camera's distance to the nucleus surface. 

\begin{figure}  
\centering
\includegraphics[width=\linewidth]{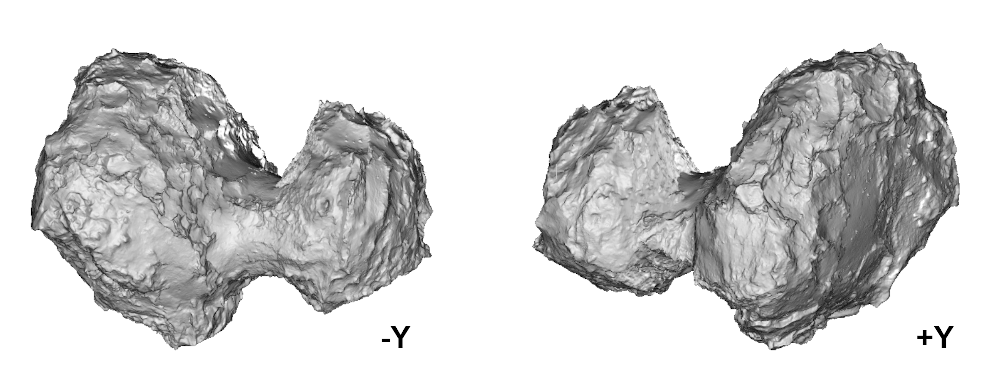}
\caption[Views of the SHAP7 shape model]{Views of the SHAP7 shape model along the negative (left) and positive (right) Y-axis of the comet-fixed coordinate system \vspace{0.2cm}}
\label{fig:shap7}
\end{figure}

In total, the WAC and NAC took 98,219 images during the entire mission, of which more than 76,000 were taken at the comet. Subsets of these images were used by several research groups to compute various 3-dimensional \textbf{shape models} of the comet nucleus \citep[e.g.][]{preusker_shape_2015,jorda_global_2016}. The most recent and most highly resolved model is commonly referred to as 'SHAP7' \citep{preusker_global_2017} (\autoref{fig:shap7}).

\begin{figure}  
\centering
\includegraphics[width=0.9\linewidth]{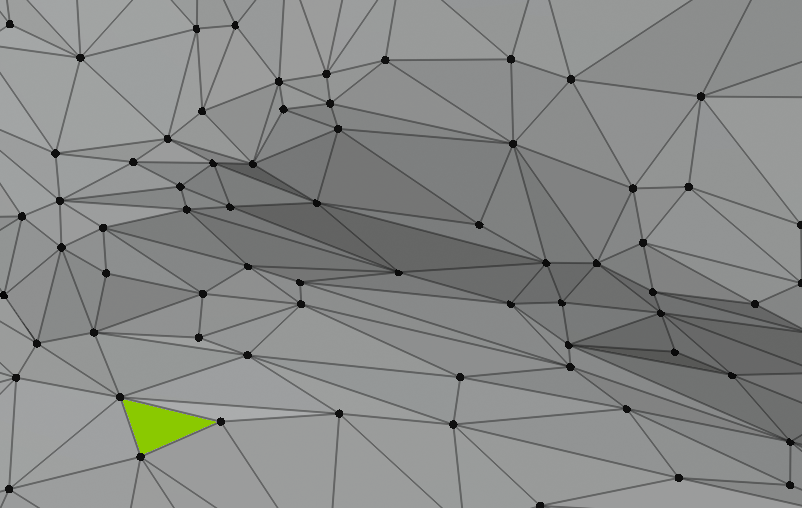}
\caption[Close-up of a shape model]{Close-up of a shape model, zoomed in until the individual vertices (black circular nodes) and facets (triangular spaces between the vertices, one shown in green for clarification) become visible. Changes in grey value are caused by the facets' orientation. The vertices are spaced only several metres apart at this scale\vspace{0.2cm}}
\label{fig:vertices}
\end{figure}

SHAP7 was created through stereo-photogrammetric (SPG) analysis of more than 1500 pre-perihelion NAC images at pixel scales between 0.2--3.0 m/px. Therefore, in contrast to some previous models, it covers the entire nucleus at high resolution. SHAP7 consists of approx. 44 million vertices and 22 million facets (\autoref{fig:vertices}). At this resolution, the shape model is expected to be accurate to the nucleus within several metres \citep{preusker_global_2017} and thus useful as a basis for geometric analyses. For some of my work, I locally further improved the resolution by aligning high-resolution NAC images onto this shape model.

\subsection{The nucleus surface of comet 67P} \label{ch_1_2_2}

In August of 2014, the Rosetta spacecraft was sufficiently close to 67P's nucleus that OSIRIS images showed the surface at decimetre- to metre-scale resolution. However, the 'southern hemisphere' of the nucleus (ca. 30\% of the surface), remained in darkness at that time, and only became illuminated by the Sun several months later. The first high-resolution images of the nucleus revealed an irregular-shaped, bilobate nucleus with a processed surface with morphologically diverse units \citep{thomas_2015_morphological}.

\begin{figure}  
\centering
\includegraphics[width=\linewidth]{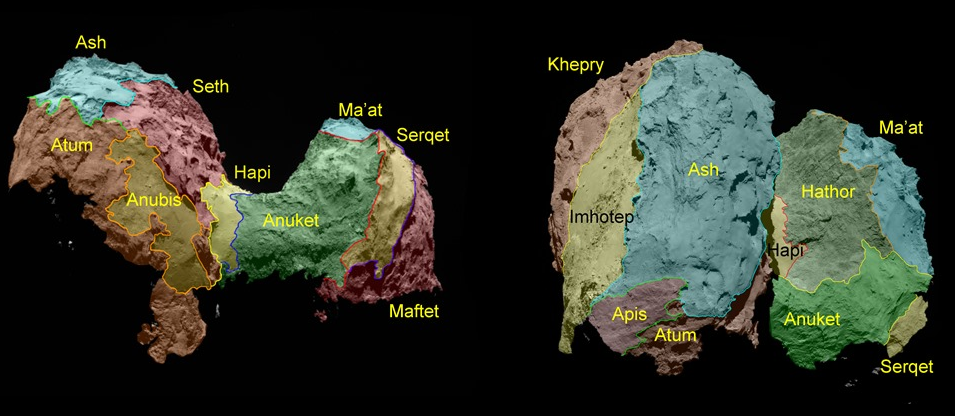}
\caption[Regions on the nucleus of comet 67P]{Regions on the nucleus of comet 67P. Image modified from \citet{regions}  \vspace{0.2cm}}
\label{fig:regions}
\end{figure}

For easier reference and orientation on the nucleus surface, \citet{thomas_2015_morphological} defined nineteen 'regions' that were named after Egyptian gods and goddesses, in accordance with the Rosetta mission's general nomenclature theme (\autoref{fig:regions}). When the southern hemisphere became visible, the total number of regions was extended to 26. These region definitions are now widely used in the literature as well as throughout this thesis. The 'Hathor' region is the focus of \autoref{ch_3}. 

\citet{thomas_2015_morphological} grouped the regions into five basic categories, based on their surface morphology: There are dust-covered terrains, brittle materials with pits and circular structures, large-scale depressions, smooth terrains, and exposed consolidated surfaces to be found on comet 67P.

The \textbf{dust-covered terrains} are understood to be 'flat' areas, i.e. areas on the nucleus that are oriented perpendicular to the local gravity vector, which are covered in fine-grained airfall material that masks most of the underlying topography. The thickness of this airfall cover varies locally, but is in the order of a few metres. It is those dust-covered areas that \citet{massironi_two_2015} and \citet{penasa_three_2017} proposed to be 'terraces' that are related to the comet's layered internal structure (cf. \autoref{ch_2}). 

Dune-like features have been observed at several locations within the dust-covered terrains, indicating an aeolian-driven surface transport of the dust \citep{thomas_shape_2007,thomas_2015_redistribution}. Near-surface winds have been proposed as transport mechanisms on cometary nuclei before, e.g. for comet 1P/Halley \citep{keller_1989_evidence}.

Surface areas classified as consolidated \textbf{brittle material}, such as in the \textit{Seth} region, show fracturing and evidence for being undercut by mass wasting of a stratum below. Overhangs in these areas were used to constrain the tensile strength to be less than 20 Pa \citep[e.g.][]{thomas_2015_morphological,attree_2018_tensile}. The brittle areas also contain structures called 'pits', which are presumed to be remnants of forceful outgassing events \citep{vincent_large_2015}. The walls of these pits are covered in a bumpy texture referred to as 'goosebumps', as well as faint horizontal lineations that \citet{massironi_two_2015} and others related to a layered internal structure of the nucleus.

To the geologically trained eye, the morphology and texture of many materials on comet 67P suggests a rock-like material, but this impression is misleading. The nucleus material's density and strength are lower by at least a factor of 5 to 10, and its composition is vastly different from any rock on Earth. To avoid the confusion that would arise by using the word 'rocky', \citet{thomas_2015_morphological} have coined the term \textbf{'consolidated cometary material' (CCM)} to use in its stead.

\begin{figure}  
\centering
\includegraphics[width=\linewidth]{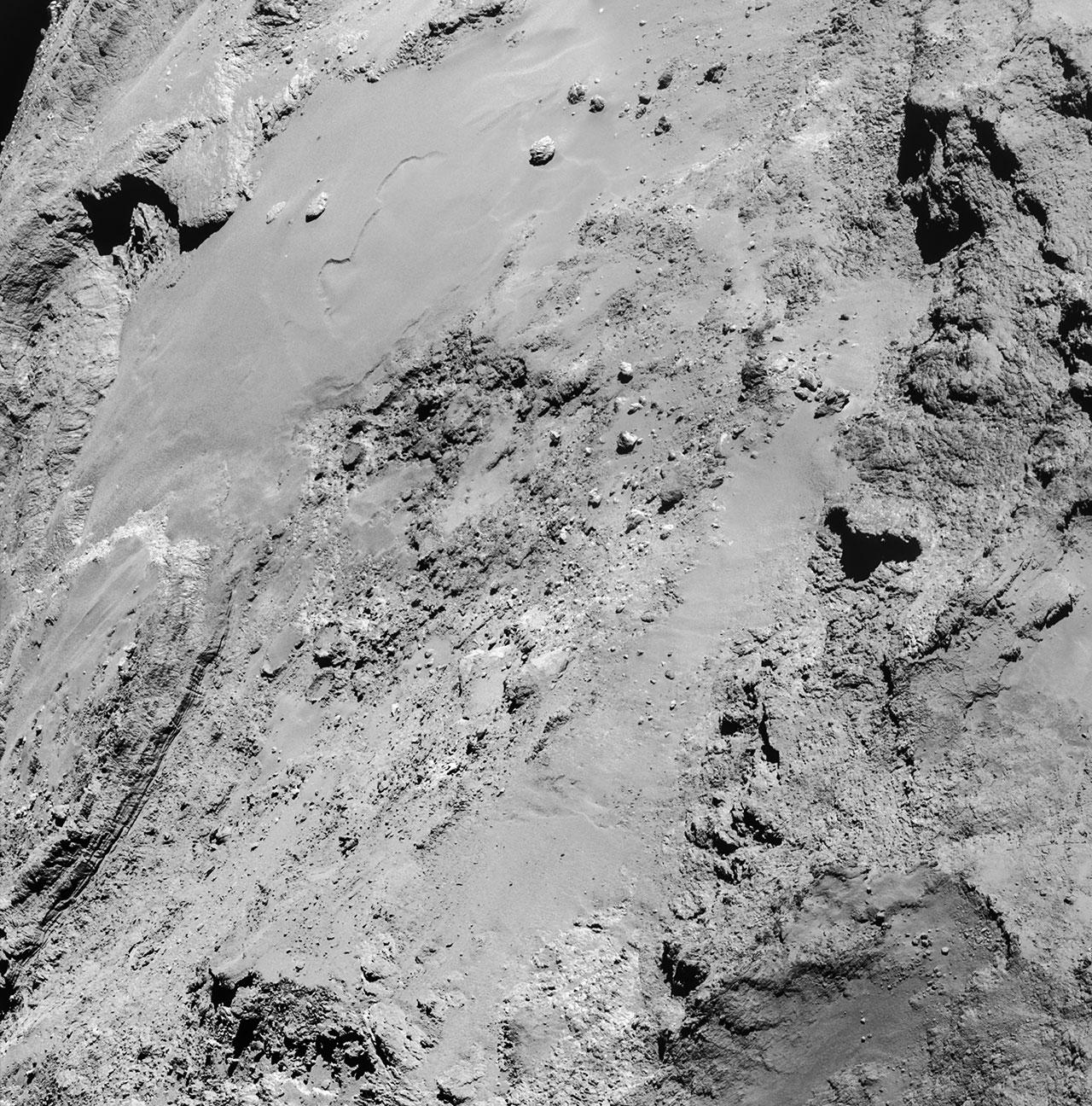}
\caption[Four image mosaic of the Imhotep region showing layering at the margin] {Four image mosaic of the Imhotep region on the comet's large lobe, comprising images taken on 14 February 2015. The image scale is 0.76 m/pixel and the mosaic measures 1.35 $\times$ 1.37 km across. All around the margin, but especially in the lower-left corner, the margin seems to consist of layered materials \citep{esa_2015_imhotep} CC BY-SA IGO 3.0  \vspace{0.2cm}}
\label{fig:smooth_margin}
\end{figure}

A typical exposure of CCM are the margins enclosing the terrain type named \textbf{smooth expanses}. These margins are relevant to this work, because the CCM gives the impression that it is layered (\autoref{fig:smooth_margin}). The smooth expanses comprise three large areas on the nucleus that are characterised by extremely smooth material (\textit{Imhotep, Anubis, and Hapi}). \citet{penasa_2017_layered} hypothesised that the smooth expanses were exposed when large packets of material were 'knocked away' from the nucleus along the internal layering boundaries, which are discontinuity surfaces and therefore constitute planes of weakness within the material. According to those authors, this removal would have happened either during the gentle collision that lead to the merging of the two lobes, or during the subsequent cycles of split and merging of the cometary body \citep{hirabayashi_2016_fission}.

It is the last type of surface category, the \textbf{exposed consolidated surfaces}, that contain the strongest hints at a layered internal structure of the nucleus. Those surfaces include cliffs (such as \textit{Hathor}) consisting of CCM that are oriented roughly parallel to the local gravity vector. On these cliff faces we find aligned, sub-parallel linear features that run perpendicular to the gravity vector. \citet{thomas_2015_morphological} and others have postulated that these linear features might suggest inner layering, and that indeed \textit{Hathor} shows the inner structure of the Small Lobe, emphasising the importance of broadening out understanding of how the linear features were produced.

\newpage

\section{Layerings} \label{ch_1_3}

I will now briefly introduce the concept of geological layerings, and give some examples of how they are formed on terrestrial planets before summarising the state of research on layerings in cometary nuclei as a basis for my work.

\subsection{Geological background} \label{ch_1_3_1}

\begin{figure} [h] 
\centering
\includegraphics[width=0.7\linewidth]{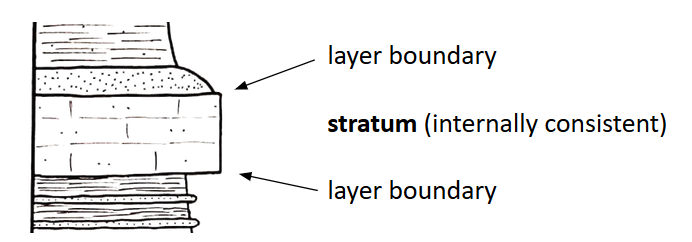}
\caption[Schematic drawing of a stratum and layer boundaries] {Schematic drawing of a stratum bounded by layer boundaries, as it is conventionally presented within a stratigraphic column.}
\label{fig:stratum}
\end{figure}

A geological layering, or 'stratum', is defined as a portion of internally consistent material that is bounded by two layer boundaries or 'stratification planes' (\autoref{fig:stratum}), which are produced by visible changes in the grain size, texture, mineralogy, composition, or other diagnostic features of the material above and below the plane \citep{encyclopaedia_2010_stratum}.

\begin{figure} [h] 
\centering
\includegraphics[width=0.82\linewidth]{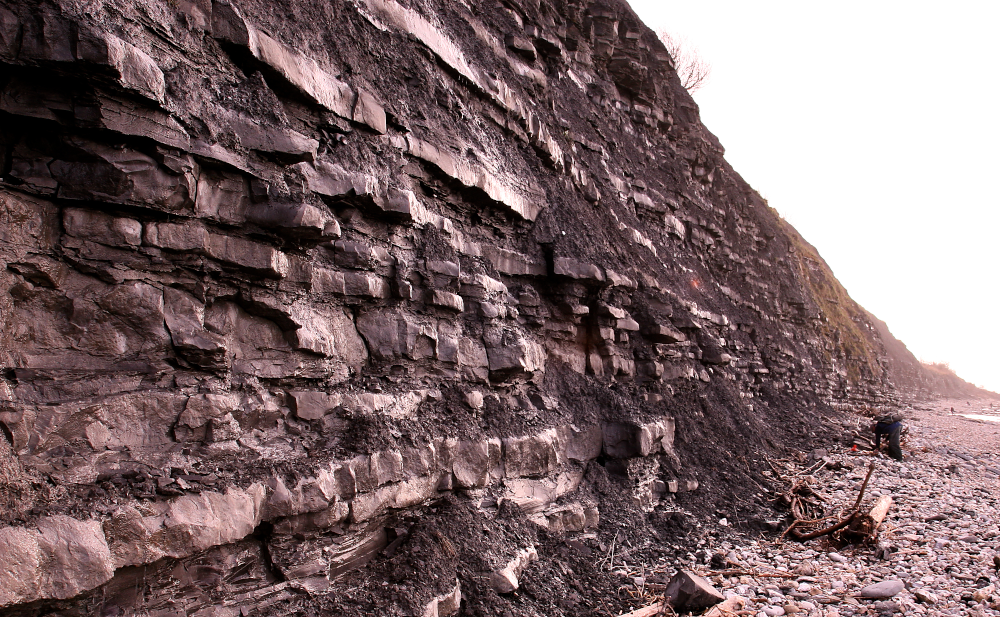}
\caption[Cliff on Earth with prime example of stratification] {Cliff on Earth, showing a prime example of stratification. The cliff is made up of an alternation of softer claystones (more strongly eroded) and harder limestones (more resistant to erosion). It is located at Lyme Regis in the United Kingdom \citep{maggs_2007_lyme} CC BY-SA 2.5, image flipped horizontally.}
\label{fig:strata_real}
\end{figure}

\begin{figure} [h] 
\centering
\includegraphics[width=\linewidth]{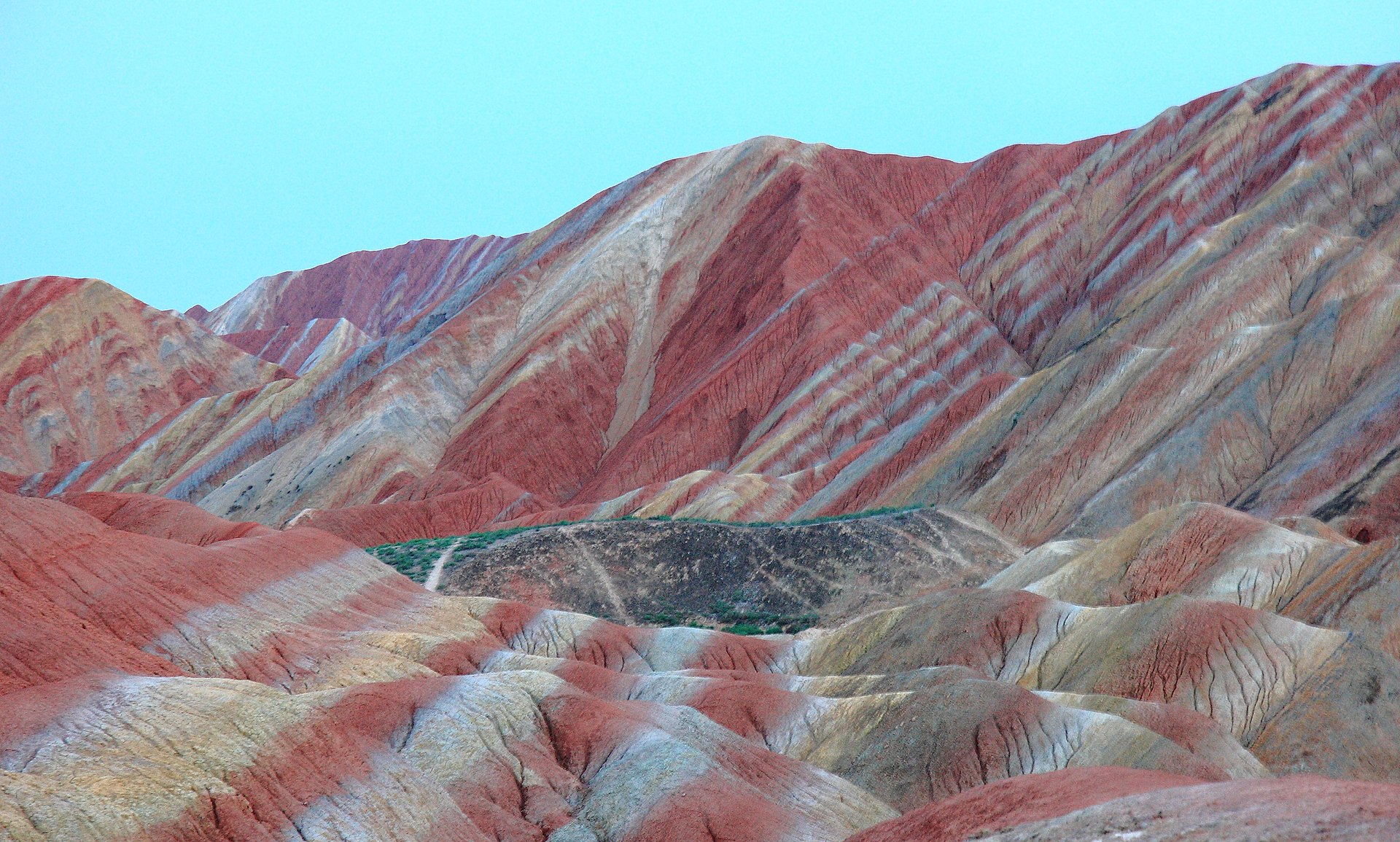}
\caption[Colourful layered sediments at Zhangye-Danxia-Geopark, China] {Colourful layered sediments at Zhangye-Danxia-Geopark, China, illustrating layerings as a primary structure \citep{lei_2013_china} CC BY-SA 3.0  \vspace{0.3cm}}
\label{fig:zhangye}
\end{figure}

On Earth, geological layerings are formed by two general types of mechanisms. The first one is formation by deposition, which results in a so-called \textbf{primary structure}. These kind of layers occur in most sedimentary and igneous rocks, glaciers, as well as other materials that are deposited sequentially. Strata have also been observed on other planets, such as in the rocks and icecaps of Mars. Layering boundaries in primary structures can also result from pauses in deposition that allow the older deposits to undergo changes before additional sediments cover them \citep{encyclopaedia_2010_stratum}. Strata in a sequence of primary layerings may be distinguishable from each other by variations in grain size or colour changes resulting from different mineral composition (\autoref{fig:zhangye}), or consist of material that is otherwise similar but is separated by distinct planes of parting. 

\vspace{0.6cm}

In order to form layerings that extend laterally at constant thickness and have smooth, parallel layer boundaries, steady environmental conditions with little tectonic movement and uniform direction of material transport are required. As such conditions are rare on Earth, parallel layerings are an exception. More frequently, we observe sedimentary structures that convey the dynamics of the depositional environments. Common structures are cross-bedding (which is common in fluvial or eolian deposits) and graded bedding (which reflects transport by currents) \citep{encyclopaedia_2010_stratum}. In turbulent environments, it frequently occurs that an underlying layer is partially or fully removed, mixed with new material, and then deposited as a newer layer.

\newpage

The second type of layerings are created when geological processes affect material that has already been deposited. These layerings are called a \textbf{secondary structure} (formed after the primary deposition). Examples include

\begin{itemize}
    \item stratification in soils, where layers are developed during pedogenesis by biochemical processes and vertical transport \citep{Buol_1973_soil}
    \item mineralogical sintering, where minerals are precipitated out of fluids permeating the material, which may lead to the formation of irregularly spaced sintering crusts within the material \citep{encyclopaedia_2018_sinter} 
    \item thermal sintering within deposits of snow and ice, where warm or hot fluids or other thermal influences melt some of the grains. When the liquid refreezes, it forms a hardened crust within the material, and
    \item foliation textures due to tectonic strain.
    
\end{itemize}{}

A sequence of layerings therefore contains a record of how the depositional or formational environment changed over time.  

\begin{figure} [h] 
\centering
\includegraphics[width=\linewidth]{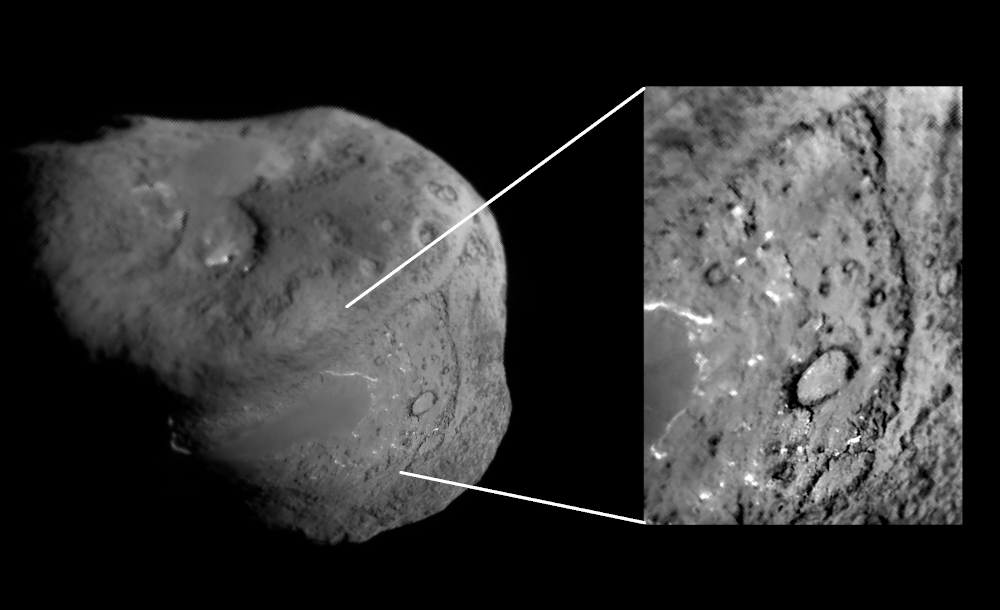}
\caption[The layerings on comet 9P/Tempel 1] {The layerings on comet 9P/Tempel 1. \textbf{Left:} Full view of the nucleus of comet Tempel 1, obtained by the Deep Space mission \citep{nasa_2005_tempel}. Image is in the public domain, modified by rotation for clearer presentation. \textbf{Right:} Close-up view of the area showing the proposed thin layerings, contrast stretched for visibility. \vspace{0.2cm}}
\label{fig:tempel}
\end{figure}

\subsection{Layerings in cometary nuclei} \label{ch_1_3_2}

Both primary and secondary mechanisms, as well as a combination of both have been proposed as the origin of the layerings in cometary nuclei before. It is worth noting that in the field of cometary research, the geological terms 'primary' and 'secondary' are frequently replaced by the synonymous terms 'primordial' and 'evolutionary'.

The first suggestive evidence for layerings was found on images from 19/Borrelly and 81P/Wild 2 \citep{thomas_shape_2007}, although data from those nuclei was limited in its spatial resolution. Much clearer images by the Deep Impact probe to comet 9P/Tempel 1 (\autoref{fig:tempel}) were interpreted to show two distinct manifestations of layerings, both as a substantial stack of layerings that each are several hundreds of metres thick, as well as exposures of strata with an estimated thickness of no more than ten metres in an adjacent region \citep{thomas_shape_2007}. Based on these observations, the 'talps' or 'layered pile' model was formulated in order to explain the formation of these layerings \citep{belton_internal_2007}. The model suggests that the interiors of cometary nuclei consist of a core overlain by a pile of randomly stacked layers, which accumulate as smaller bodies impact the nucleus surface. The impactors are believed to be fragmented and deposited on the surface, where they combine with the impact-ejecta to form a new layer. \citet{belton_internal_2007} upholds that cometary nuclei are primordial remnants of the early agglomeration phase and that the layerings observed on 9P/Tempel 1 must constitute an essential element of the internal structure of all Jupiter family comets.

The layerings on comet 67P were first mentioned in a high-level overview by \citet{thomas_2015_morphological}. \citet{massironi_two_2015} conducted a detailed study of the layerings' orientation by using the orientation of morphological terraces on the nucleus surface as a proxy. They mapped the normal vectors of the terraces on an early shape model of the nucleus, and used the normals to create a series of geologic cross-sections through both its lobes. In this way, they established that the layerings in both lobes of 67P's nucleus are geometrically independent from each other. They proposed that the lobes were formed separately with an 'onion-shell'-type layered inner structure, and at a later point merged into the bilobate shape we see today. These results were later corroborated and refined by \citet{penasa_three_2017} through modelling internal, concentric ellipsoidal shells to the terraces. \autoref{ch_2} describes how I applied that method to layering-related linear features to confirm that the airfall accumulated on top of the terraces did not skew the results of the two aforementioned works. Both \citet{massironi_two_2015} and \citet{penasa_three_2017} presume that the two lobes of comet 67P were formed from the solar nebula as rubble-piles of primordial pebbles. \citet{massironi_two_2015} proposes variations in the relative abundances of volatile materials as the source of the stratification and concludes that stratification is a primary structure within the nucleus material.

It remains a much debated matter whether the cometary nuclei structures as observed today are pristine and preserve a record of their original accumulation, or are a result of later collisional or other processes \citep{jutzi_how_2017}. The key argument against pristine primordial nuclei is that statistically, an object of the size of these nuclei would have experienced a high number of catastrophic collisions since its formation \citep{jutzi_how_2017}, making it exceedingly unlikely that a nucleus like 67P could have survived in its primordial configuration. The main argument against comets being collisionally processed are the physical and chemical properties of the nucleus material, such that its low density, high porosity, weak strength, and high contents of supervolatiles and amorphous water ice rule out an origin as collisional rubble piles \citep{davidsson_primordial_2016}. \citet{fulle_2017_fractal} affirm that the properties of the fractal dust observed by the Micro-Imaging Dust Analysis System (MIDAS) instrument onboard Rosetta preclude that the nucleus has experienced any catastrophic collisions. 

The most recent contribution to this debate was made by \citet{belton_origin_2018}, who presents an alternate, secondary mechanism for the formation of layerings and also requires that the initial nuclei were not extensively collisionally processed. Their model begins at a primordial nucleus containing a high amount of water bound as amorphous ice. During the Centaur orbital phase, the surface is warmed above 115 K where amorphous water ice becomes unstable, initiating a phase-change from to crystalline water ice. The phase-change is exothermic and therefore self-sustaining while it propagates from the surface towards the centre of the nucleus. The proposed front-propagation is bi-modal, with an 'active mode' (moving rapidly, which produces the intra-strata material) and 'quiescent mode' (essentially stationary, which produces the strata boundaries). The globally coordinated strata ('onion shells') observed on comet 67P are be achieved by controlling the direction of phase-change propagation via the radial outflow of CO, as well as the existence of a coarsely layered structure in the primordial material below the front.

It is the aim of this thesis to contribute to the understanding of cometary formation. For this purpose, I used methods of structural geology, statistical image processing, and solar system science on images of comet 67P in order to constrain the geometry and spacing of the layerings on cometary nuclei.

\chapter{Interactive mapping of layering- related linear features on comet 67P}\label{ch_2}

The work presented in this chapter has been published in the 'Monthly Notices of the Royal Astronomical Society' titled "Analysis of layering-related linear features on comet 67P/Churyumov-Gerasimenko" \citep{ruzicka_mnras_2018}. I am the first author of this publication and contributed all research presented in it except for the part described in chapter 2.2, subsection "Ellipsoidal model fitting". I also wrote the entire manuscript except for the aforementioned subsection. The manuscript was accepted for publication after major revisions, which I wrote entirely by myself.

The content of this chapter is identical to text of the publication aside from correcting minor mistakes of spelling and grammar. I adapted the layout to fit the format of this thesis by changing the size and position of figures and tables. \autoref{fig:figure_2_3} was replaced by an identical figure with higher resolution. The publication's 'supplementary materials' are listed in \autoref{app:ch2}.

\section*{Abstract}

We analysed layering-related linear features on the surface of comet 67P/Churyumov-Gerasimenko (67P) to determine the internal configuration of the layerings within the nucleus. We used high-resolution images from the OSIRIS Narrow Angle Camera onboard the Rosetta spacecraft, projected onto the SHAP7 shape model of the nucleus, to map 171 layering-related linear features which we believe to represent terrace margins and strata heads. From these curved lineaments, extending laterally to up to 1925\, m, we extrapolated the subsurface layering planes and their normals. We furthermore fitted the lineaments with concentric ellipsoidal shells, which we compared to the established shell model based on planar terrace features. Our analysis confirms that the layerings on the comet's two lobes are independent from each other. Our data is not compatible with 67P's lobes representing fragments of a much larger layered body. The geometry we determined for the layerings on both lobes supports a concentrically layered, `onion-shell' inner structure of the nucleus. For the big lobe, our results are in close agreement with the established model of a largely undisturbed, regular, concentric inner structure following a generally ellipsoidal configuration. For the small lobe, the parameters of our ellipsoidal shells differ significantly from the established model, suggesting that the internal structure of the small lobe cannot be unambiguously modelled by regular, concentric ellipsoids and could have suffered deformational or evolutional influences. A more complex model is required to represent the actual geometry of the layerings in the small lobe.
\section{Introduction} \label{ch_2_1}

In-situ images of the nucleus surfaces of Jupiter-family comets (e.g., 9P, 81P, 103P) have previously been used to speculate about a layered structure of these nuclei \citep[e.g.,][]{thomas_shape_2007,bruck_syal_geologic_2013,cheng_surface_2013}. \citet{belton_internal_2007} proposed a model in which their interior consists of a core overlain by layerings that were locally and randomly piled onto the core through collisions (the `talps' or `layered pile' model). More recently, \citet{belton_origin_2018} suggested that the layerings could have been formed by fronts of self-sustaining amorphous to crystalline ice phase-change propagating from the nucleus surface to the interior. 

The ongoing debate about the origin of layerings in cometary nuclei might benefit from a more comprehensive understanding of their geometry and orientation. This understanding was greatly improved through the data collected by ESA's Rosetta mission to comet 67P/Churyumov-Gerasimenko (67P). Images taken by the OSIRIS camera system, surpassing the spatial resolution of previous missions by more than an order of magnitude, resolved features exposed across most of the nucleus surface which we interpret as layerings. Repetitive staircase patterns are formed by laterally persistent cliffs separating planar `terrace' surfaces. The cliff faces display parallel linear grooves, which are reminiscent of sedimentary outcrops on Earth where differential erosion carves such grooves into layerings of alternating hardness. The dust-free walls of deep pits reveal quasi-parallel sets of lineaments, oriented roughly perpendicular to the local gravity vector and extending to depths of at least a hundred meters below the present-day nucleus surface \citep{vincent_large_2015}.

A first systematic analysis of the orientation of 67P's layerings was conducted by \citet{massironi_two_2015}. They created a series of geologic cross-sections of the comet nucleus based on the orientation of planes fitted to morphologically flat areas (`terraces') on a shape model of the nucleus surface. \citet{massironi_two_2015} concluded that the two lobes of the nucleus are independently-formed bodies with an `onion-shell' layered inner structure that formed before the two bodies merged in a gentle collision to form the nucleus of 67P. 

Using a similar approach, albeit on a much smaller number of measurements, \citet{rickman_comet_2015} suggested that morphological ridges and other features on opposing sides of the comet's lobes can be connected by planar features. They interpreted this correlation as evidence for a semi-planar, pervasive internal layering. This would suggest that the two lobes of 67P are fragments of a much larger body.

\citet{penasa_three_2017} modelled the inner layered structure of comet 67P by fitting concentric ellipsoidal shells to a total of 483 terraces on both lobes, providing a first simplified 3D geological model of their inner structure. By comparing the orientation of the surface planes with the two model ellipsoids, they suggest that the inner structure of the two lobes can be explained by a set of concentric ellipsoids.

\begin{figure}[h]
	\centering
	\includegraphics[width=0.65\linewidth]{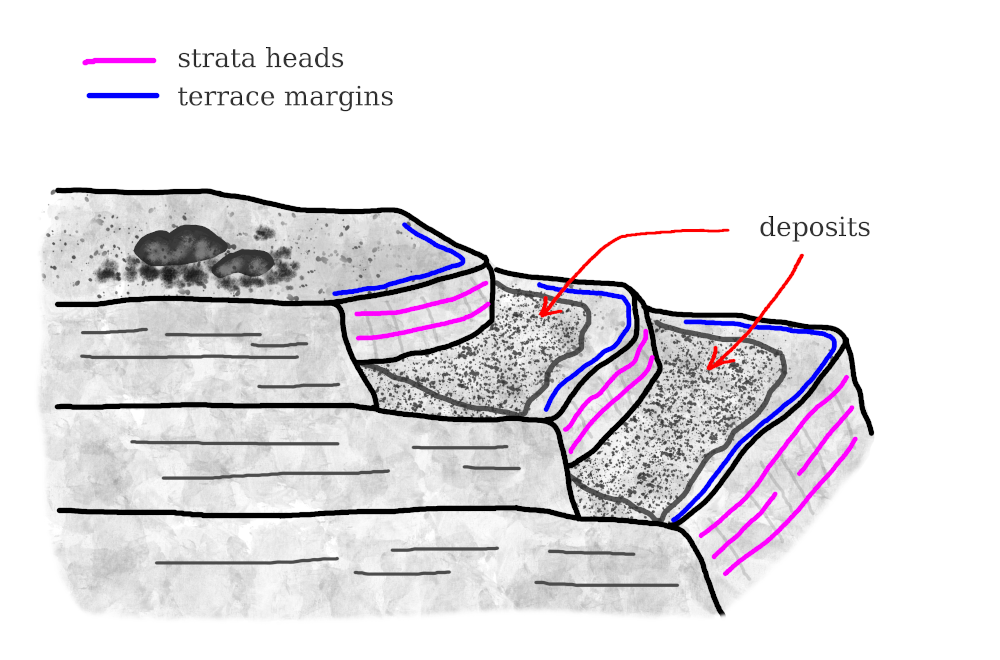}
	\caption[The two types of layering-related linear features analysed in this chapter: Terrace margins and strata heads]
	{Schematic illustration of the two types of layering-related linear features we analysed in this paper: i) morphological terrace margins (red curves) and ii) strata heads visible as lineaments on cliff faces and hill slopes (blue curves). Bedding orientations derived from these linear features are not affected by deposits on the terraces.}
	\label{fig:figure_2_1}
\end{figure}

The studies by \citet{massironi_two_2015}, \citet{rickman_comet_2015}, and \citet{penasa_three_2017} made use of terraces as a proxy for the underlying orientation of the layers. The advantage of such an approach is that terraces are ubiquitous on the cometary body, that they can easily be mapped, and that their orientation can be easily estimated by means of a best fitting procedure applied to the vertices of a shape model \citep{massironi_two_2015}. A downside of the terrace approach is that results may be biased due to airfall and mass wasting processes from nearby cliffs (\autoref{fig:figure_2_1}). \citet{penasa_three_2017} acknowledge this limitation and estimate that this might introduce an error of up to $\sim$20\textdegree{} to their normals. 

Here we use layering-related linear features instead of terraces to avoid possible bias due to depositional processes. We studied two types of linear features (\autoref{fig:figure_2_1}): 

i) morphological edges along terrace margins, adjacent to cliffs (referred to as `terrace margins') and 

ii) lineaments on hill slopes and cliff faces \citep[referred to as `strata heads', in accordance with][]{lee_geomorphological_2016} produced by the intersection of the layers with the topographic surface. Both types of features appear to be erosional consequences of discontinuities between the individual layerings, which are possibly related to planes of different physical properties within layered materials. The linear traces mark the locations where the inner bedding planes intersect with the topography at a high angle \citep{massironi_two_2015}.

\section{Data and Methods} \label{ch_2_2}

\subsection*{Data base}

For the three-dimensional representation of the nucleus of 67P we used the SHAP7 shape model, which is obtained from a stereo-photogrammetric (SPG) analysis of images taken by the OSIRIS Narrow Angle Camera \citep[NAC;][]{keller_osiris_2007} onboard Rosetta. The model covers the whole nucleus and consists of about 44 million facets, has a mean accuracy of 0.3\,m at a horizontal sampling of about 1-1.5\,m, and vertical accuracy at the decimetre scale \citep{preusker_global_2017}. 

Most of our mapping was conducted on high-resolution images of the nucleus surface. We selected suitable OSIRIS NAC images, publicly available and retrieved from the ESA Planetary Science Archive \citep[''PSA'',][]{besse_psa_2018}, according to these criteria: Images taken with the OSIRIS NAC orange filter F22 (which have a high signal-to-noise ratio); images taken at a spacecraft-comet distance of <\,30\,km (resulting in a pixel resolution of 0.5 to 1.5\,m at the image centre); images where the illumination is both sufficiently bright and also in a suitable direction to show layering-related features in good contrast. Initially, we selected only images that were calibrated for geometrical distortion due to the internal camera geometry \citep[CODMAC level 4;][]{tubiana_scientific_2015}. We later decided to supplement these with geometrically uncalibrated images (level 3) in order to improve coverage of the nucleus. We minimised the impact of this distortion on our data by restricting our mapping efforts to the central area of uncalibrated images, where the distortion decreases to near-zero.

\subsection*{Mapping of linear features}

Large-scale morphological terrace margins were mapped directly on the shape model by manually tracing the features as polylines in \textsc{CloudCompare} \citep{girardeau-montaut_2014_cloudcompare}.  

Mapping the finer morphological details of strata heads required a higher resolved mapping medium, for which we projected two-dimensional OSIRIS images onto the three-dimensional shape model. We used a customised version of the software \textsc{philae localisation workshop} \citep[PLW,][]{remetean_philae_2016} for the measurement process of the linear features. For each image, we manually selected a set of reference points, consisting of prominent landmarks visible on both the image and the shape model (e.g., large boulders, sharp corners), and used the software to spatially align the image with the shape model by minimising the distance between corresponding reference points. Using a set of 20 reference points we achieved a root-mean square error (RMSE) between 3.1 and 9.9\,m (7.0\,m on average) for the alignment, depending on how many clearly defined landmarks are visible on the image that is to be aligned. Increasing the number of reference points did not further improve the RMSE. The point of view onto the projected image can then be changed in 3D to allow mapping from an optimum viewing angle.

Subsequently, we manually selected nodes along each feature of interest. The nodes each have three-dimensional XYZ point-coordinates in the comet-fixed Cheops reference frame. For ease of handling we exported the nodes as one polyline per feature. To aid visualisation of the layering orientations, we determined best matching local plane solutions for each linear feature. The plane solutions consist of the normal vector ($\bm{n}$) and the reference point of the plane ($\bm{nb}$), which is the centroid of the nodes used to fit the plane. It also corresponds to the location of the base of $\bm{n}$. We found these plane solutions by applying a weighted least squares plane-fitting routine to the nodes of the feature \citep[Planefit,][]{schmidt_planefit_2012} in \textsc{matlab} \citep[release 2017a]{matlab_2017}). 

We assessed the uncertainty of the plane normal vectors $\bm{n}$ by means of a Monte Carlo simulation: The values of coordinates $X$, $Y$, and $Z$ of each node were varied, by a normally-distributed random value within the error of the nodes, around the measured coordinates $X_m$, $Y_m$, and $Z_m$. Those planes where $\bm{n}$ was poorly defined (variance of Monte Carlo results in at least one direction of $\geq$\,90\textdegree) were discarded from the pool of mapped features.

\subsection*{Ellipsoidal model fitting}

To evaluate whether linear features can be used as a stand-alone product to produce three-dimensional geological models, we made use of the same model defined by \citet{penasa_three_2017}. The model represents the layering of each lobe as a scalar field:
\begin{equation}
f(c_x,c_y,c_z) = k
\end{equation}
Function $f()$ is defined such that for any constant value of $k$ a single contour surface with ellipsoidal shape is determined, while the value of the scalar field represents a metric in the stratigraphic space ($0$ in the centre of the ellipsoidal model and increasing toward the outermost layers).

Function $f()$ is completely defined by eight parameters: $c_x, c_y$, and $c_z$ for the centre of the ellipsoidal model, $b$ and $c$ for the axial ratios and finally $\alpha, \beta$, and $\gamma$ for orienting the concentric ellipsoids in space. The parameters can be determined by maximising the accordance of the model with the provided constraints. The orientation of the terraces, were used by \citet{penasa_three_2017} to provide observations of the gradient of the function $f()$ in a specific point $\bm{p}$, thus providing a modelling constraint of the type:
\begin{equation}
\label{eq:constraint_grad}
\nabla f(\boldsymbol{p})  = \boldsymbol{n} 
\end{equation}
where $\bm{n}$ is the normal to the surface element located in the point $\bm{p}$.

In this work we instead tested the use of polylines for modelling purposes. Each polyline is formed by segments which are expected to lie on a contour surface of the model and are thus tangential to the ellipsoidal shell passing for that location. Each segment can be defined by a pair of points $\bm{p}_1$ and $\bm{p}_2$ describing a direction in space, which can be represented as a unit vector:
\begin{equation}
\label{eq:segment}
\boldsymbol{\hat {t}} =\dfrac{\boldsymbol{p}_2 - \boldsymbol{p}_1}{\|\boldsymbol{p}_2 - \boldsymbol{p}_1\|} 
\end{equation}
where $\boldsymbol{\hat {t}}$ represents a constraint of tangent type \citep[e.g.][on this subject]{hillier_2014_threedimensional}:
\begin{equation}
\label{eq:constraint}
\left\langle\nabla f(\boldsymbol{p}), \boldsymbol{t} \right\rangle  = 0
\end{equation}
where $\bm{p}$ is the location of the observation (i.e. a reference point for the location of the segment). From these observations an angular misfit of the observation in respect to any model can be obtained by computing:
\begin{equation}
\label{eq:angle}
\vartheta =  \arccos\left(\left\Vert\dfrac{\nabla f(\boldsymbol{p})}{\| \nabla f(\boldsymbol{p}) \|} \times \boldsymbol{\hat t}\right\Vert\right)
\end{equation}

By minimising the squares of the angles provided by \autoref{eq:angle} for each segment composing the mapped polylines, we were able to determine the most-likely parameters for the ellipsoidal model approximating the observations in this work. We employed a weighting strategy to ensure that each polyline contributes equally to the obtained solution. For this we divided each squared residual by the total number of segments of the specific polyline. Finally, we used a bootstrap strategy, based on the resampling of the polylines, to estimate the standard errors associated with each parameter.
\section{Results} \label{ch_2_3}

We used the PLW software to align a total of 34 OSIRIS images (\textbf{Table A1} in the supplementary material), covering most of the nucleus surface. On these images we mapped 171 linear features, of which 31 are terrace margins and 140 are strata heads. The mapped features are distributed approximately evenly between 67P's big and small lobe. The features extend laterally for 863\,m on average, ranging from 185 to 1925\,m. The feature length is calculated from the cumulative length of all segments connecting the polyline nodes. Most features contain between 9 and 20 segments (14 on average), and the average segment length is 38\,m. 

The uncertainty of each node results cumulatively from the error of the shape model ($\pm$ 0.3\,m), the resolution of the images (ca. 0.2\,\,m/px on average), and the image-alignment uncertainty in the PLW software (7\,m on average). Considering that the image resolution, and the error introduced by the alignment procedure, vary depending on the observation geometry and the distance from the camera, the overall uncertainty of each mapped node cannot be precisely quantified. However, based on the aforementioned considerations, it might be expected to be <\,10\,m.

The Monte Carlo error analysis showed that affecting the node coordinates $X$, $Y$, and $Z$ by random amounts between zero and 10\,m results in a variance of the normals $\bm{n}$ by less than 10\textdegree{} for most features (\textbf{Figure A1} in the supplementary material). For a small number of features, whose polylines have a low curvature, $\bm{n}$ is more strongly affected. As expected in such cases, the uncertainties show significant directional asymmetry and have large amplitudes only perpendicular to the main extension of the lineament. As shown in \textbf{Figure A1(B)}, these cases are the exception in our data, and the node uncertainty does not have a substantial effect on the bulk of the layering orientations we reconstructed from the linear features.

\begin{figure}
	\centering
	\includegraphics[width=0.95\linewidth]{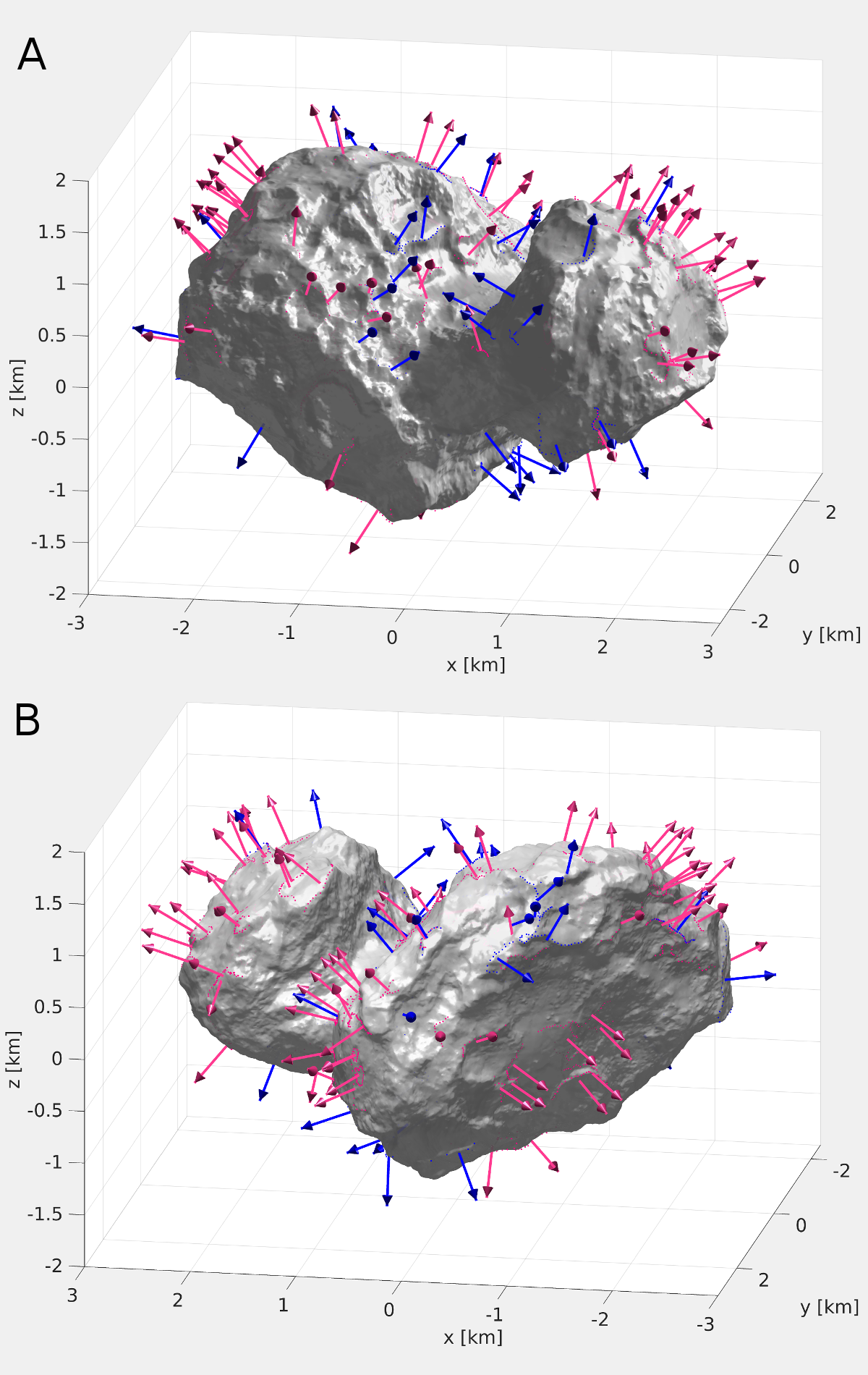}
	\caption[Distribution and orientation of feature normals, shown on the SHAP7 shape model]
	{The SHAP7 shape model of comet 67P, showing the feature normals $\bm{n}$ at the bases $\bm{nb}$ for all mapped terrace margins (blue arrows) and strata heads (pink arrows). The dotted lines represent the mapped nodes along each feature. Coordinates are in the Cheops reference frame. \textbf{A:} `Front' view towards positive y-values; \textbf{B:} `Back' view rotated around the z-axis by 180\textdegree. Full-resolution images are available in the supplementary materials.}
	\label{fig:figure_2_2}
\end{figure}

An exemplary excerpt of the parameters of the plane solutions to the mapped features are listed in \autoref{tab:table1} (the complete table is available in the supplementary material). The 3D orientation of vectors $\bm{n}$ is illustrated in \autoref{fig:figure_2_2}. Qualitatively, by visual impression the normals are oriented perpendicular to the nucleus surface. For the big and small lobe separately, the normals are pointing outward from the respective lobe's gravitational centre.

For the purpose of comparison, we fitted our own set of ellipsoids to the polylines of our linear features. The parameters we obtained for the best-fitting ellipsoidal model are summarised in \autoref{tab:table2}, next to the parameters achieved by \citet{penasa_three_2017}. The parameters are consistent for the big lobe within the achieved uncertainties, but we found a notable a misfit of the results for the small lobe. Our small lobe model is offset by 0.32\,km from their model (which amounts to ca. 40\% of the lobe's semi-minor axis, according to \citet{jorda_global_2016}), there is a minor difference in the ellipsoidal axis ratio, and a major mismatch in the rotational angles.

\begin{figure}
	\centering
	\includegraphics[width=\linewidth]{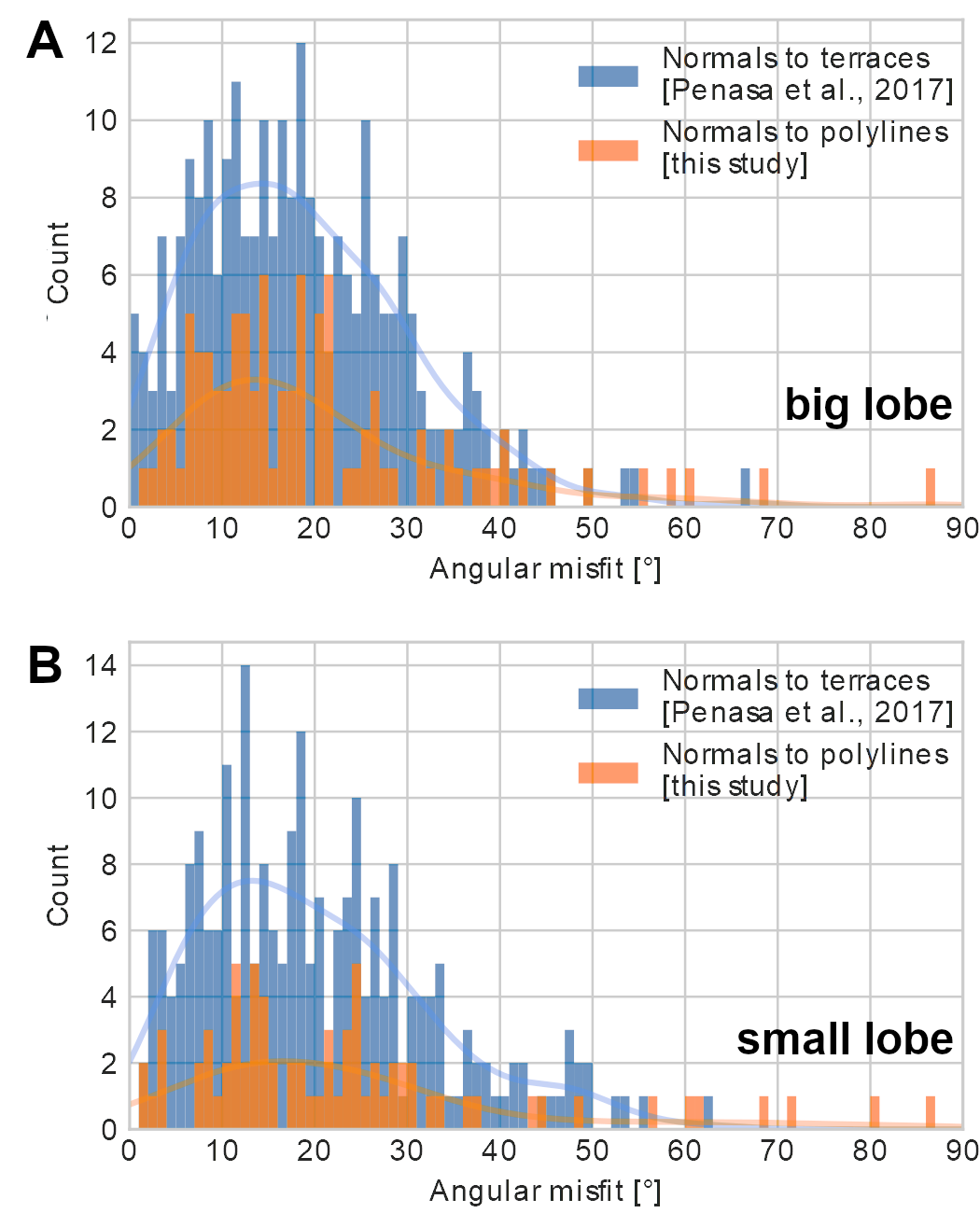}
	\caption[Histograms showing the angular misfit between the feature normals, and the normal vectors to the ellipsoid model in the same location]
	{Histogram of the angular misfit between normal vectors $\bm{n}$ to the surface features and corresponding normal vectors $\bm{n_{ell}}$ of the ellipsoid model. The misfit for our linear features is plotted in orange, the misfit for the terraces in \citet{penasa_three_2017} is plotted in blue. \textbf{A:} Features on 67P's big lobe, \textbf{B:} Features on the small lobe.}
	\label{fig:figure_2_3}
\end{figure}

Finally, we compared the orientation of the plane normals $\bm{n}$ to the orientation of the ellipsoid surface for both \citet{penasa_three_2017} and our model. \autoref{fig:figure_2_3} shows the angular misfit between $\bm{n}$ and the corresponding normal to the ellipsoid surface $\bm{n_{ell}}$, both at location $\bm{nb}$. Again, we find an overall low angular misfit for the big lobe (median 16.7\textdegree{}, panel A), and a larger misfit for the small lobe (median 19.6\textdegree{} with larger percentiles, panel B), indicating that the linear features are less congruent with the ellipsoid on the small lobe. \citet{penasa_three_2017} did not observe this dichotomy.

\begin{table} [h]
	\centering
	\caption[Exemplary excerpt of the normal vectors $\bm{n}$ and their bases $\bm{nb}$]{Exemplary excerpt of the normal vectors $\bm{n}$ and their bases $\bm{nb}$ to the plane solutions of the mapped strata heads on the big lobe. r is the radial distance of each vector base from the gravitational centre of the lobe. Values are in [km] in the comet-fixed Cheops reference frame. The complete tables are available in the supplementary material.}
	\label{tab:table1}
	\begin{tabular}{rrrrrrr}
		\hline
		$\bm{nb_x}$ & $\bm{nb_y}$ & $\bm{nb_z}$ & $\bm{n_x}$ & $\bm{n_y}$ & $\bm{n_z}$ & r\\
		\hline
		0.316 & 0.695 & 0.616 & 0.632 & -0.425 & 0.648 & 1.3 \\
		-0.311 & 0.740 & 0.696 & 0.840 & -0.117 & 0.530 & 1.4 \\
		0.473 & 0.981 & -1.320 & 0.013 & 0.131 & -0.991 & 1.9 \\
		0.489 & 0.536 & -1.257 & 0.919 & 0.025 & -0.393 & 1.7 \\
		0.290 & -0.074 & -1.237 & 0.776 & -0.084 & -0.625 & 1.5 \\
		... & ... & ... & ... & ... & ... & ... \\
		\hline
	\end{tabular}
\end{table} 

\begin{table*} [h]
	\centering
	\caption[Maximum likelihood estimates of the ellipsoidal model parameters]{Maximum likelihood estimates of the ellipsoidal model parameters and relative 2$\sigma$ errors, modelled to our data and compared to \citet{penasa_three_2017}. Distances in km in the Cheops reference frame, and Tait-Bryan angles in degrees. BL means 67P's big lobe, SL is the small lobe; $b$ and $c$ are the axial ratios with respect to the a-axis ($a$\,=\,1).Values for which this study differs from \citet{penasa_three_2017} in excess of the uncertainties are highlighted in bold in the table.}
	\label{tab:table2}
	\begin{tabular}{lrrrrrrrrr} 
		\hline
		Parameter & \multicolumn{4}{l}{This study (171 linear features)} & & \multicolumn{4}{l}{\citet{penasa_three_2017} (483 terraces)} \\
		&  \  BL \ & \ 2$\sigma$ \ & \ SL \ & \ 2$\sigma$ \ & 
		&  \  BL \ & \ 2$\sigma$ \ & \ SL \ & \ 2$\sigma$ \ \\
		\hline 
		$c_x$   & \ -0.55 \ & \ 0.12 \ & \ \textbf{1.35} \ & \ 0.09 \ & & \ -0.47 \ & \ 0.08 \ & \ 1.06 \ & \ 0.13 \ \\
		$c_y$   & \ 0.20 \ & \ 0.13 \ & \ -0.40 \ & \ 0.10 \ & & \ 0.32 \ & \ 0.08 \ & \ -0.35 \ & \ 0.07 \ \\
		$c_z$   & \ -0.15 \ & \ 0.11 \ & \textbf{0.13} \ & \ 0.08 \ & & \ -0.17 \ & \ 0.07 \ & \ 0.01 \ & \ 0.06 \ \\[1ex]
		b 	    & \ 0.80 \ & \ 0.08 \ & \ \textbf{0.85} \ & \ 0.08 \ & & \ 0.81 \ & \ 0.04 \ & \ 0.76 \ & \ 0.07 \ \\
		c  	    & \ 0.48 \ & \ 0.04 \ & \ 0.71 \ & \ 0.08 \ & & \ 0.55 \ & \ 0.03 \ & \ 0.70 \ & \ 0.07 \ \\[1ex]
		$\alpha$& \ 47.6 \ & \ 6.5 \ & \ \textbf{55.2} \ & 9.3 \ & & \ 44.8 \ & \ 4.3 \ & \ 28.1 \ & \ 9.3 \ \\
		$\beta$ & \ 7.3 \ & \ 10.7 \ & \ 4.2 \ & 18.0 \ & & \ 15.0 \ & \ 6.7 \ & \ -11.2 \ & \ 5.7 \ \\
		$\gamma$& \ 63.2 \ & \ 4.1 \ & \ \textbf{71.5} \ & 13.8 \ & & \ 66.3 \ & \ 3.9 \ & \ -7.3 \ & \ 34.4 \ \\
		\hline
		
	\end{tabular}
\end{table*} 

\newpage
\section{Discussion and Conclusions} \label{ch_2_4}

While the layering-related linear features are not affected by sedimentation, we cannot rule out that the subset of terrace margins might be influenced by erosion and cliff collapse \citep[cf. e.g.][]{pajola_pristine_2017}. However, we took care to minimise this effect by accurately following the external border of the mapped edges, including any niches resulting from local breakoffs, thus preserving the orientation of the underlying layering.

The orientation of our feature normals $\bm{n}$  particularly in the `neck' region of the nucleus (\autoref{fig:figure_2_2}), supports the widely accepted findings of previous works that the orientation of the layerings on the big and small lobe of 67P are independent from each other \citep[e.g.][]{massironi_two_2015, davidsson_primordial_2016}; it does not support a common envelope structure surrounding both lobes, nor the interpretation that both lobes represent fragments of a much larger, layered body \citep[as proposed by][]{rickman_comet_2015}.

We find that the internal structure of the nucleus, as deduced from the orientation of layerings mapped at the surface, is in agreement with the `onion-shell' model proposed by \citet{massironi_two_2015} and the concentric ellipsoidal shell model by \citet{penasa_three_2017}. Particularly for the big lobe, our results match those of \citet{penasa_three_2017} closely. We understand this as confirmation that the big lobe has a regular, concentric inner structure that follows a generally ellipsoidal configuration. Nevertheless, our data cannot confirm that the layerings are indeed connected into globally coherent shells. Approximating the minimal lobe circumferences as 6\,km and 8.6\,km, respectively, our measurements (including polylines of almost 2\,km length) might be mistaken to span a substantial portion of the nucleus surface. However, most of our polylines intentionally have a strong curvature and represent features with a continuous lateral extent of no more than a few hundred metres. This leaves room for the possibility of a discontinuously layered structure, as would be a consequence of e.g. the `talps' model \citep{belton_internal_2007} or layering by thermal processes \citep{belton_origin_2018}. Our findings would be also compatible with either concept.

For the small lobe, our results differ significantly from those of the other authors. Neither are the orientations of our proposed layerings clearly compatible with the ellipsoidal shell model based on its terraces, nor do the parameters of our own ellipsoidal model agree with those of the terrace-based model. This disagreement could either be explained by the circumstance that \citet{penasa_three_2017} mapped their terraces exclusively on the shape model, and thereby might have included some planar areas that are unrelated to the layerings. Another possible explanation is that the small lobe's inner structure has been affected by processes of evolution or deformation, and thus cannot be unambiguously modelled by regular, concentric ellipsoids. In this case, a more complex model is required to represent the real geometry of the layerings.

\chapter{Fourier-based detection of layering- related lineaments on comet 67P}\label{ch_3}

\section{Introduction} \label{ch_3_1}

Traditionally, the creation of geological maps involves physically visiting the target area, observing features in the field, and noting their location and orientation (strike and dip) on a topographic base map for later analysis. Today, the mapping is increasingly done remotely from digital photographs, aircraft or spacecraft imagery, extending the scope and speed at which maps can be produced. In addition to mapping these images manually (using geospatial information software (GIS), or a customised approach like the one described in \autoref{ch_2}), a trend exists towards automating the mapping process. In this context, 'automated' means using a software or an algorithm that, once set up, conducts parts or all of the mapping process with minimal involvement of a human.

There are several advantages to automated mapping: Once the process is set up (i.e. the coding is done), it saves a lot of time because the algorithm takes mere minutes to run, is easily repeatable with different parameters, and is applicable to most types of images with minor adjustments to customise the code for each context.

Manual mapping generally results in a map that is biased, mainly by the scientist's previous experience and interpretation, but also by the normal daily fluctuations in attention from time of day. If more than one human works on a map, inconsistencies are bound to arise. Manual mapping is particularly vulnerable to 'confirmation bias', i.e. mapping what one expects or wants to see rather than what is objectively visible. All of this can be improved or solved by automating parts or all of the mapping. Additionally, an algorithm can be programmed to work at pixel- or sub-pixel-resolution, which means that it can pick up much finer details than a human eye in the same images.

The goal in this work was to develop an approach that works with as little human intervention as possible, provides the layout of the lineaments on the mapping area, and analyses their structural properties such as their orientation and spacing. Ideally, such a process consists of an algorithm that receives an image, detects the features of interest (i.e. lineaments), and produces a file containing their locations and properties to be overlaid on the original image to create the visual map. Analogously to the approach described in \autoref{ch_2_2}, this file would then be exported for further analysis. 

In an ideal-case, fully automated mapping process, the critical step is the automated detection of lineaments, which is commonly approached by detecting discontinuities in intensity values by using edge detection algorithms. These algorithms work by using derivatives \citep{gonzalez_2004_digital} such as the gradient vector. For a two-dimensional image function $ f(x,y) $ it is defined as

\begin{equation}
\nabla \textbf{f} = 
\begin{bmatrix}
G_x\\ 
G_y
\end{bmatrix} = 
\begin{bmatrix} 
\frac{\partial f}{\partial x}\\
\frac{\partial f}{\partial y}
\end{bmatrix}
\end{equation}

and points in the direction of the maximum rate of change of $ f $ at the coordinates $ (x,y) $. Its magnitude is zero in areas of constant intensity and changes proportionally to the degree of intensity change in an area. An edge point is a point whose intensity is a local maximum in the direction of the gradient  \citep{gonzalez_2004_digital}. The directional gradient components for each part of the image are estimated by applying a filter mask, which is also called an 'edge operator' \citep{burger_2010_principles}. From the components, the strength and local direction of an edge can be computed.

\begin{figure} [h]	
	\centering
	\includegraphics[width=\linewidth]{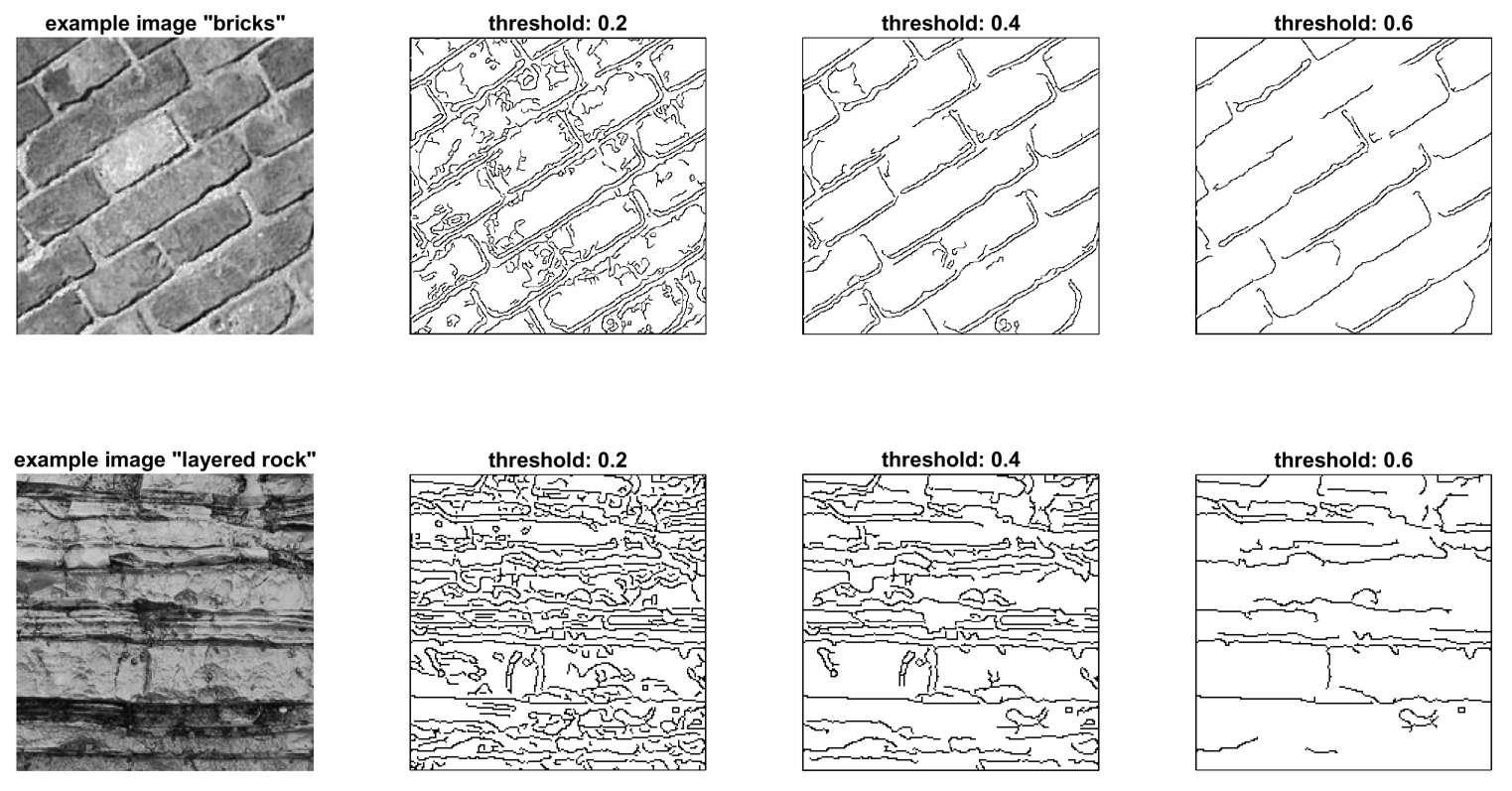}
	\caption[Applying edge detection to images containing linear structures]
	{Example of applying edge detection to images containing linear structures (using MATLAB function 'edge' with 'Canny' algorithm and three different, manually set thresholds). Output images are inverted for visibility.}
	\label{fig:edgedetectiondemo}
\end{figure}

A powerful edge operator is the Canny detector \citep{canny_1987}. It pre-processes the image by smoothing it with a Gaussian filter to reduce noise, before computing the local gradient and edge direction at each point to find edge points, which together form ridges. Using the Canny edge detector in MATLAB (contained in the image processing toolbox), ridge pixels are thresholded either automatically or with user-given threshold values. The algorithm then sets all pixels to zero that are not on the top of these ridges (a process known as \textit{nonmaximal suppression}) and performs edge-linking to produce an output of a thin white line on a black background \citep{gonzalez_2004_digital}. An example of how Canny edge detection might be applied to images containing layerings is shown in \autoref{fig:edgedetectiondemo}.

\newpage

\begin{figure}	
	\centering
	\includegraphics[width=\linewidth]{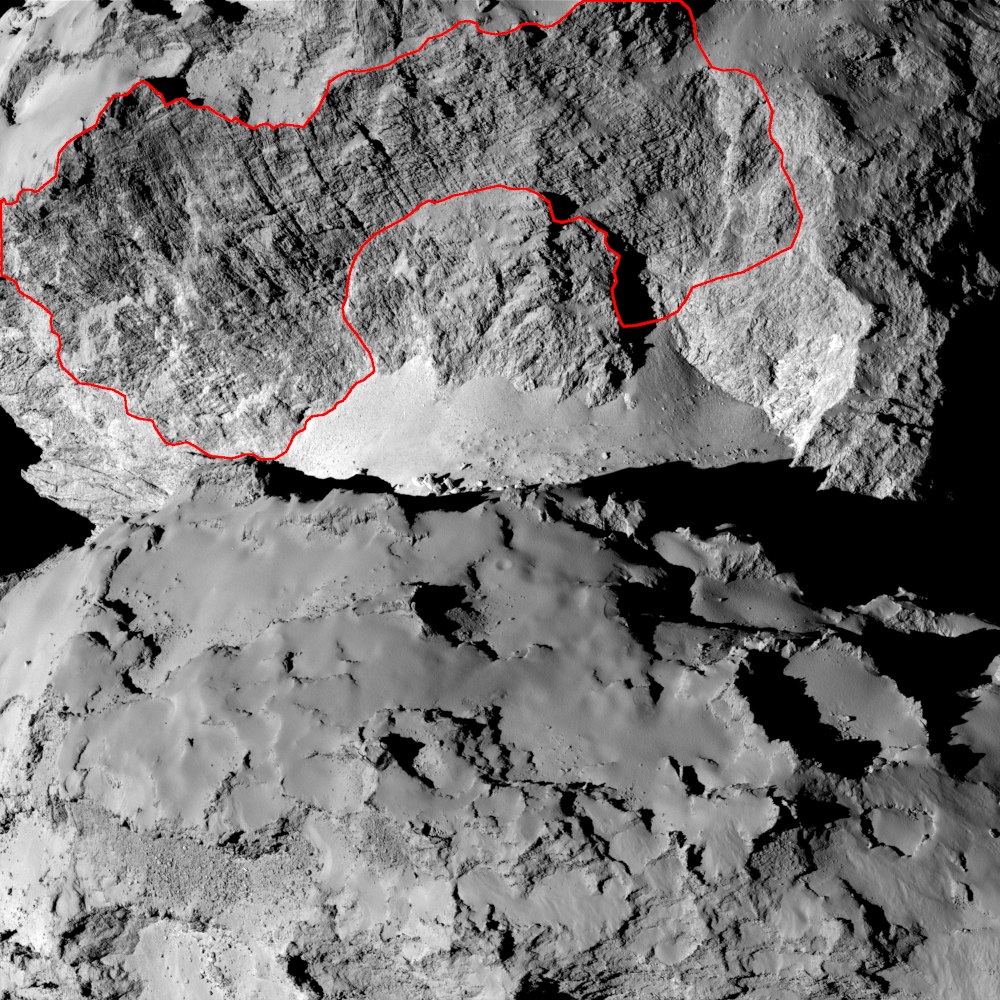}
	\caption[Location of the Hathor wall on the Small Lobe of comet 67P]
	{Location of the Hathor wall on the side of the Small Lobe of comet 67P that is facing the neck. The red line marks an area where the comet's internal structure seems to be exposed. OSIRIS image ID: N20140828T124254563ID30F22. Modified from ESA/Rosetta/MPS for OSIRIS Team MPS/UPD/LAM/IAA/SSO/INTA/UPM/DASP/IDA.}
	\label{fig:hathor_trace}
\end{figure}

On comet 67P, the Hathor wall is a promising target for high-resolution automated mapping. Hathor is a 900\,m high and almost 2000\,m wide cliff located on the Small Lobe (\autoref{fig:hathor_trace}), and was likely created by a hang collapse followed by a large landslide event \citep[e.g.][]{basilevsky_geologic_2017}. Hang collapses move a lot of material in a short amount of time, exposing a fresh view into the inner structures of the hang. For an example of a hang collapse observed during the Rosetta mission, cf. \citet{pajola_pristine_2017}. As they have not been exposed to space weathering for long, the comet's cliff faces are among its most pristine surfaces. Their steep angle relative to the local gravity vector also prevents the collection of dust from airfall on the cliff face, which might smoothen the morphology. 

\begin{figure} [h]	
	\centering
	\includegraphics[width=\linewidth]{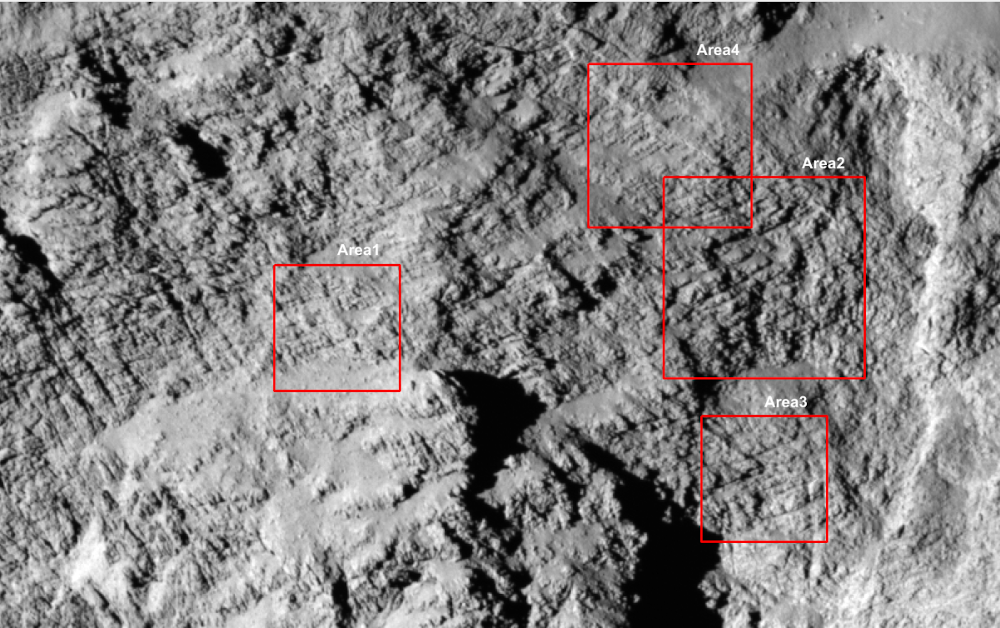}
	\caption[Locations of the example areas used in this chapter]
	{A cropped version of the OSIRIS image from \autoref{fig:hathor_trace}, showing the locations of the example areas used in this chapter. These areas mark exemplary patches of the Hathor wall where layering-associated lineaments are exposed particularly clearly.}
	\label{fig:areas}
\end{figure}

The Hathor wall shows two sets of lineaments that are roughly perpendicular to each other: one set of downslope lines, and a second set of subhorizontal features. The downslope lineaments are interpreted as shadows cast by narrow furrows that might be scars produced in the landslide \citep{basilevsky_geologic_2017}, or an expression of pre-existing vertical jointing in the material. The subhorizontal lineaments are interpreted as expressions of the comet's internal layerings \citep[e.g.][]{belton_origin_2018}. A third direction that needs to be considered in this image is the direction of sunlight, which causes additional shadows that are unrelated to the two sets of lineaments. \autoref{fig:areas} highlights some areas where layerings are exposed particularly clearly on Hathor; Areas 3 and 4 will be referenced as examples throughout this chapter.

\begin{figure}	
	\centering
	\includegraphics[width=\linewidth]{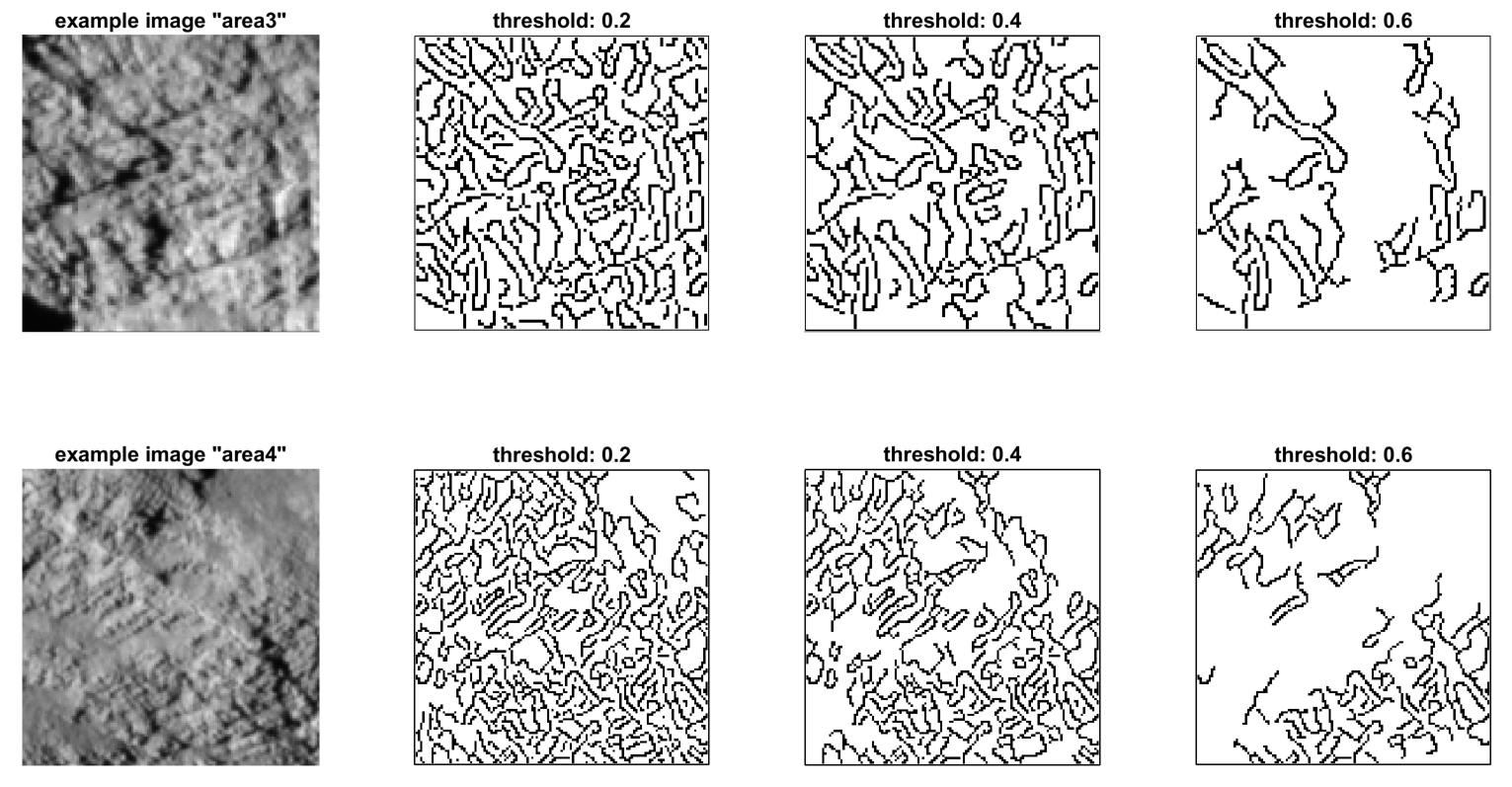}
	\caption[Unsuccessful attempt at edge-detection on areas of an OSIRIS image]
	{Unsuccessful attempt at edge-detection on example areas of an OSIRIS image (using MATLAB function 'edge' with 'Canny' algorithm). }
	\label{fig:edgedetectionexample}
\end{figure}

Unfortunately, not every type of image is suitable for fully automated mapping via edge detection. For geoscientific applications, favourable image properties include sufficient spatial resolution, a suitable viewing angle of the target area, supportive lighting conditions, surface texture and topography (e.g. coverage with boulders or vegetation), and a surface in relatively fresh condition (e.g. not too blurred by weathering or hang collapses). On the Hathor wall, lighting conditions (which cast harsh shadows and produce strong contrast), and the knobby 'goosebumps' texture of the cliff introduce a large amount of visual noise into the OSIRIS images. This results in an inconsistent brightness distribution such that layers which are perceived as cohesive by a human are not recognised as such by the edge detection algorithm. At any threshold, the algorithm produces a segmented, chaotic pattern instead of a systematic layering map (\autoref{fig:edgedetectionexample}).

Even if fragmentation of the layerings was not a problem, the variable illumination across the width of the image would require using different binarisation thresholds across the image, so the algorithm would have to be adaptive and recursive. Therefore, unassisted automated detection of lineaments on the OSIRIS images fails due to intrinsic image properties.

Nevertheless, the images proved suitable for a more statistical approach. I developed an algorithm that reliably finds dominant orientation(s) of linear features in an image by analysing its Fourier domain. The Fourier domain of an image, also called the 'frequency domain', is a complete representation of the amplitudes and frequencies that make up the 2D brightness distribution in the image; as the features of interest are sub-parallel repetitive lines, it is rather straightforward to get the location of layerings, their extent, orientation, and statistical information about their spacing from analysing the Fourier domain. Conversely, Fourier analysis can be used to curb the over-interpretation of structures by the brain, as a signal that is not contained in the frequency domain is not unambiguously contained in the image.

This method, which will be described in detail in \autoref{ch_3_2}, has the advantage that it can be applied to a wide range of images of objects at various scales (outcrops to thin-sections), and works even for lighting- and surface-conditions that are unsuitable for automated mapping via edge detection. Unlike the approach described in \autoref{ch_2}, it is also not limited to lineament features with significant curvature, e.g. along hill slopes and the edges of mesas. This makes it applicable also to planar cliff faces, where recently-exposed layering-boundaries are distinguishable at a spacing of no more than a few meters apart. Knowing the thickness and number of layerings in the cometary nucleus would be a key parameter in modelling potential mechanisms of their formation.

\newpage

\section{Methods}  \label{ch_3_2}
\subsection{Introduction to the Fourier Transform}

The Discrete Fourier Transform (DFT) is an image processing tool which is used to decompose the grey-values in an image into sine and cosine periodic wave-functions. Thus, the DFT is useful for deriving dominant length scales and angles in an image, but computing it is impractically slow. I therefore use a variety of this tool called the 'Fast Fourier Transform' (FFT), which rapidly computes the Fourier Transforms by factorising the DFT matrix into a product of sparse, mostly zero factors \citep{vanloan_1992_fft}. This substantially reduces the computational effort of computing a Fourier Transform for $n$ points from $n^2$ to $n \times \log n$. 

The analysis described in this chapter was conducted in MATLAB \citep[release 2018a]{matlab_2018}. The FFT of a greyscale image \texttt{f} of size M $\times$ N is obtained using

\vspace{0.6cm}

\texttt{F = fft2(f);} 

\vspace{0.6cm}

which returns an array \texttt{F} of the same size. Each element of this array is called a 'mode', which contains a real and an imaginary part. To ease the graphical display, \texttt{F} is thus commonly converted to its magnitude, i.e. the square root of the sum-of-squares of the real and imaginary parts, such that

\vspace{0.6cm}

\texttt{F1 = abs(F);}

\begin{figure}[h]	
	\centering
	\includegraphics[width=\linewidth]{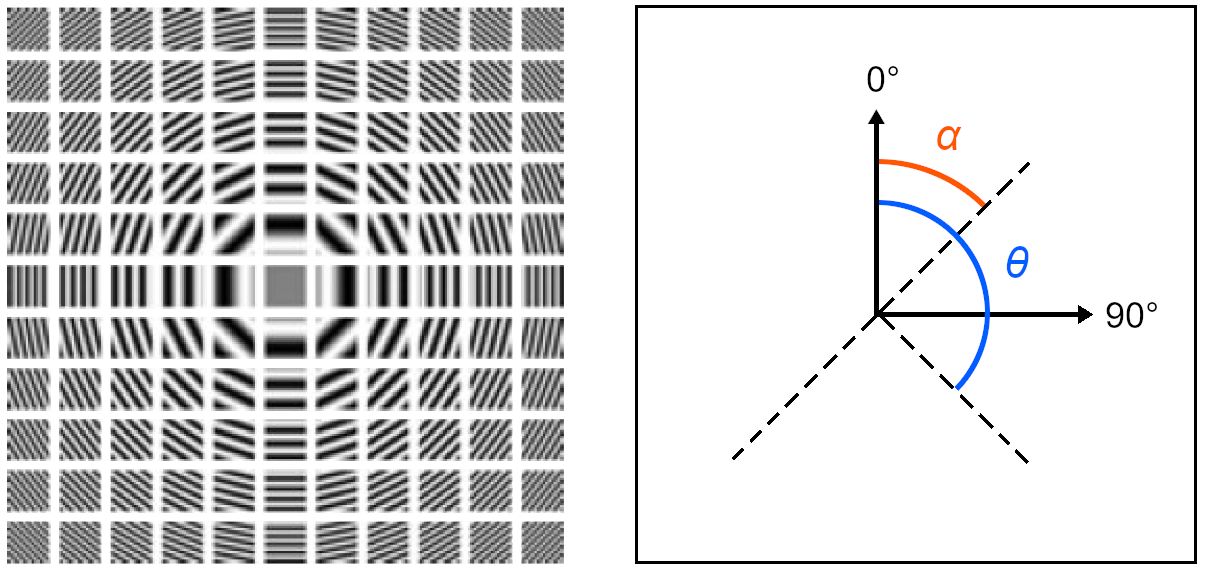}
	\caption[Fourier modes and directionality in the Fourier domain; illustration of \autoref{eq:theta}]
	{\textbf{Left:} Each tile of this image represents a mode, illustrating what an image would look like that consists of only this mode (and its symmetrical partner across the centre). The 'zero mode' $m_0$, located in the image centre, represents the mean grey value of \texttt{f}. With increasing distance to $m_0$, increasingly smaller wavelength are represented. The angular location $\theta$ of a mode contains the orientation $\alpha$ of a linear signal represented by the mode \citep[modified from][]{DIP_2019}. \textbf{Right:} Illustration of \autoref{eq:theta}.}
	\label{fig:fouriermodes}
\end{figure}

Each mode $m$ corresponds to a wavelength $\lambda$ contained in the image \texttt{f}, and the information of \texttt{f} is fully contained within the modes. The mode's location encodes the specific wavelength it represents in \texttt{f}, such that its radius $r$ (which is the distance between $m$ and $m_0$) is inversely proportional to $\lambda$ (\autoref{fig:fouriermodes}, left). The angle $\theta$ (which denotes the clockwise-positive angle between the line connecting $m$ and $m_0$, and a vertical line arising upwards from $m_0$) indicates the direction $\alpha$ in which the signal represented by the mode occurs in the image \texttt{f}, such that 

\begin{equation} \label{eq:theta}
\alpha = \theta - 90^\circ
\end{equation}

\vspace{0.4cm}

\begin{figure}	
	\centering
	\includegraphics[width=\linewidth]{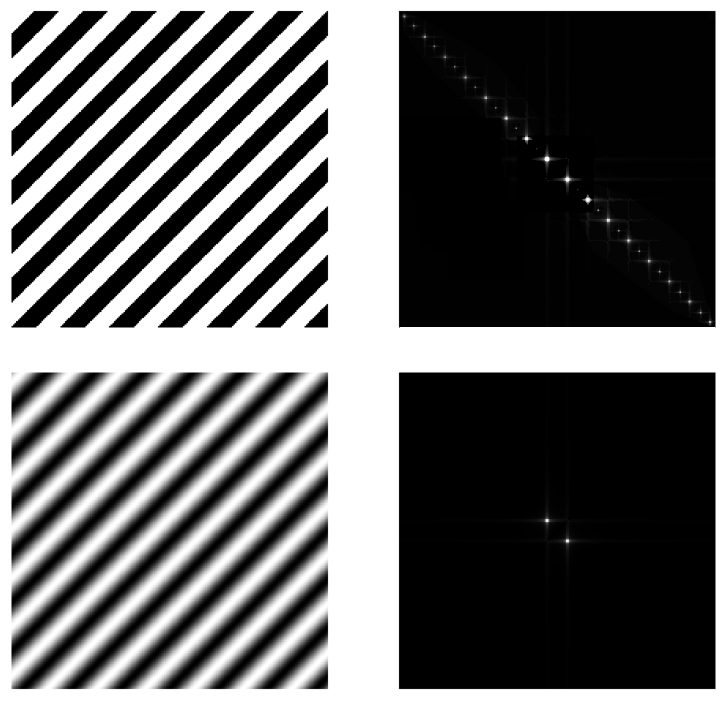}
	\caption[Directionality in the Fourier domain, demonstrated on synthetic images]
	{Demonstration of directionality in the Fourier domain. \textbf{Left:} Synthetic images of black-and-white stripes, oriented diagonally in the image. The direction of the lines in the image is $\alpha= 45^\circ$. The upper panel shows lines with hard edges where the brightness changes suddenly, while the lower panel shows the same lines with a gradual, sinusoidal change in brightness. \textbf{Right:} The modes representing these striped signals in the Fourier domain (i.e. frequency space), located along an imagined line in the direction of $\theta = 180^\circ - \alpha = 135^\circ$ (\autoref{eq:theta}). The upper panel shows the higher harmonics caused by the hard edges, which are absent in the lower image. Brightness of the white areas has been artificially exaggerated for visibility.}
	\label{fig:example_stripes}
\end{figure}

This angular relationship is illustrated in \autoref{fig:fouriermodes} (right), an example is shown in \autoref{fig:example_stripes}. 

A mode's intensity is proportional to the amount by which the mode contributes to \texttt{f}. The origin of the Fourier domain is called the 'zero mode', it has by far the highest intensity of all modes because it represents the image's mean grey value (\autoref{fig:fouriermodes}, left).

\autoref{fig:example_stripes} further illustrates directionality within the Fourier domain of an image \texttt{f}: Sudden changes in brightness in \texttt{f} will appear in the Fourier domain as a row of modes of decreasing intensity along $\theta$, where mode with the smallest radius $r$ represents the signal, and the accompanying modes represent the signal's higher harmonics. Gradual changes in brightness (e.g. a sinusoidal signal) will not produce higher harmonics.

When the greyscale-values at opposite image boundaries of \texttt{f} are notably dissimilar, several parallel lines appear in the FFT that are crossing the centre in horizontal and vertical direction (cf. also \autoref{fig:example_preprocessing2}, right). This effect is called 'leakage' and happens when the image boundaries are wrongly recognised as edges, as the Fourier transform algorithm is expecting a periodic input signal and therefore repeats the image \texttt{f} infinitely. The modes represented by the leakage lines are thus not truly part of the signal, they only appear because energy has 'leaked' into them. Measures need to be taken during image pre-processing which remove the disturbance caused by the leakage.

\newpage
\subsection{Pre-processing the input image}
\subsubsection*{Rearranging the FFT quadrants}

Before an image can be meaningfully analysed with the Fourier transform, it needs to be pre-processed. To begin with, the four quadrants in \texttt{F1} are rearranged to compensate that MATLAB uses an unconventional and counter-intuitive arrangement:

\vspace{0.6cm}

\texttt{F2 = fftshift(F1);} 

\vspace{0.5cm}

\begin{figure} [h]	
	\centering
	\includegraphics[width=\linewidth]{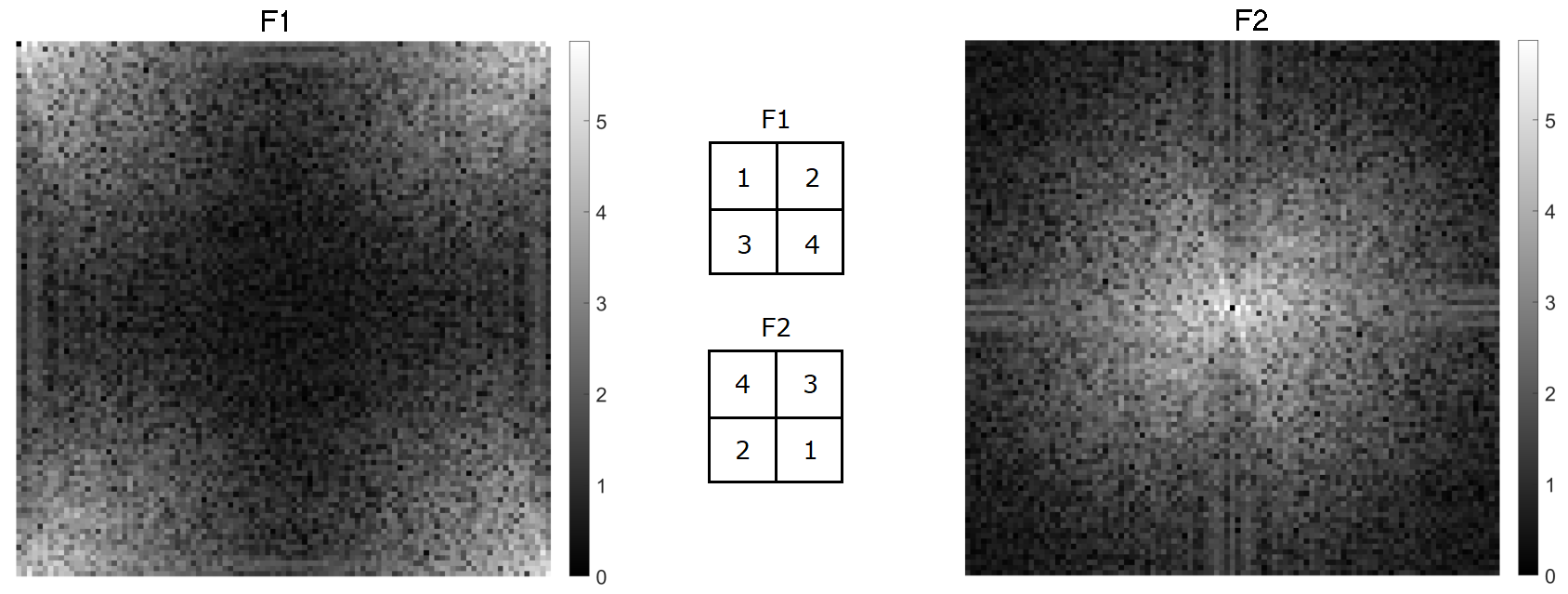}
	\caption[Re-arranging the quadrants of the Fourier domain during pre-processing]
	{\textbf{Left:} Native output in MATLAB for \texttt{F1}, $m_0$ is located in the top-left corner. \textbf{Centre:} Re-arrangement of quadrants. \textbf{Right:} Logarithmic display of \texttt{F2}, after the quadrants have been re-arranged using \texttt{fftshift}. $m_0$ is now located in the image centre. Note the horizontal and vertical lines crossing the image, which are caused by leakage.}
	\label{fig:example_preprocessing2}
\end{figure}

\autoref{fig:example_preprocessing2} illustrates how this step improves the graphical output. All figures from here on will show the FFT in logarithmic display to brighten them sufficiently to discern details.

\subsubsection*{Reducing leakage}
\begin{figure} [h]	
	\centering
	\includegraphics[width=\linewidth]{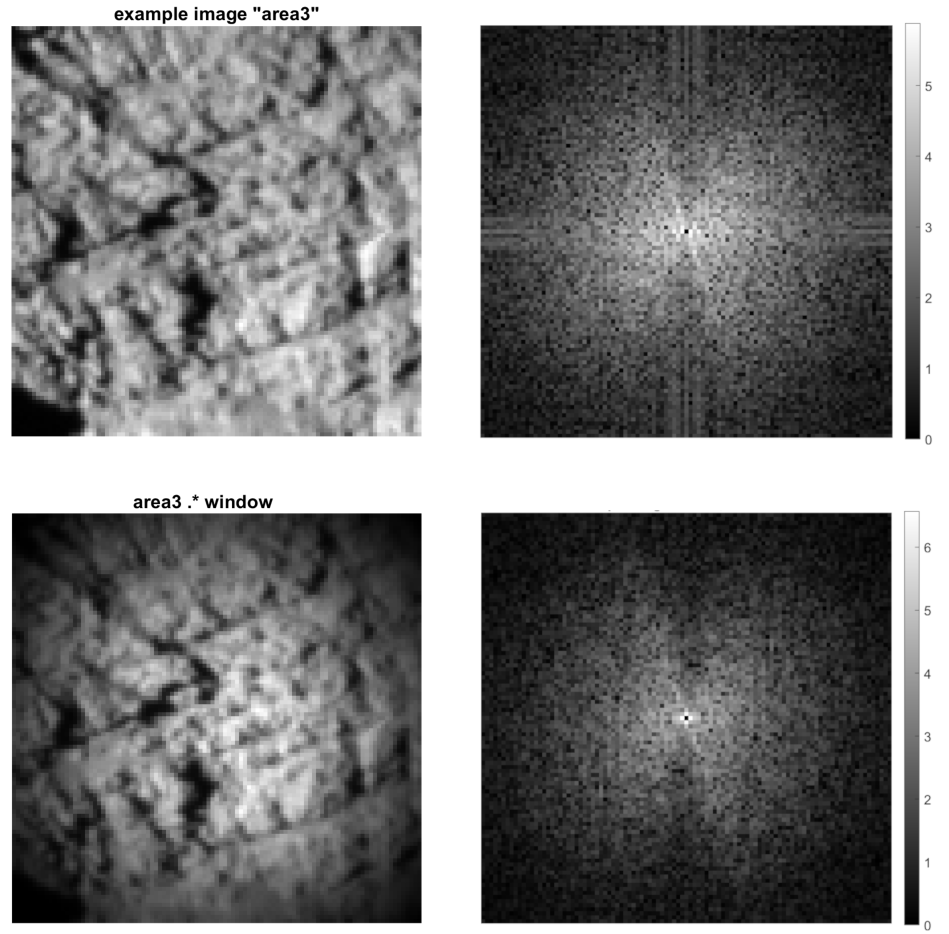}
	\caption[Effect of windowing an image on the image's FFT]
	{Demonstration of the effect of applying a window mask to an image. \textbf{Top row:} An example input image (area3) and the corresponding Fourier domain, the leakage lines are clearly visible and interrupt the signal. \textbf{Bottom row:} Image 'area3' after multiplication with a Gaussian window mask ($\sigma = 40$ pixels). The image boundaries are now all of notably similar brightness. Also note the reduction in area that is 'usable' for performing the FFT. The leakage lines are removed in the corresponding FFT.}
	\label{fig:example_preprocessing3}
\end{figure}

After the previous step, several parallel lines become visible in the Fourier domain that are crossing the centre in horizontal and vertical direction. As explained in the previous subsection, the effect of these 'ghost lines' is caused by leakage. The leakage noise can be reduced in the FFT by masking the input image \texttt{f} with a 2D window function before computing the FFT of \texttt{f} (\autoref{fig:example_preprocessing3}). The window function has the value '1' at the centre and tapers off to zero towards the edges. 

\vspace{0.6cm}

\texttt{window = mat2gray(fspecial('Gaussian',101,40));}

\vspace{0.3cm}

\texttt{F3 = fftshift(abs(fft2(f .* window)));}

\vspace{0.6cm}
\noindent This process is also referred to as 'windowing'. I determined that a Gaussian mask with $\sigma = 40$ pixels works well for removing the leakage lines and produces an FFT with the clearest structure out of a range of available masks, while sustaining the highest possible degree of transmission.

Without the leakage noise, it becomes even clearer that the intensity of the modes is not distributed equally in all directions of the Fourier domain, which is the key property that I am exploiting for my analysis. The directions whose modes contain the greatest cumulative intensity represent the directions in which the image contains the strongest linear features. 

\newpage
\subsection{Finding the directions of structures in an image}
\subsubsection*{Determining the angular intensity spectrum} \label{ch_3_2_3}

\begin{figure} [h]  
	\centering
	\includegraphics[width=0.99\linewidth]{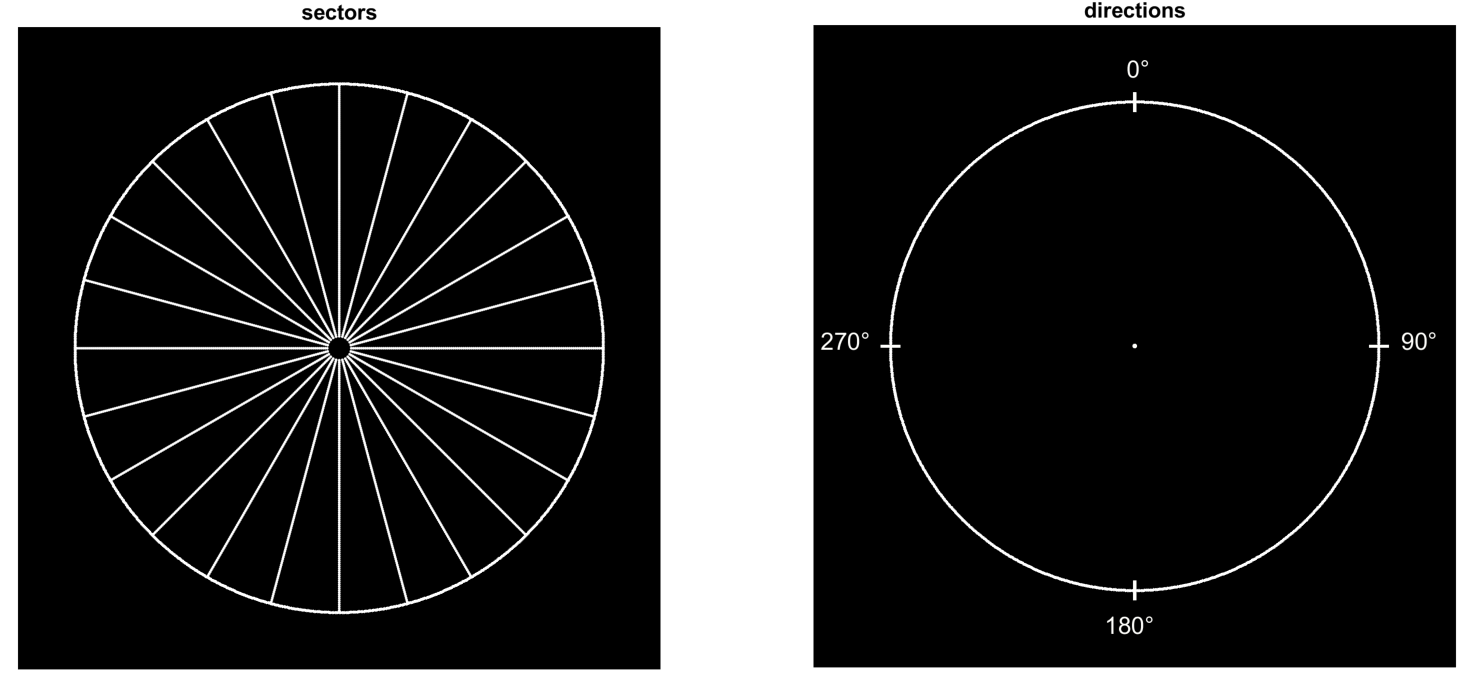}
	\caption[Concept of dividing the FFT into sectors; convention for directions $\theta$]
	{\textbf{Left:} Exemplary display of FFT division into sectors, in this example the sector width is 15\textdegree{} and there is no overlap. \textbf{Right:} Convention for directions in this thesis. 0\textdegree{} is up, angles increase clockwise.}
	\label{fig:directions_sectors}
\end{figure}

As was previously established, I am looking for the directions $\theta$ of maximum mode intensity, which represent the directions $\alpha$ of strongest structures in the Fourier domain. I determined the cumulative intensity for each direction $\theta$ in the Fourier domain by dividing \texttt{F3} into 'sectors' (\autoref{fig:directions_sectors}, left) and summing up the intensities of all modes contained in each sector. The resulting 'intensity sum of the sector' ($I_s$) is normalised by the count of pixels in each sector, which varies for small input images due to limited resolution and the square nature of pixels.

The sectors have a minimum and a maximum radius (\autoref{fig:directions_sectors}). $r_{max}$ helps normalise the number of modes per sector, as otherwise sectors facing a 'corner' (e.g. at 45\textdegree) would include more modes than sectors facing an edge of the domain (e.g. at 90\textdegree). Using a minimum radius ($r_{min}$) reduces the issue of modes being attributed to more than one sector, as the sector width narrows below the pixel size towards the centre.

The cumulative intensity $I_s$ for each sector is found as follows:

\begin{equation}
\forall m(i,j) \text{ such that } \theta_{min} \leq \theta(i,j) \leq \theta_{max} \text{ : } I_s = \sum_{r_{min}}^{r_{max}} I(i,j)
\label{eq:sectorsum}
\end{equation}

where $m(i,j)$ is the mode represented by the FFT pixel with coordinates $(i,j)$ and $I(i,j)$ means the intensity of this mode. After completing this step for all sectors, $I_s$ is a vector of size 1 $\times$ \texttt{nsec}, where \texttt{nsec} is the number of sectors. $I_s$ thus contains the angular intensity spectrum of \texttt{F3}, an example spectrum is shown in \autoref{fig:intensities_bar}. I saved computational resources by using a previously prepared lookup-table to assign a $\theta(i,j)$ to each pixel.
\subsubsection*{Finding and labelling peaks in the intensity spectrum}

In the next step, a peak-fitting algorithm (included in MATLAB's signal processing toolbox) is applied to this 'intensity spectrum' to identify the sectors that objectively stand out from the other sectors, i.e. that have the highest local signal-to-noise ratio.

\vspace{0.55cm}

\texttt{[pks,locs] = findpeaks(sector\_intensities,sector\_theta)} 

\begin{figure} [h]	
	\centering
	\includegraphics[width=0.99\linewidth]{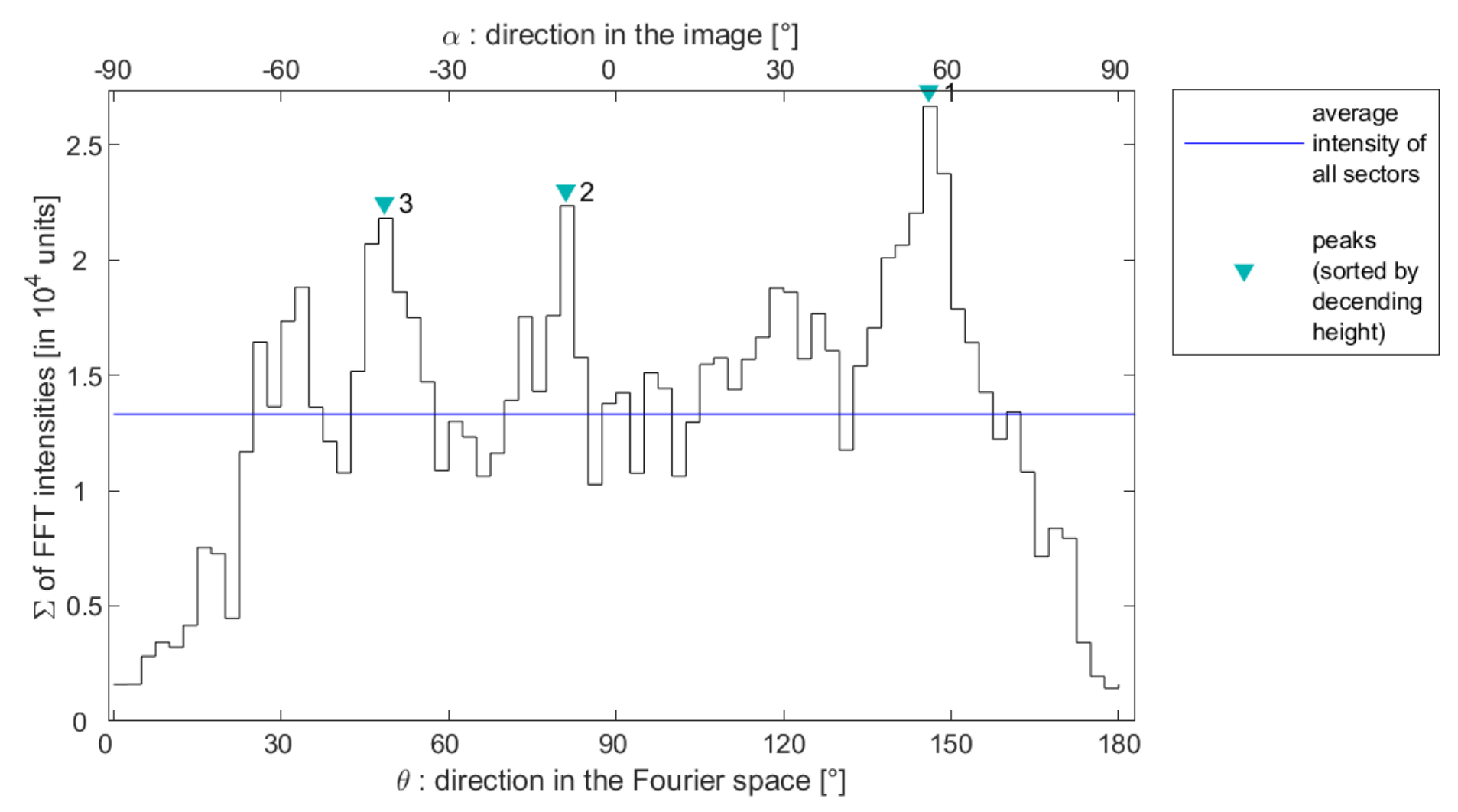}
	\caption[Stair plot of FFT-intensities per sector for an example image]
	{Stair plot of FFT-intensities per sector in image 'area4', where each step represents one sector. A peak-fitting algorithm was run with the following parameters: Minimum peak height is 1.5 $\times$ average intensity of sectors (which excluded the peaks at $\theta=35$\textdegree{} and $\theta=125$\textdegree); minimum peak distance is 15\textdegree{}; minimum peak width is 5 sectors.}
	\label{fig:intensities_bar}
\end{figure}

\vspace{0.55cm}

gives the height (pks) and location (locs) of peaks in the signal. The input parameters for the algorithm are a vector containing the cumulative intensities for all sectors (\texttt{sector\_intensities}), and a vector containing the mean theta values of all sectors (\texttt{sector\_theta}). This process is illustrated in \autoref{fig:intensities_bar}. 

One of the key aspects of this work lies in determining a useful combination of parameters (image size, sector width, $r_{min}$, $r_{max}$, minimum accepted height and width of peaks) that produces a significant and reliable detection of lineament structures in the OSIRIS images of comet 67P. 

\begin{figure} [h] 
	\centering
	\includegraphics[width=\linewidth]{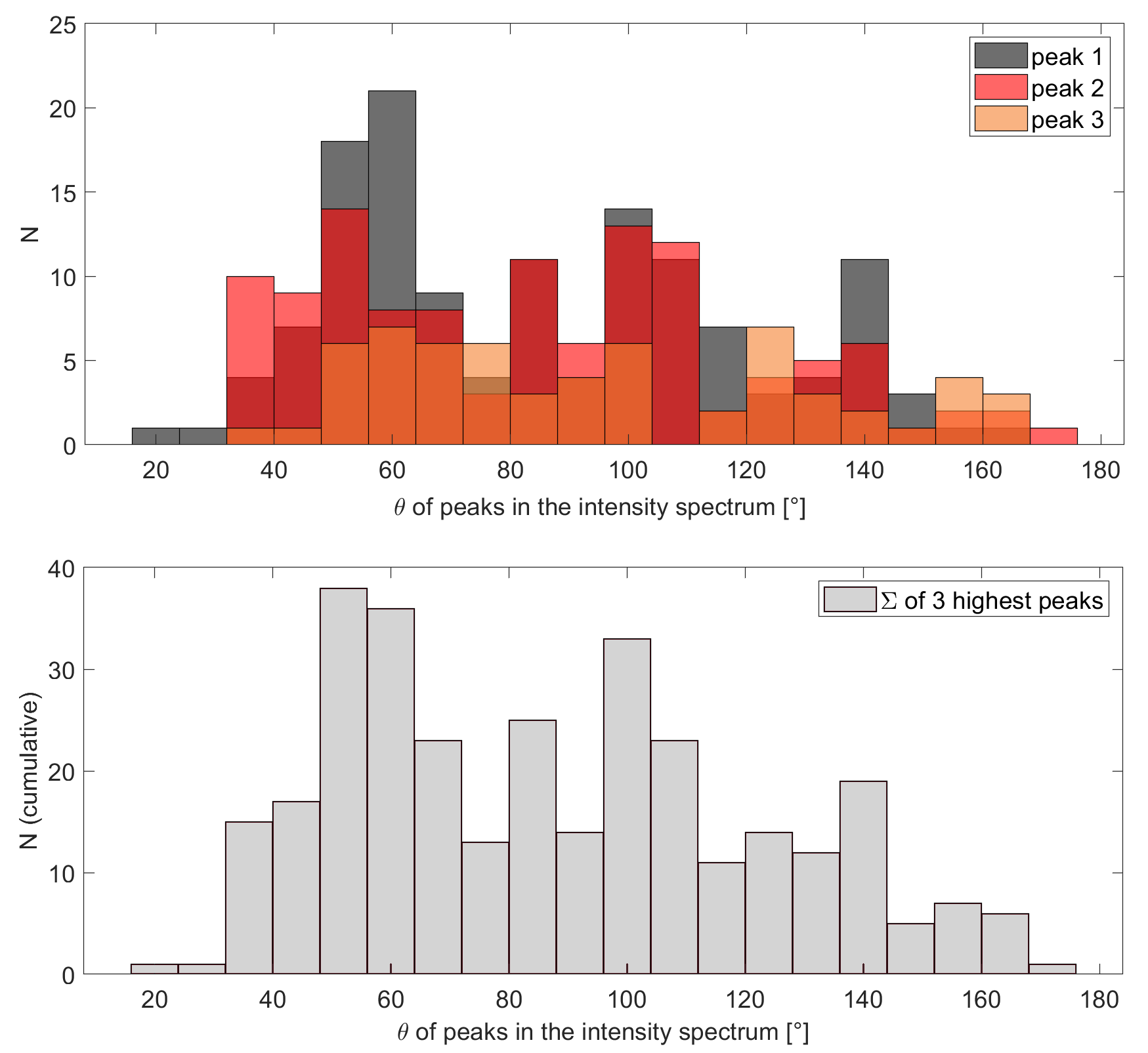}
	\caption[Histogram of $\theta$ angles for intensity peaks of an example image]
	{\textbf{Top:} Histogram distribution of $\theta$ angles for peaks in the intensity spectra of 135 frames of the Hathor wall (see \autoref{ch_3_3} for definition of 'frames'). \textbf{Bottom:} Cumulative occurrence of peaks per direction (sum of N per bin in the top panel).}
	\label{fig:histogram_theta}
\end{figure}

\autoref{fig:histogram_theta} shows a histogram of the distribution of directions for the highest peak, taken from the intensity spectra of 135 frames of \autoref{fig:areas}
(see \autoref{ch_3_3} for a definition of 'frames'). The histogram has a local maximum in the vicinity of 50\textdegree{}, a second local maximum around 110\textdegree{}, and a third local maximum around 140\textdegree. Considering the geomorphological situation on the Hathor wall, a high probability exists that these directions correspond to, in order of increasing $\theta$, the downslope lineaments, the direction of shadows, and the layering-associated lineaments. They will be respectively labelled as such through this work.

\newpage

\subsubsection*{Visualisation of FFT peaks}

The linear structures in the image which are represented by the peaks in FFT intensity can be visualised by creating a backtransformation of \texttt{f} based solely on the modes along a specific direction. This is done by taking \texttt{F3} and blackening out (i.e. setting to zero) all modes except those located within the specific sector. Next, an inverse Fourier Transform is performed on this reduced matrix \texttt{F4}:

\vspace{0.6cm}

\texttt{iF = ifft2(F4)}

\vspace{0.6cm}

\begin{figure} [h]	
	\centering
	\includegraphics[width=\linewidth]{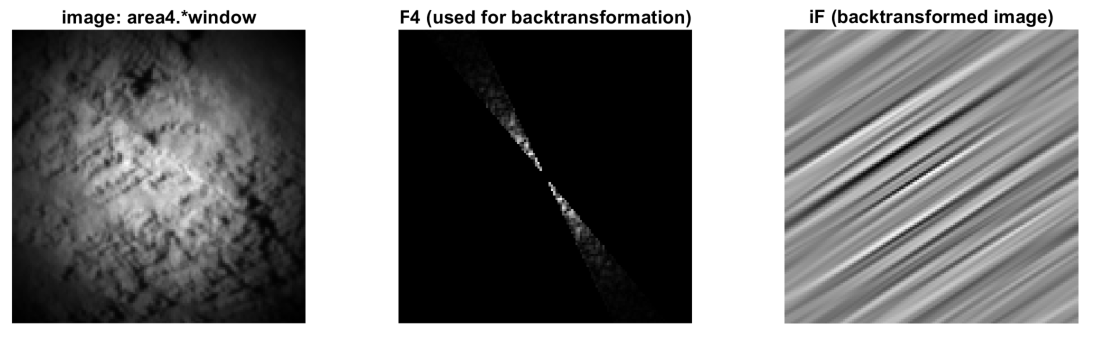}
	\caption[Inverse FFT ('backtransformation') on a windowed example image]
	{Demonstration of performing an inverse Fourier Transform ('backtransformation') on a windowed example image (\textbf{left}) using an \texttt{F4} where all modes were set to zero except those in the direction of interest (\textbf{centre}). This results in an image composed only of signals in this direction (\textbf{right}).}
	\label{fig:backtransformarea4}
\end{figure}

The resulting image \texttt{iF} (\autoref{fig:backtransformarea4}) is composed only of signals in the direction of interest. Comparing \texttt{iF} visually with the original image \texttt{f} clarifies which structures the peak finding algorithm detected in the intensity spectrum.

\subsubsection*{Determining the dominant wavelengths}

In the final step of the individual image analysis, the power spectrum of the Fourier domain is plotted along the direction of the peaks (if they exist) in order to determine the dominant wavelength in the image. The power spectrum is a plot of the modes' wavelengths against their intensities, where the wavelength $\lambda(i,j)$ represented by a mode is inversely proportional to the mode's radius $r(i,j)$ (i.e. its distance from $m_0$) such that

\begin{equation}
\lambda(i,j) = w \cdot \frac{1}{r(i,j)}
\end{equation} \label{eq:lambda}

\begin{figure} [h]	
	\centering
	\includegraphics[width=\linewidth]{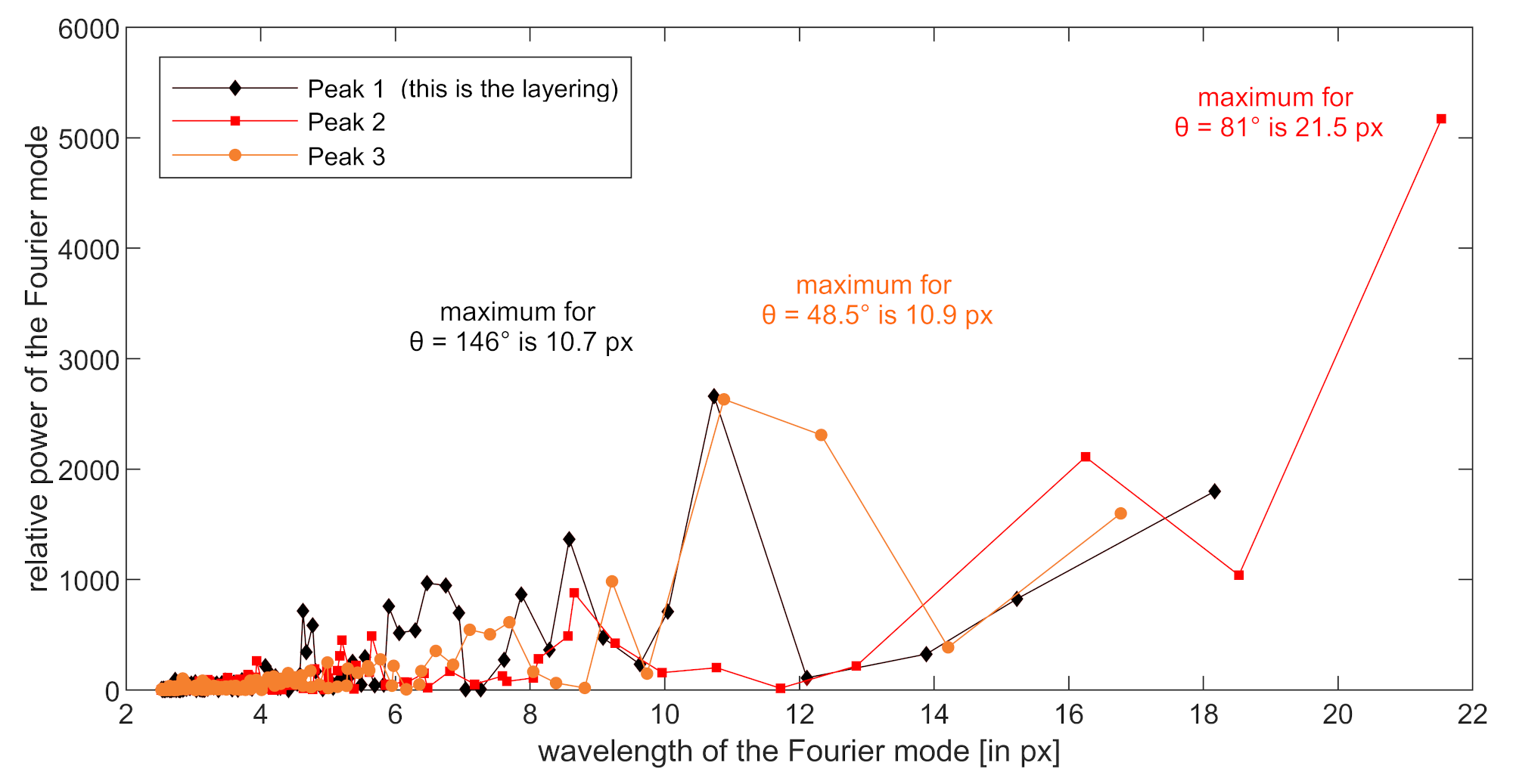}
	\caption[Exemplary FFT power spectra along the directions of highest intensity]{Demonstration of FFT power spectra along the directions of the three highest intensities in \texttt{F3} of example image 'area4'.}
	\label{fig:powerspectrum}
\end{figure}

\vspace{0.4cm}

where $w$ is the width of the image in pixels. The power spectra for the three highest peaks of 'area4' (\autoref{fig:intensities_bar}) are shown in \autoref{fig:powerspectrum}. The maxima of this graph correspond to the dominant wavelength in the respective peak-direction. The magnitudes of the dominant wavelengths are declared in the legend, and the signal most likely associated with the layerings is labelled accordingly.

\newpage

All of the aforementioned steps are executed by the MATLAB function \texttt{fftdir.m} that I wrote (see \autoref{app:ch2} for the full code). To summarise, the function generates the following output parameters:

\begin{itemize}
	\item \texttt{peak\_max\_x, peak\_max\_y}
	\item[] Location and height of the three highest peaks in the intensity graph (if at least one peak was detected, otherwise these are empty arrays)
	\item[] \texttt{peak\_lay\_x, peak\_lay\_y}
	\item Location and height of the layering-associated peak (if a peak was detected in the layering-associated range of directions, it will correspond to one of the three highest peaks in the image. Otherwise these are empty arrays) 
	\item[] \texttt{sectorsum\_avg}
	\item The image FFT's average intensity, used to calculate the factor by which a peak's height surpasses the average intensity
	\item[] \texttt{wavelength\_max, wavelength\_lay}
	\item The dominant wavelengths in the FFT power spectrum along the directions of the highest peak, and the layering-associated peak
\end{itemize}

For an intuitive overview of the steps of pre-processing and the results of the Fourier analysis, these output parameters can be displayed in a composite figure (cf. Appendix \autoref{app:ch3_5}).

\vspace{10cm}

\begin{figure} [h]	
	\centering
	\includegraphics[width=\linewidth]{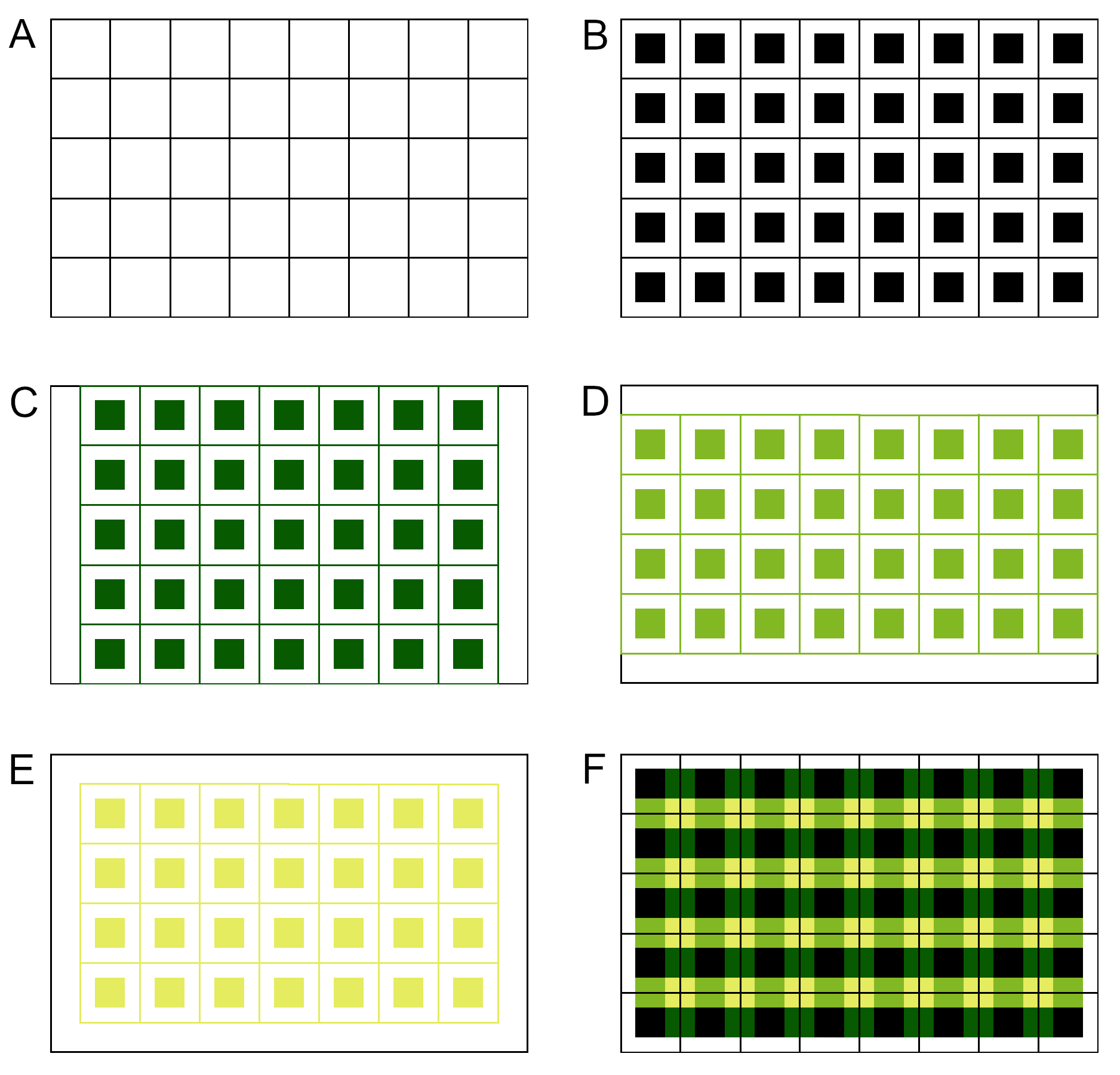}
	\caption[Illustration of the method by which tiles are defined in a 501 $\times$ 801 image]
	{Step-by-step illustration of how tiles are defined in a 501 $\times$ 801 image. \textbf{A:} In the first step, a grid is laid over the image, denoting the frame boundaries. \textbf{B:} The central 50\% of the area of each frame makes up the 'tile' that will represent the frame in the final map. \textbf{C:} A second grid of frames and tiles is defined, offset by 50 pixels to the right. The last column is left off in order to cut off any tiles. \textbf{D:} A third grid is defined, offset 50 pixels down from B. \textbf{E:} The fourth grid is offset 50 pixels to the right relative to D. \textbf{F:} A composite figure of the tiles shown in B to E, demonstrating that collectively the tiles cover the original image entirely, except for a 50 pixels wide margin.}
	\label{fig:frames_composite}
\end{figure}

\newpage
\subsection{Analysing the entire Hathor cliff wall} \label{ch_3_2_4}

Now that I can derive the direction and wavelength of the strongest signals in an image, I can analyse how these parameters are expressed across the Hathor wall. For this purpose, I wrote a MATLAB script \texttt{fullcliffscan.m} (see \autoref{app:ch3}) that splits an image (e.g. cropped from \autoref{fig:hathor_trace}) into frames and then runs the \texttt{fftdir.m} function on each frame, displaying the function's output parameters on a map of the original image. The maps, meant to serve as the basis for further analysis, are composed of 'tiles' that are graphical representations of the output for each frame, replacing the frames in their location. The concept is illustrated in (\autoref{fig:frames_composite}), which also shows that the frames overlap each other by 50\%. The overlap improves the resulting map's spatial resolution, and compensates the loss-of-area caused by windowing. 
\newpage
\section{Results and Discussion}  \label{ch_3_3}

I ran \texttt{fullcliffscan.m} multiple times on the cropped image of the Hathor wall, adjusting the free parameters each time until I found a configuration that brings out the layerings as distinctly and reliably as possible. While my method can in principle be used for analysing any kind of image, it is necessary to find a suitable configuration of parameters for each specific target. A good choice of parameters is not obvious, the selection cannot be automated but requires human assessment. Parameters are chosen suitably if they help answering the following questions: Are there repetitive structures in the image? Where in the image are they found? What is their orientation and typical spacing? Are there distinguishable sets of structures based on orientation, and where is each set found? 

\vspace{0.3cm}
In this chapter, I will list the parameters I found to deliver reliable results for the structures on the Hathor wall, and show the maps I created with them. 

\subsection{Experimentally constrained input parameters for the algorithm} \label{ch_3_3_1}

\autoref{tab:parameters} gives an overview of the parameters I determined. They affect how reliable the algorithm detects linear structures and their wavelengths in the frames. In this subsection, I will describe and discuss in detail how I selected those values for the parameters.

\begin{table} [h] 
	\centering
	\caption[Input parameters, determined by trial-and-error]{Values used for the input parameters, determined by trial-and-error.}
	\label{tab:parameters}
	\begin{tabular}{ll}
		\hline
		Parameter & Value \\
		\hline
		Frame size              & 101 $\times$ 101 pixels\\
		Sector width            & 5\textdegree{}\\
		Minimum peak height     & 1.5 $\times$ average intensity\\
		Minimum peak width      & 5 sectors\\
		Minimum peak distance   & 15\textdegree{}\\
		Minimum radius $r_{min}$ & 6 pixels\\
		Maximum radius $r_{max}$ & 52 pixels\\
		\hline
	\end{tabular}
\end{table} 

\subsubsection{Parameters relevant to the detection of structure in the frames}

\begin{figure} [h]   
	\centering
	\includegraphics[width=\linewidth]{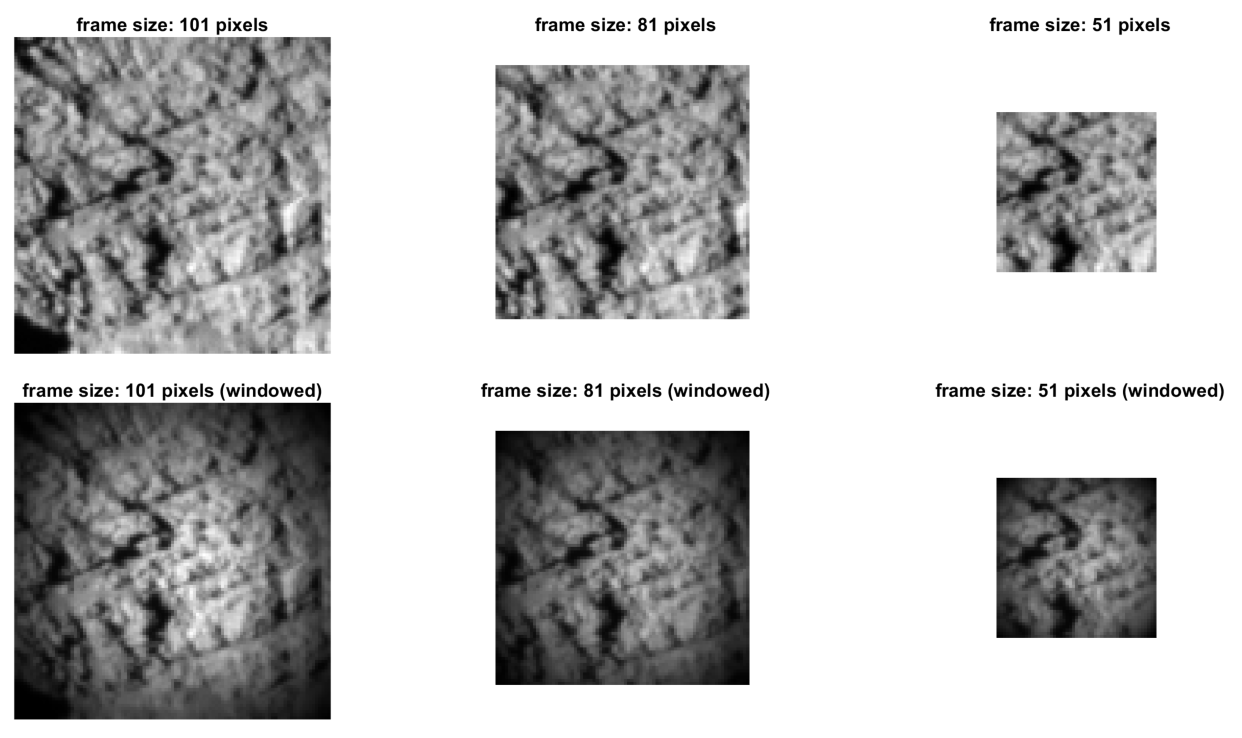}
	\caption[Effect of frame size on the usable area of a frame]{Effect of frame size on the usable area of the frame (demonstrated on 'area4'). Left: 101 pixels. Centre: 81 pixels (limits ability to detect  structures in their full extent). Right: 51 pixels (structures become virtually indistinguishable from noise).}
	\label{fig:framesizes_windowed}
	
\end{figure}

\noindent\textbf{Frame size.} A suitable frame size for detecting linear structures on the Hathor wall, for this specific OSIRIS image (\autoref{fig:hathor_trace}), was determined through trial and error to be 101 by 101 pixels. A larger frame size lowers the resolution of the resulting map so that trends across an image are harder to detect, and also leads to less distinct output parameters for the individual frames (e.g. blurring from slightly curved lineaments). A smaller frame size does not show layering-related lineaments in their entirety, nor allows it the detection of an appropriate range of wavelengths (\autoref{fig:framesizes_windowed}). An appropriate frame size needs to take into account that not the entire area of a frame is usable due to windowing. A frame size of uneven number of pixels in both directions was found to simplify finding the frame's central pixel. \\

\begin{figure} [h]   
	\centering
	\includegraphics[width=\linewidth]{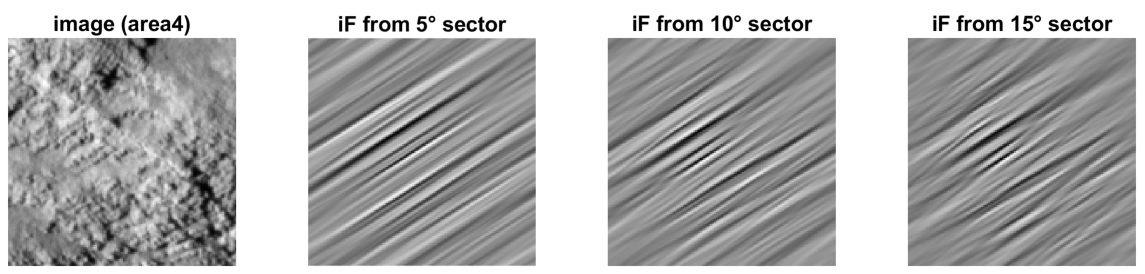}
	\caption[Effect of sector width on the backtransformed image]
	{Effect of sector width on the backtransformed image \texttt{iF}. Larger sectors result in less clarity of \texttt{iF}, as larger angular range is considered for backtransformation.}
	\label{fig:sectorwidth_iF}
\end{figure}

\noindent\textbf{Sector width.} The angular width of the sectors used to divide the Fourier domain into directions is inversely proportional to the number of sectors in the circle. The sector width has two main effects: 

Smaller sectors result in sharper backtransformed images, as a smaller angular range of frequencies is considered for the backtransformation (\autoref{fig:sectorwidth_iF}). Using small sectors strongly improves the usefulness of \texttt{iF}. However, the sectors cannot be set too small either: They need to be selected at least wide enough to allow for some variation in direction (bending) that is bound to occur in natural features (or can result from the camera perspective).

 

\begin{figure}  
    \centering
	\includegraphics[width=0.92\linewidth]{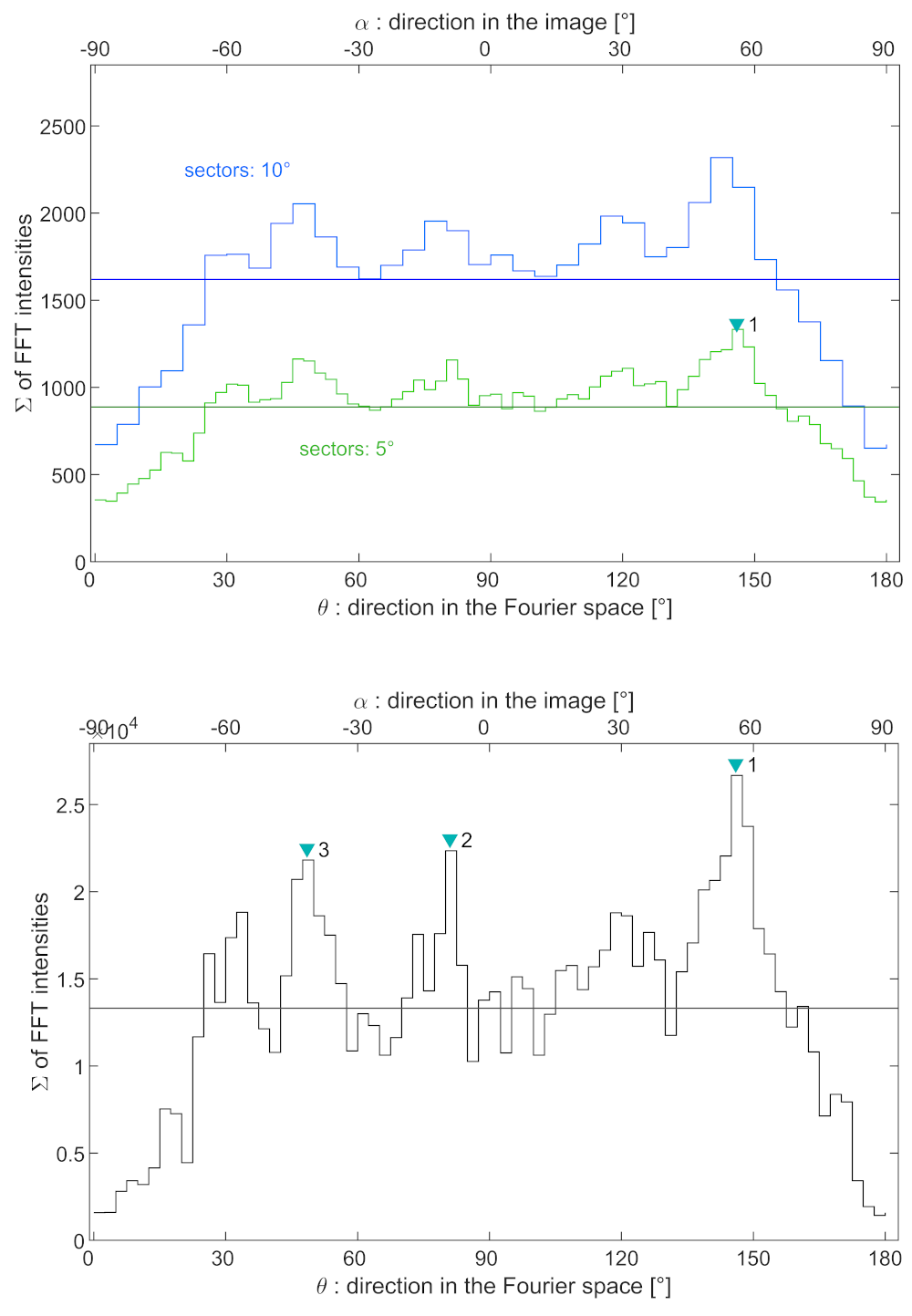}
	\caption[Effect of sector width on peak detection in the FFT intensity spectrum]
	{Effect of sector width on peak detection in the FFT intensity spectrum (area4). \textbf{Top:} Intensity spectra for \texttt{F3} (intensities not squared), for two different sector widths. The horizontal lines indicate the respective average intensity, the teal triangle indicates a peak detection. \textbf{Bottom:} Intensity spectrum for \texttt{F3'}. Intensities are squared, and now two additional peaks are detected. Sector width is 5\textdegree. Units on the vertical axis are an order of magnitude larger than in the top panel.}
	\label{fig:sectorwidth_peaks}
\end{figure}

The sector width also affects the detection of peaks in the intensity graph (\autoref{fig:sectorwidth_peaks}). Choosing smaller sectors results in a higher angular resolution of the intensity spectrum. But it also lowers the cumulative intensities (\autoref{eq:sectorsum}) in the sectors (as pixel modes get distributed to a larger number of bins), as well as the average intensity of sectors (blue horizontal line in \autoref{fig:sectorwidth_peaks}). For some frames, this effect balances out and the same peaks are detected as for larger sectors, but in most cases the increased angular resolution means that pixels get binned such that a sector's intensity is amplified enough to be recognised as a peak (compare \autoref{fig:sectorwidth_peaks}, left and centre panels). However, this peak-lowering effect of small sectors is easily compensated by squaring the FFT intensities, using 

\vspace{0.4cm}
\texttt{F3' = fftshift(abs(fft2(image.*window))).$\hat{}$ 2}
\vspace{0.4cm}

which strongly enhances the peaks (\autoref{fig:sectorwidth_peaks}). Taking all of this into account, a sector with of 5\textdegree{} was found to yield the best results for the Hathor cliff.

\vspace{0.3cm}
Squaring the FFT intensities enhances the amplitudes, which results in a spectrum that shows several groupings of sectors with above-average intensities. The selection of true peaks within these groupings is affected by the following parameters:\\

\noindent\textbf{Minimum peak height.} Peak identification is strongly influenced by the minimum required intensity, or 'height', of a sector. A frame that shows chaotic surface texture instead of aligned linear features, will have an intensity spectrum that fluctuates only slightly around the mean, but squaring the intensities (cf. discussion on 'Sector width') will even produce moderately positive local maxima within the spectrum. It is important to not misclassify those local maxima as peaks, and they are dismissed by setting a minimum peak height. This critical height can be set as either a fixed intensity value, or as a factor by which a peak needs to surpass the average intensity of all sectors. I used the latter, as the value for minimum required intensity is then specific to each frame and thus can account for varying illumination across an image. For convenience, I henceforward refer to this factor as the 'peak factor'. A suitable factor needs to be determined by trial-and-error for each image; it will depend largely on the lighting conditions and the surface texture. For the Hathor wall, a factor of 1.5 above the average intensity was found to yield the best results.\\

\noindent\textbf{Minimum peak width.} Selecting a minimum peak width reduces the risk of wrongful peak identification in noisy spectra, by excluding outlier values. If the value is chosen too large, valid peaks might be dismissed if their flanks are not steep enough and thus broader than the cutoff value. For the Hathor wall, a minimum width of 5 sectors was found to be most helpful. Accounting for sector overlap, this corresponds to 12.5\textdegree.\\

\noindent\textbf{Minimum peak distance.} Linear features on the Hathor wall are found to run in one of three directions, which are separated from each other by approximately 30\textdegree{} or more (cf. \autoref{ch_3_2_3}, and \autoref{fig:histogram_theta}). In the intensity spectrum, potential peaks are occasionally located much closer to each other than this (e.g. \autoref{fig:sectorwidth_peaks}, bottom, shows two potential peaks at $\sim$35\textdegree{} and $\sim$50\textdegree). It follows then that not both of those potential peaks are representative of the feature direction and that one peak must be chosen. By selecting a minimum peak distance, \texttt{findpeaks} will automatically select the higher one as the peak. This was found to yield reliable results for the Hathor wall, and works best with a minimum peak distance of 15\textdegree. 

\subsubsection{Parameters relevant to the determination of wavelengths}

\vspace{0.3cm}

\noindent\textbf{Minimum peak height.} A stronger peak includes modes of higher intensities, and the resulting FFT power spectrum in this direction will have a better signal-to-noise ratio. An example is found in \autoref{app:ch3_5} (frame 'An'): The strongest peak (1) has the clearest power spectrum; the maxima in the spectra of peaks 2 and 3 are less pronounced. However, the threshold must not be set too high that valid signals are discarded. \\

\noindent\textbf{Minimum and maximum radii.} The minimum and a maximum radius that were set for practical reasons (cf. \autoref{ch_3_2_3}) influence which wavelengths can be detected by \texttt{findpeaks}). As the radius of a mode is inversely proportional to the wavelength it represents (\autoref{eq:lambda}), the maximum radius $r_{max}$ determines the minimum wavelength that can be found. This reduces noise by modes which do not contribute strongly to \texttt{f} anyway. The minimum radius $r_{min}$ determines the maximum wavelength that can be detected. Suitable values have to be found by trial and error, and I determined the best $r_{min}$ to be 6 pixels, and the best $r_{max}$ to be 52 pixels, for a frame of size 101 $\times$ 101 pixels.\\

\noindent\textbf{Frame size.} The frame size limits the wavelengths that can be detected in two ways: Firstly this effect works directly, as a signal needs enough space to occur at least twice in a frame to even have a wavelength. Secondly, the frame size affects the range of detectable wavelengths indirectly, as $r_{max}$ cannot be larger than half the frame size.

\vspace{1cm}
\subsection{Effect of photometric properties} \label{ch_3_3_2}

Any feature measured on a photographic image receives its physical meaning only by considering the context within which the image was taken. This means that I need to consider photometric properties such as the relative locations of the light source (i.e. the Sun), the target (i.e. the nucleus), and the camera, to rule out that my results were skewed by optical effects. 

The information needed to constrain the photometric properties is contained in data sets called 'SPICE kernels', which hold ancillary information such as when an instrument was acquiring data, where it was located, and where it was pointed \citep{acton_1996_ancillary}. For the OSIRIS camera, SPICE data includes the distance of the camera to the nucleus surface, as well as the photometric angles during each millisecond of the mission (\autoref{fig:emission_angle}). OSIRIS image files of CODMAC calibration level 4 and higher also contain some of this information on a pixel-by-pixel basis in the form of additional layers. I will discuss the effect of these parameters on my results now.

\newpage

\textbf{Distance of camera to nucleus surface.} The distance predominantly depends on the surface topography. For the image analysed in this chapter, i.e. a section of OSIRIS image N20140828T124254563ID30F22 that shows the Hathor wall, topography is most strongly expressed along the left border (\autoref{fig:distance_map}). The distance-to-camera, according to the DISTANCE\_IMAGE layer, is 56.9131 km on average and ranges from 56.7131 to 57.0923 km. The variability in distance therefore is 379 metres, or 0.66\% of the mean distance. 
As the ground resolution is directly proportional to the distance to the target surface, it also varies by 0.66\%. For images taken with by the OSIRIS narrow angle camera, ground resolution (in m/px) is calculated as

\vspace{0.4cm}
\noindent\texttt{ground\_resolution = distance\_to\_camera [m] $\times{}$ 18.6 $\times{} 10^{-6}$ [1/px]}
\vspace{0.4cm}

\begin{figure} [h]  
	\centering
	\includegraphics[width=0.85\linewidth]{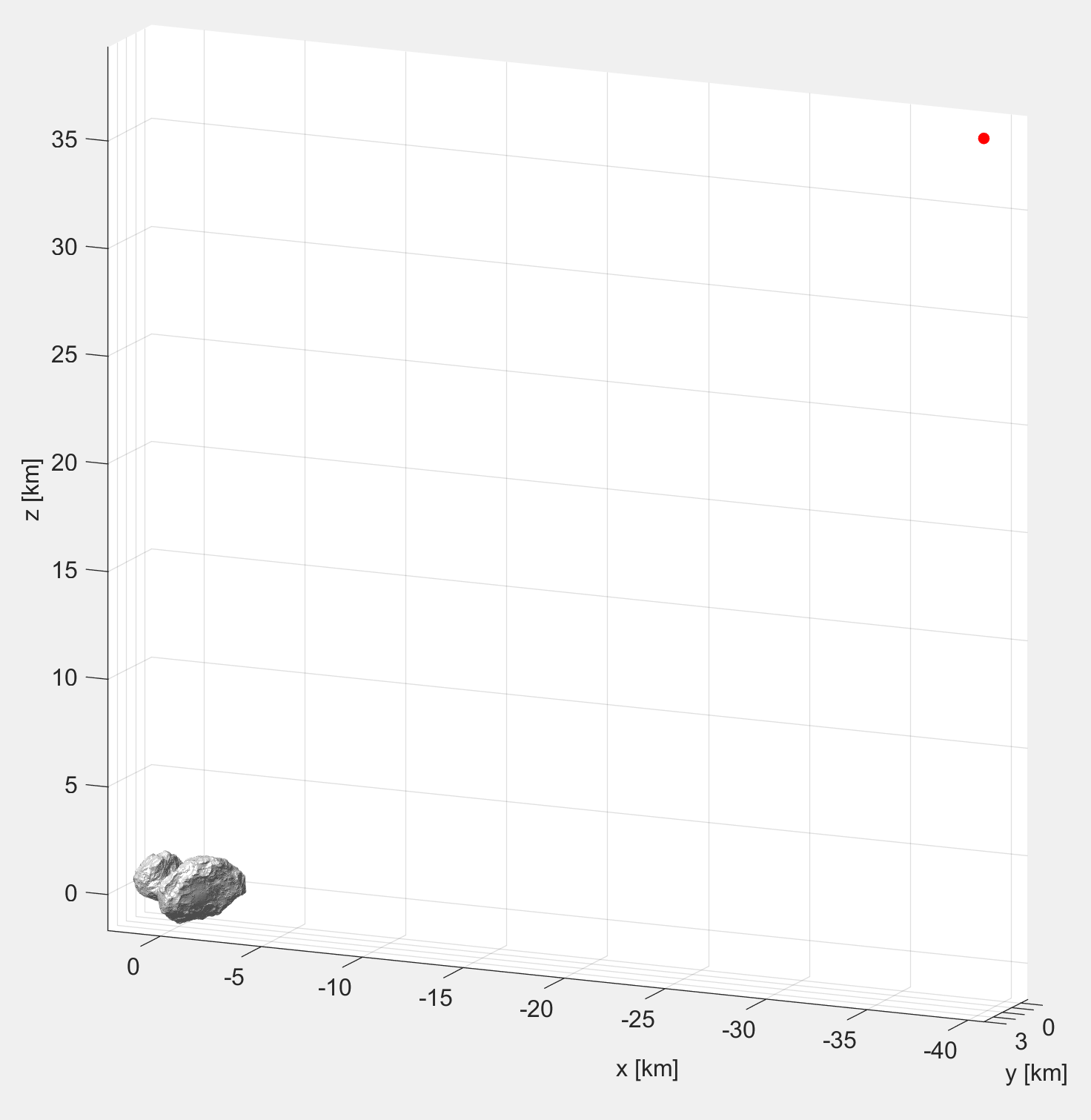}
	\caption[Distance between the nucleus and the spacecraft, to scale]
	{Distance between the nucleus and the spacecraft, illustrated to scale of comet. The red dot representing the Rosetta spacecraft is not to scale but shown enlarged for visibility.}
	\label{fig:distance}
\end{figure}

\begin{figure} [h]  
	\centering
	\includegraphics[width=0.98\linewidth]{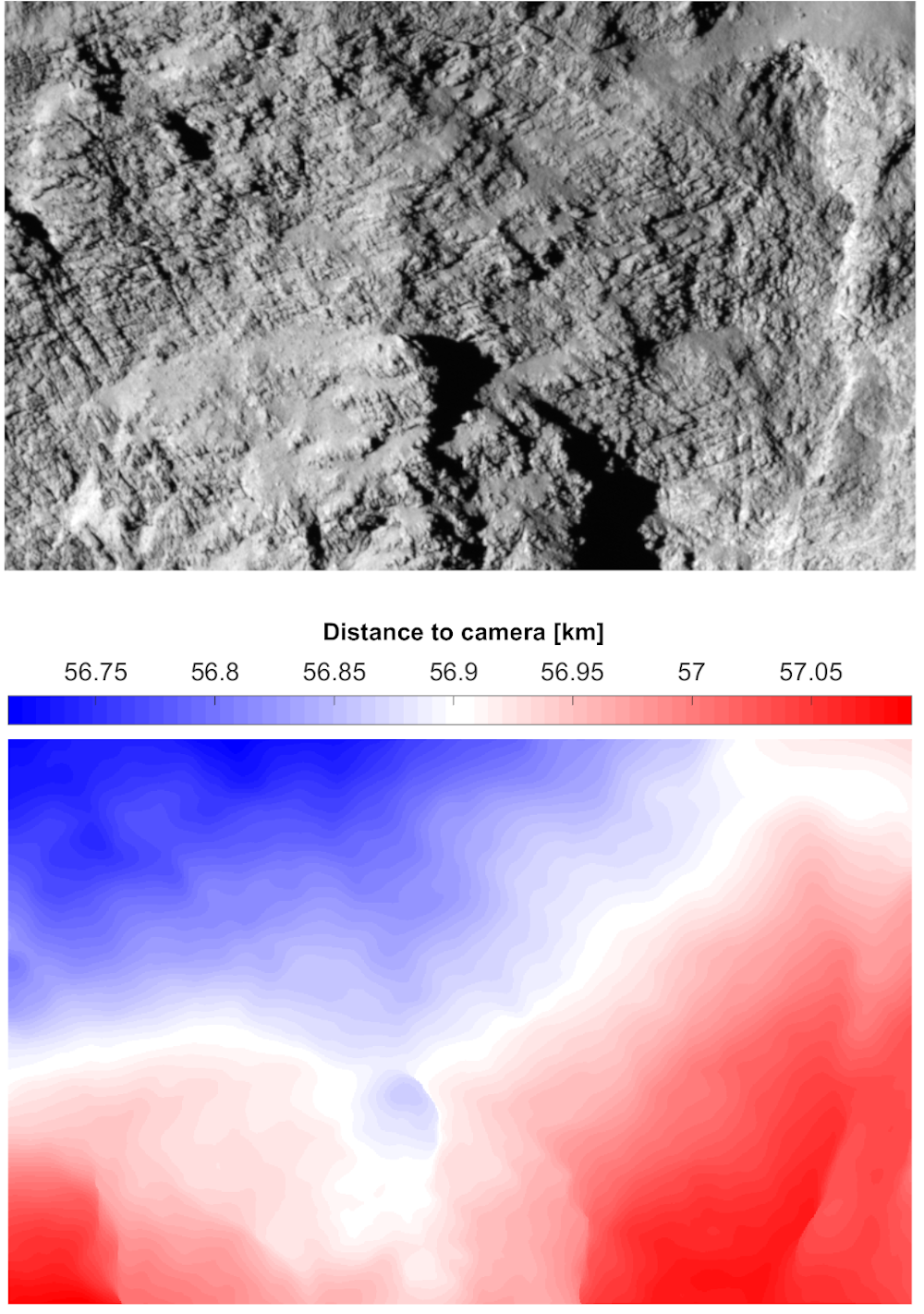}
	\caption[Map of distance between the camera and the nucleus surface for each pixel]
	{Map of distance between the OSIRIS camera and the nucleus surface represented by each image pixel. Blue indicates areas closer to the camera, red areas are further away. Input image of the Hathor wall shown on top for comparison.}
	\label{fig:distance_map}
\end{figure}

\begin{figure} [h]  
	\centering
	\includegraphics[width=\linewidth]{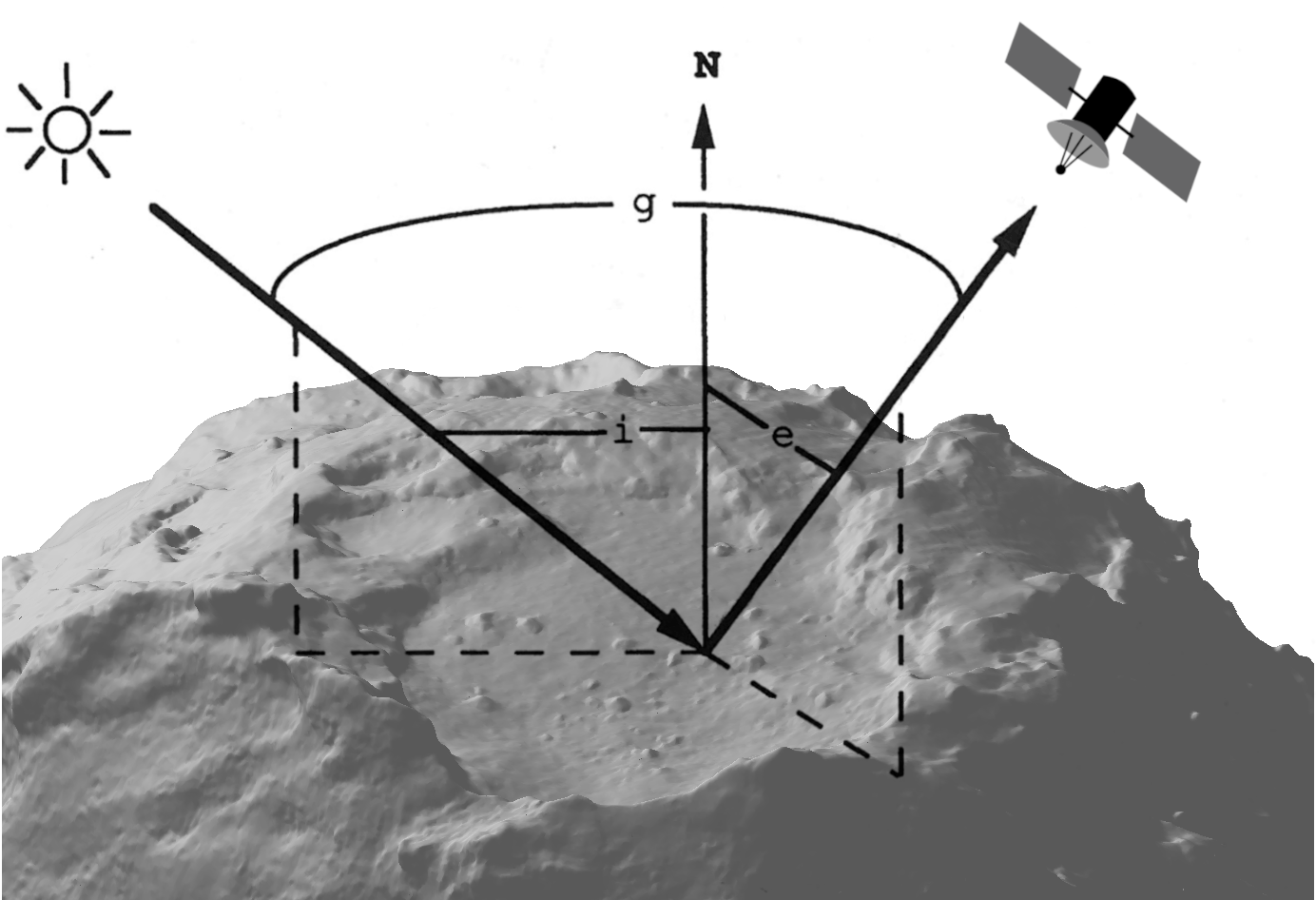}
	\caption[Photometric angles between the Sun, the nucleus surface, and the camera]
	{Photometric angles. \textbf{N} is the normal vector to the nucleus surface, \textbf{g} is the phase angle between the Sun and the camera, \textbf{i} is the incident angle, and \textbf{e} is the emittance or emission angle. Figure is based on fig. 8.1 in \citet{hapke_2005_reflectance}.}
	\label{fig:emission_angle}
\end{figure}

\citep{keller_osiris_2007}. Ground resolution for this image varies between 1.0549 and 1.0619\,m, which has a negligible effect on my results. \autoref{fig:distance} illustrates the distance of nucleus and camera to scale, which is larger than intuitively expected.

\vspace{0.4cm}

\textbf{Phase angle.} The phase angle describes the angle between incident sunlight hitting a location on the surface and the direction of the camera relative to that location on the surface (\autoref{fig:emission_angle}). A phase angle of 0\textdegree{} means that the Sun is located behind the camera, whereas 180\textdegree{} mean that the camera is looking into the direction of the Sun. At the time of image acquisition, the vector of the spacecraft as seen from the origin of the comet-fixed reference frame (in km) was \\

\texttt{[-40.78008583 3.02932856 39.35459972]}\\

whereas the vector of the Sun was \\

\texttt{[-162404434.65031720 341089248.32254330 356052659.99622524]}\\

(see \autoref{app:ch3_4} for details). According to the relationships shown in \autoref{fig:emission_angle}, the phase angle at time of image acquisition is 42.6\textdegree. It is readily observed in nature which phase angles are conductive to studying features that express themselves through topography: Angles below $\sim$20\textdegree{} result in minimal to no shadows, and angles above $\sim$60\textdegree{} produce overly pronounced and lengthened shadows. The phase angle of 42\textdegree{} was one of the main reasons why I selected this OSIRIS image for my analysis.

\newpage

\begin{figure} [h]  
	\centering
	\includegraphics[width=\linewidth]{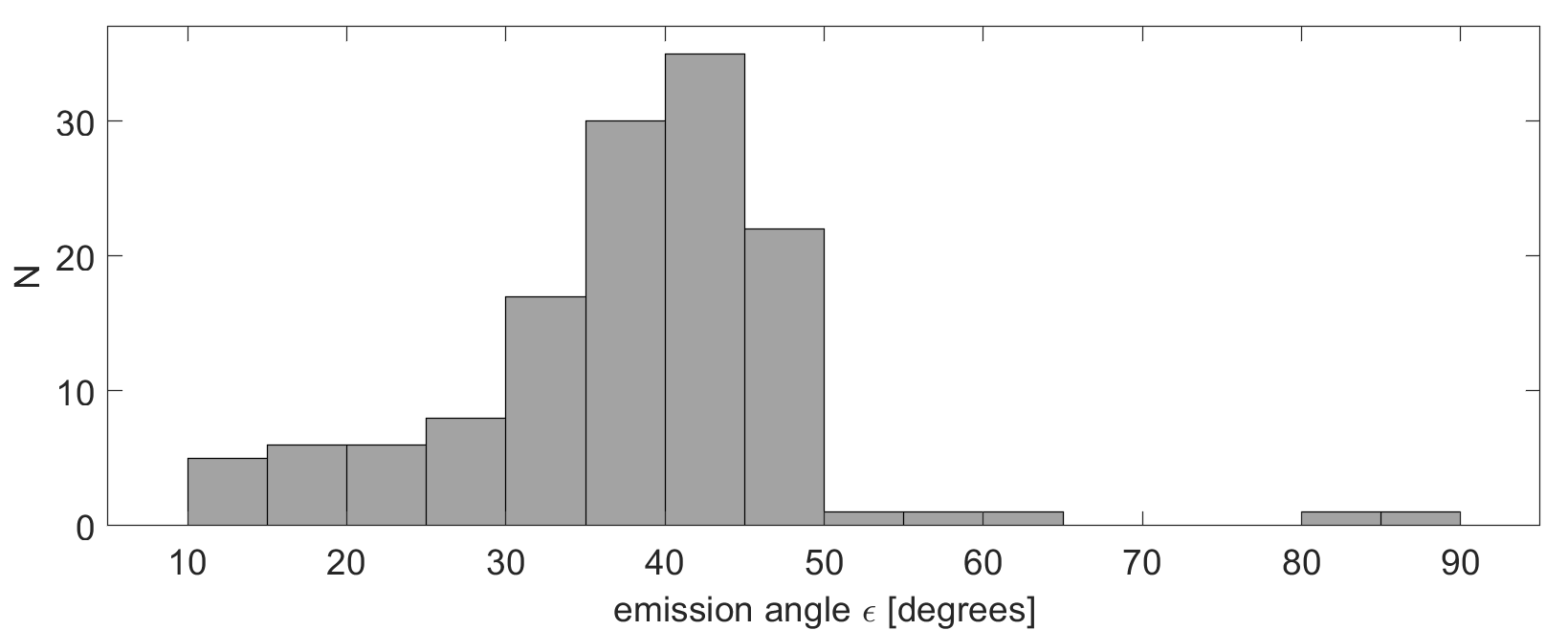}
	\caption[Emission angles for the Hathor wall, averaged per frame]
	{Emission angles for the Hathor wall, averaged per frame. The histogram peak is located at 45\textdegree{}, most values are lower. The mean of their distribution is 37.7\textdegree{}}
	\label{fig:emission_histogram}
\end{figure}

\textbf{Emission angle.} The emission angle $\epsilon$ is defined as the angle between the normal vector to a surface and the position vector of an observer relative to that surface (\autoref{fig:emission_angle}). It thus depends on the surface normal, which, due to interpolation effects, varies depending on the resolution at which the surface is sampled. On a pixel-by-pixel scale, normal vectors scatter widely because the nucleus surface on the Hathor cliff is uneven and bumpy. However, this level of resolution is neither helpful nor necessary for my analysis, which is based on frames that are 101 pixels wide. 
To receive meaningful emission angles for each frame, I fitted a plane through the XYZ coordinates of all 101x101 pixels in the frame, effectively down-sampling the topography. I then used the normal to that plane to calculate the emission angles. The majority of angles are at or below 45\textdegree, with a mean of 37.7\textdegree (\autoref{fig:emission_histogram}).

A high emission angle is disadvantageous for measuring distances on the surface, as it leads to an oblique point of view in which features appear to be closer to each other than they really are. This effect is strongest for features oriented perpendicular to the viewing direction, and approaches zero for features oriented parallel to the viewing direction. Therefore I will consider the first case to constrain the worst-case error introduced by the emission angle. From the geometric relation

\begin{equation}
    \frac{\text{apparent spacing}}{\text{true spacing}} = \cos{\epsilon}
    \label{eq:epsilon}
\end{equation}{}

follows that the error is directly dependent on the cosine of the emission angle, which also means that it declines at an above-linear rate with decreasing $\epsilon$. According to \autoref{eq:epsilon}, the true spacing of features observed at an emission angle of 45\textdegree{} is larger than the apparent spacing by a factor of 1/cos(45\textdegree), or approximately 1.4 (error of 40\%), whereas for an exemplary $\epsilon=22.5^\circ$ the factor is only 1.08 (error of 8\%).

\newpage

\subsection{Parameter-maps of the Hathor wall} \label{ch_3_3_3}

\begin{figure}[h] 
	\includegraphics[width=\textwidth,trim={0 4cm 0 0},clip]{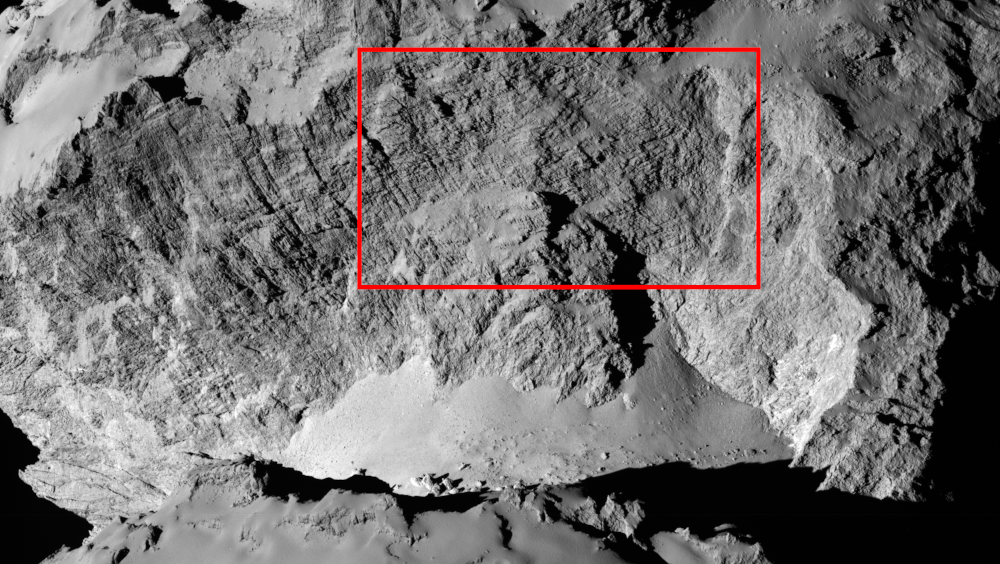}
	\caption[Location of the cropped section on the Hathor wall]{The red quadrangle indicates the location of the cropped section that was used as an input image for the analysis. For the full OSIRIS image, please cf. \autoref{fig:hathor_trace}.}
	\label{fig:crop_location}
\end{figure}

Having determined a suitable configuration of input parameters for my \texttt{fullcliffscan.m} algorithm, we are now ready to analyse the linear features on the Hathor wall. The input image is a 501 $\times$ 801 pixel section cropped from the aforementioned OSIRIS image N20140828T124254563ID30F22, chosen to show the highest number and spatial concentration of the proposed layering-related lineaments. \autoref{fig:crop_location} shows where the cropped section is located on the image and gives its structural context on the cliff. 

To visualise the output of \texttt{fullcliffscan.m}, I created several composite images which I refer to as 'maps'. Each map illustrates one of the following output parameters of the algorithm: Orientation theta of the lineaments, their peak-factor, backtransformations along the direction of the peaks, and wavelength of the signal along the direction of the peaks. The maps are made up of tiles, which show the results for the respective frame in its location (cf. \autoref{ch_3_2_4}). In the maps, empty tiles represent areas of the input image where no appropriate peak was detected, whereas filled tiles represent areas where either layering-related lineaments, or downslope lineaments, respectively, were detected. In order to easily refer to specific tiles, the maps are equipped with a coordinate system, where columns are designated with lower case letters a to o and rows designated with upper case letters A to I.

Although images of the Hathor wall contain linear structures along three main orientations (cf. \autoref{ch_3_2_3}), I focused my analysis on the two orientations that are linked to physical features in the cliff (i.e. the proposed layerings and downslope lineaments), and disregarded the orientation of shadows cast by the oblique sunlight. 

\subsubsection*{The directions of layering-associated lineaments across the Hathor wall}

\begin{figure} [h] 
	\centering
	\includegraphics[width=\linewidth]{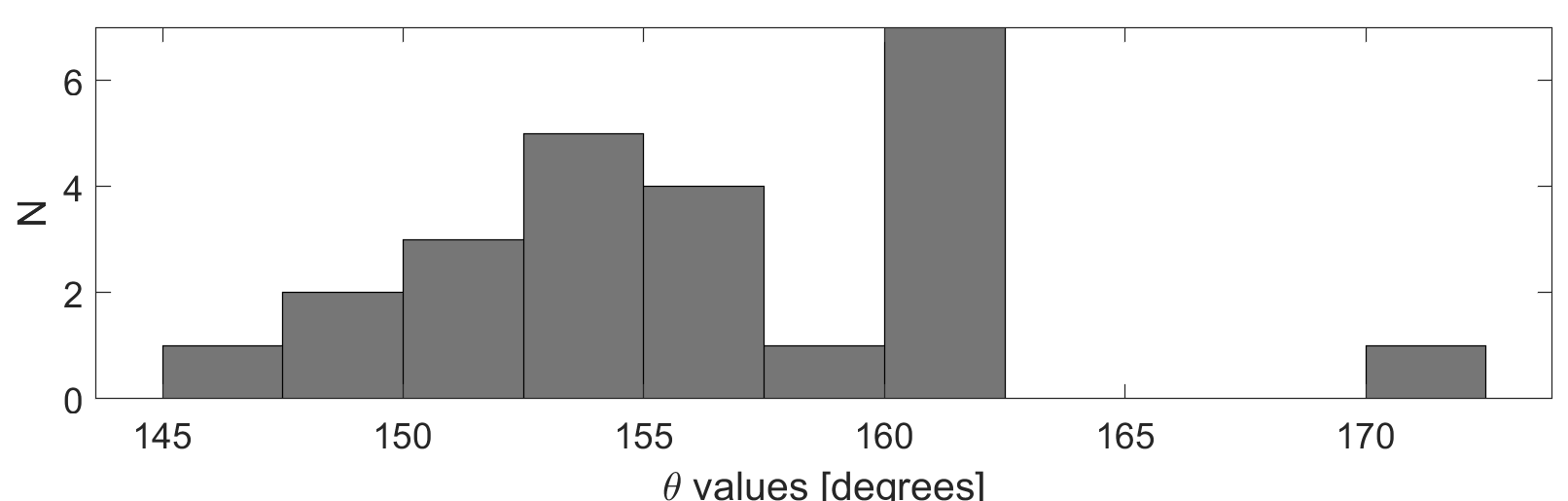}
	\caption[Histogram of $\theta$ values for layering-associated lineaments]
	{Histogram of $\theta$ values for layering-associated lineaments} 
	\label{fig:histogram_lay}
\end{figure}

\begin{figure} [h]  
	\centering
	\includegraphics[width=\linewidth]{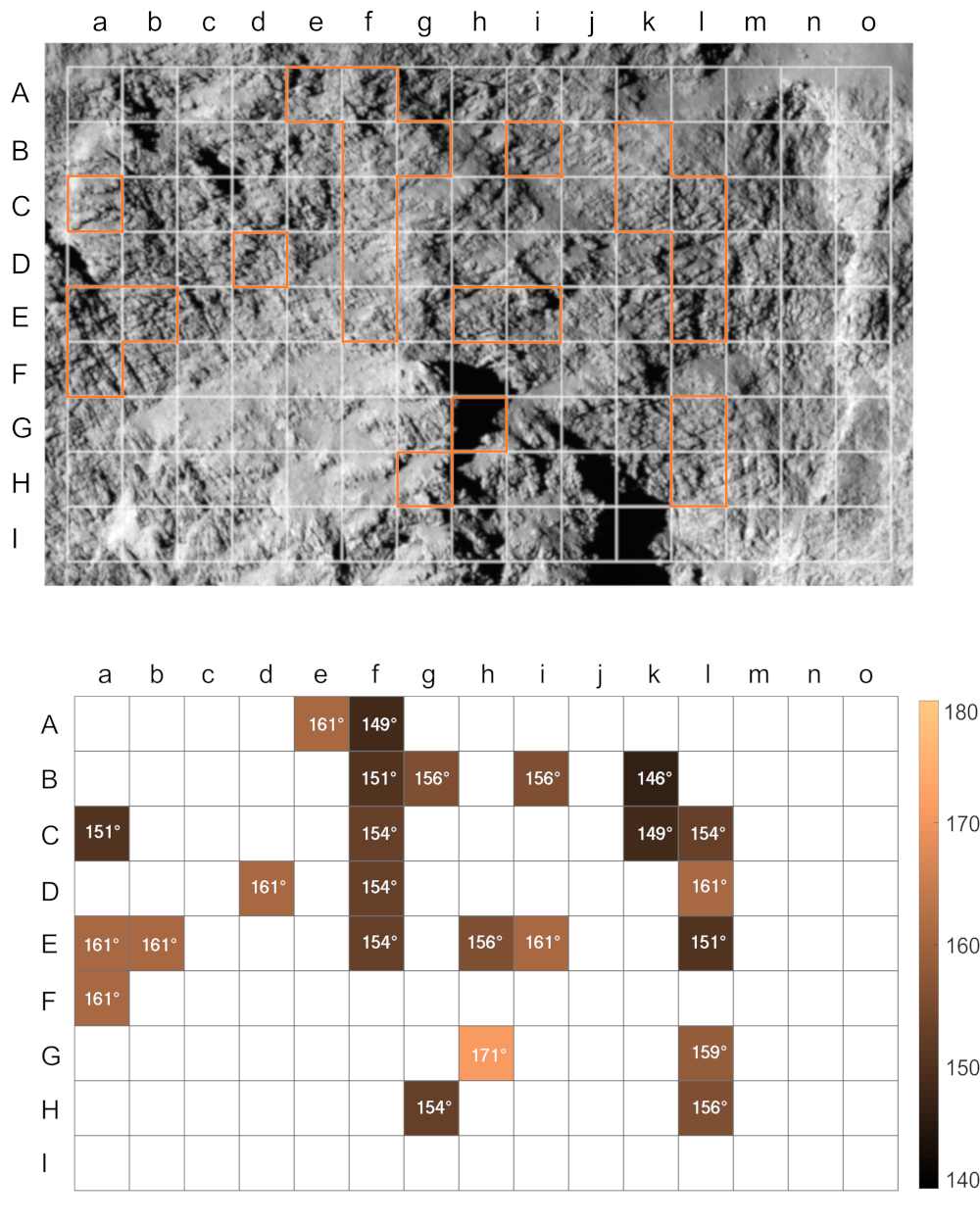}
	\caption[Map of $\theta$ values for the intensity-peaks of layerings]
	{Distribution of $\theta$ values (in degrees) for the intensity-peak associated with the \underline{layerings}. 24 peaks were found, their mean $\theta$ value is 156\textdegree. Input image of the Hathor wall shown on top for comparison.}
	\label{fig:lay_theta}
\end{figure}

\textbf{\autoref{fig:lay_theta}} shows the \textbf{orientation of $\theta$ values} for the intensity peak associated with the proposed layering-related lineaments. The $\theta$ values range from 146\textdegree{} to 171\textdegree{}, with a mean value of 156\textdegree{} and a standard deviation of 5.6\textdegree (\autoref{fig:histogram_lay}). It bears repeating that $\theta$ represents orientations in the Fourier domain that are offset by 90\textdegree{} from the corresponding orientations $\alpha$ in the image domain, such that the orientation of the layering-associated lineaments varies from 56\textdegree{} to 81\textdegree{} (mean of 66\textdegree{}). There appears to be no trend within the data, as both high and low $\theta$ values are found on all parts of the studied section of the Hathor wall.

24 tiles show a positive detection, they are distributed across most of the image, but notably no layerings were detected in the lower-left quadrant nor in columns m, n, and o (see also \autoref{fig:no_peaks}). The absence of detections in the lower-left quadrant is explained by the fact that the upper half of the ca. 500 $\times$ 600 metres landslide body resting at the foot of the Hathor wall is largely covered in airfall deposits. These deposits smoothen the surface and hide the underlying topography, which explains why no layering-related lineaments were detected in tiles F(b-g), G(a-f), H(a-f), and I(a-f). Column o  shows the margin of an area covered in a seemingly substantial layer of sediment which hides the structure of the cliff to the right of the cropped section. There is no sharp boundary to this sediment cover, which also covers part of column n. 

The lack of detections in column m is not as straightforward to explain, as the eye discerns a layered structure in tiles Cm, Dm, Em, Gm, and Hm, but the algorithm does not. In tiles Gm and Hm, this could be explained by the low difference in brightness between layering boundaries and intra-strata material, which is insufficient to produce an intensity peak that passes the required Minimum Peak Height. 

Removing the Minimum Peak Height requirement increases the overall detection count in the image from 24 to 51, and in column m it leads to detections in tiles Am, Cm, Dm, Gm and Hm. While this seems like an improvement, a higher number of detections does not equal a better result. Instead, the result now includes a considerable number of false positives, as for example in tile Am. Presumably, the algorithm identified the linear boundary between dust-covered surface and not-covered bumpy wall texture as a lineament whose $\theta$ lies within the range of appropriate $\theta$ values for layering-associated lineaments. As false detections introduce unnecessary noise into the data and thereby make interpretation of the results more difficult, I decided to err on the side of caution and uphold the Minimum Peak Height requirement of $\geq 1.5 \times$ the mean intensity value. Even with this requirement in place, false detections happened. The 171\textdegree{} $\theta$ value in tile Gh appears to be unrelated to actual structure in the wall, and instead originates from the border of the shadow cast by the raised landslide body. When removing this value from the results, the mean $\theta$ is lowered to 155\textdegree{} and the standard deviation is lowered to 4.2\textdegree.

The matter remains why, while upholding the requirement, several tiles do not show positive detection where the eye recognises layerings. A possible explanation might lie in the human brain's supreme ability to recognise, complete and even extend patterns tends to create false positives in ambiguous situations. The bumpy surface texture on this part of the cliff, in combination with the strongly expressed layering pattern in column l, lures the brain into seeing structure where there is none. However, pattern is not actually present in image, and this becomes clear when covering the rows a-l with a sheet of paper. This type of involuntary confirmation bias, and other kinds of cognitive bias, are what I am hoping to curtail with the Fourier-based, i.e. statistical approach to layering analysis.\\

\begin{figure} [h] 
	\centering
	\includegraphics[width=\linewidth]{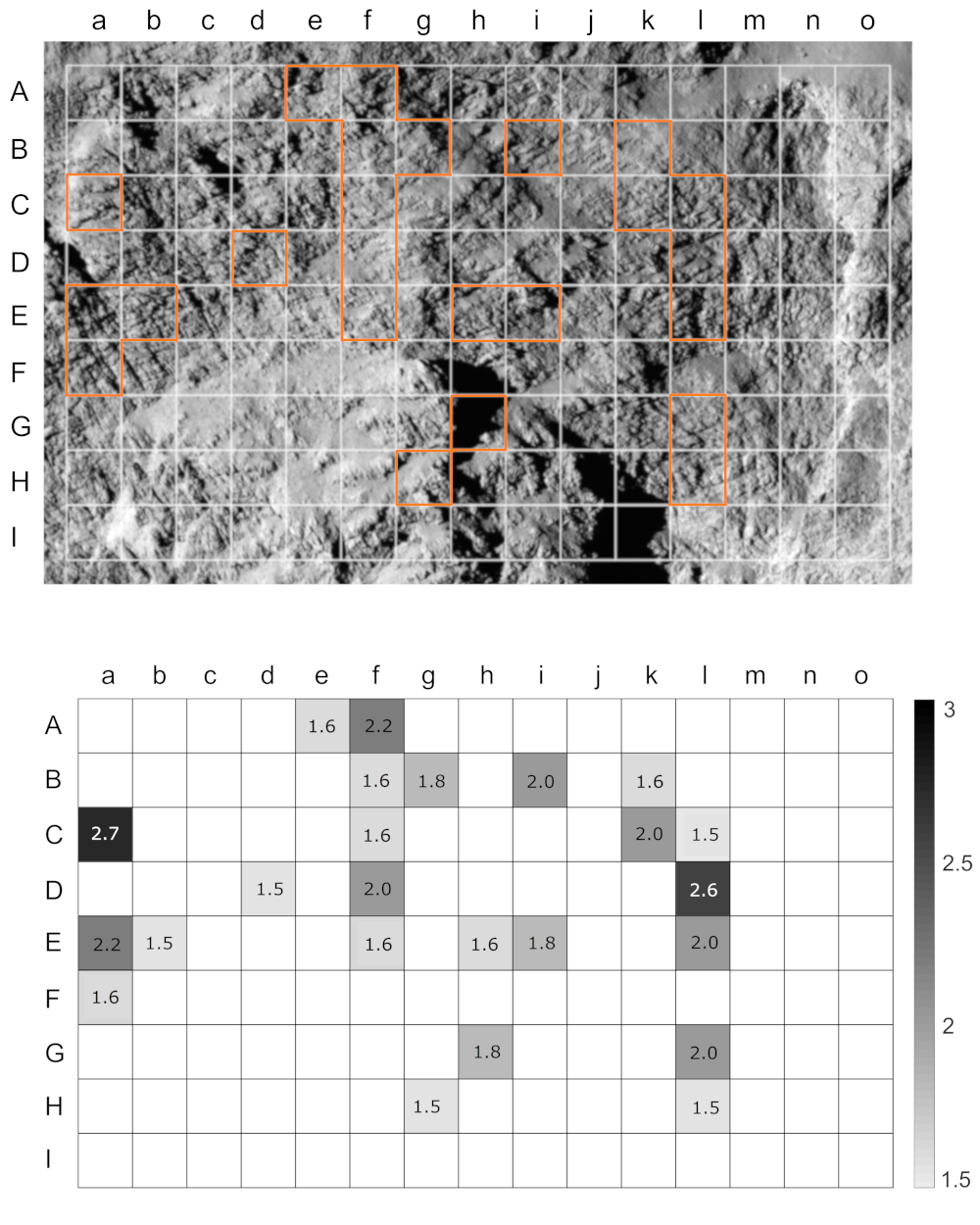}
	\caption[Map of peak-factors for the intensity-peaks of layerings]
	{Distribution of peak-factors for the intensity-peak associated with the \underline{layerings}. Input image of the Hathor wall shown on top for comparison.} 
	\label{fig:lay_peakfactor}
\end{figure}

\noindent\textbf{\autoref{fig:lay_peakfactor}} shows the \textbf{peak-factor} for each tile where layering-related lineaments were detected. Higher values mean that the peak was more distinctly defined in its prominence above the average intensity of all sectors, but the lower values also passed the cutoff criterion of 1.5 and are thus considered reliable. Values range from 1.5 to 2.7 with a mean of 1.9. The two frames with the highest peak-factors are 'Ca' and 'Dl'. When inspecting these two frames in the original image, the impression arises that in those areas, the contrast in brightness between the dark layer boundaries and the lighter intra-strata material is most strongly expressed. For a given location on the wall, the brightness of the intra-strata material is controlled by the phase angle and the roughness of the surface, whereas the darkness of the layer boundaries depends on how well they are shielded from the light source, considering that they have a negative topographic expression. The greater their depth, the more perpendicular their orientation is relative to the incoming sunlight, and the greater the phase angle is in their location, the darker they appear in the image.\\

\begin{figure} [h]  
	\centering
	\includegraphics[width=\linewidth]{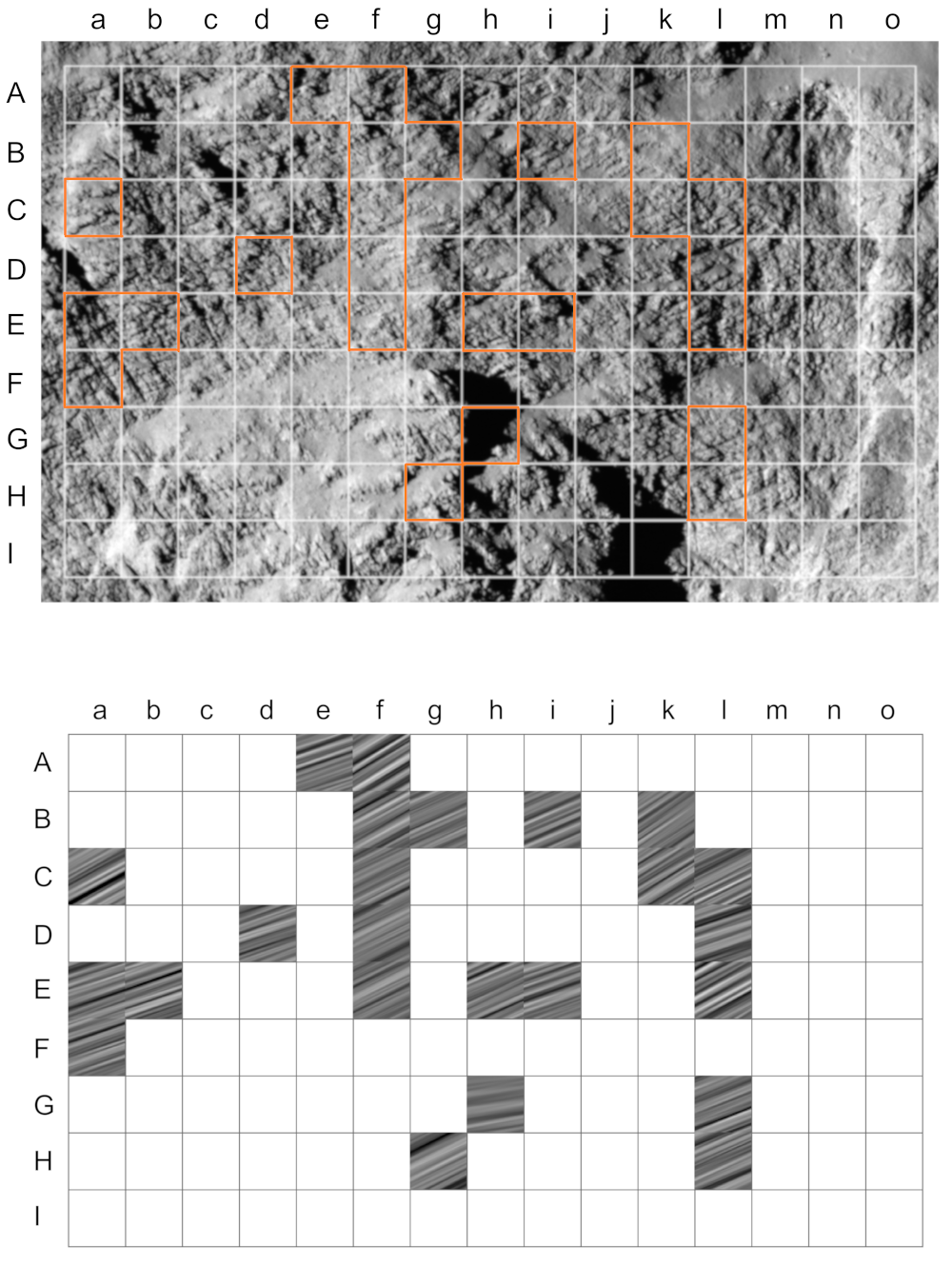}
	\caption[Map of backtransformations along the directions of layerings]
	{Backtransformation along the direction of the \underline{layering-associated} peak. Input image of the Hathor wall shown on top for comparison.}
	\label{fig:lay_backtransforms}
\end{figure}

\noindent\textbf{\autoref{fig:lay_backtransforms}} shows the \textbf{backtransformations} of frames, where layering-associated peaks were found. This map is a visual representation of the quantitative results shown in the previous two figures. \\

\newpage 

\newpage

\newpage

\newpage

\subsubsection*{The directions of downslope lineaments across the Hathor wall}

\begin{figure} [h] 
	\centering
	\includegraphics[width=\linewidth]{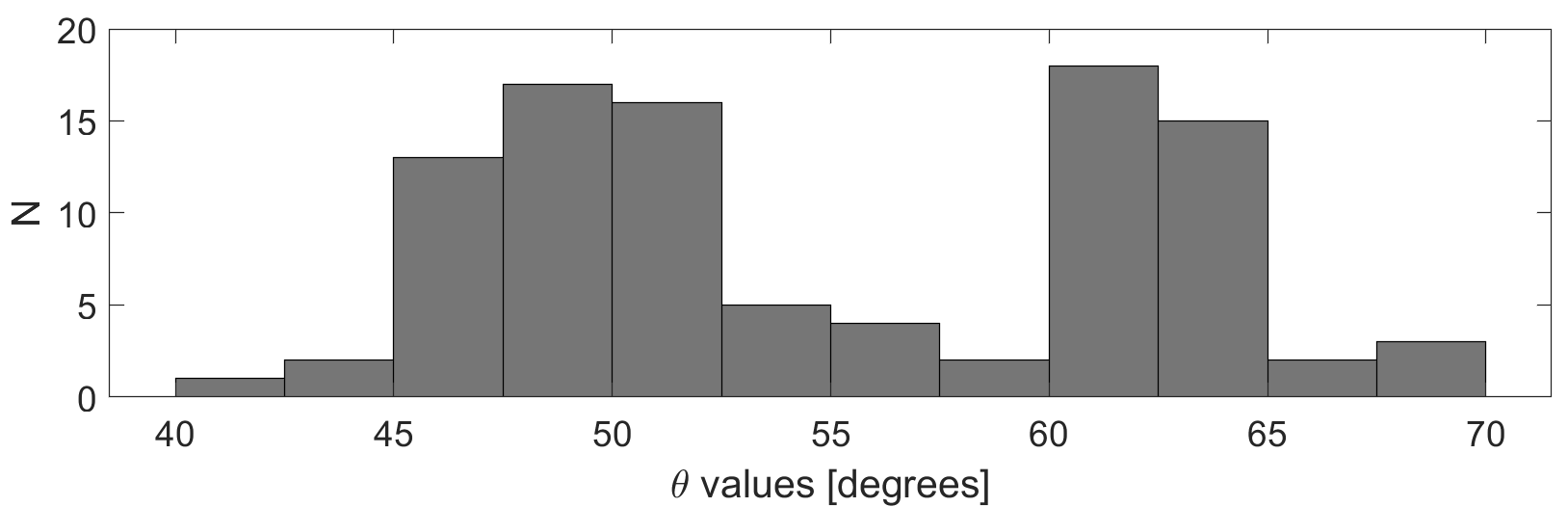}
	\caption[Histogram of $\theta$ values for downslope lineaments]
	{Histogram of $\theta$ values for downslope lineaments} 
	\label{fig:histogram_down}
\end{figure}

\textbf{\autoref{fig:down_theta}} shows the \textbf{orientation of $\theta$} values for the intensity peak associated with the proposed downslope lineaments. 89 peaks were found, so 89 tiles are filled in these maps. The $\theta$ values range from 41\textdegree{} to 69\textdegree{}, with a mean value of 54\textdegree{} and a standard deviation of 7.2\textdegree. The orientation $\alpha$ of the downslope lineaments thus varies from 131\textdegree{} to 159\textdegree{} (mean of 144\textdegree{}). The distribution of the orientations is bi-modal (\autoref{fig:histogram_down}), with one mode around 47\textdegree{} and one mode around 62\textdegree. There is no spatial trend in the distribution of values across the studied section of the Hathor wall. 

Downslope lineaments were detected on many areas of the input image. Areas where no downslope lineaments were detected mostly correspond to areas where fine-grained airfall-material covers up the underlying topography. No significant airfall-cover is distinguishable for most of the empty tiles in columns k, l, and m, but I propose that the lack of downslope lineaments in those areas is due to their location near the outer edge of the Hathor cliff. The bulk of the landslide mass that is believed to have formed the downslope lineaments (\autoref{ch_3_1}) might have missed these marginal areas - or at least affected them less strongly. This interpretation is supported by the observation that the remaining landslide mass resides centrally at the foot of the cliff, and only a small amount of material has come to rest towards its sides.\\

\noindent\textbf{\autoref{fig:down_peakfactor}} shows the \textbf{peak-factor} for each tile where downslope lineaments were detected. Values range from 1.5 to 4.8, with a mean of 2.3. Peak-factors for the downslope lineaments are much higher than for the layering-associated lineaments, which is likely explained in the same way as high peak factors in the layering-map: The downslope lineaments are furrows that appear to have even greater depth than the layering-boundaries, and are oriented at a high angle to the incoming sunlight. The greatest accumulation of high peak-factors is found in the lower-left corner of \autoref{fig:down_peakfactor}, where illumination of the matrix material appears to be brightest which creates a strong brightness contrast to the dark downslope furrows. \\

\noindent\textbf{\autoref{fig:down_backtransforms}} shows the \textbf{backtransformations} of frames, where downslope-associated peaks were found. This map qualitatively synthesises the quantitative results shown in the previous two figures. Within this map, the eye is drawn to some tiles whose $\theta$ values differ strongly from those of their immediate neighbours. Upon closer inspection, some of those tiles appear to show false detections, for example tile Dd whose $\theta$ is the lowest value in the distribution ($\theta = 41$\textdegree). There are barely any lineaments discernible by eye in tile Dd, and certainly none that strongly differ in orientation from the neighbouring tiles. Further examples are found in the tiles to the right of the landslide body, where the edges of shadows cast by the raised body were wrongfully classified as downslope lineaments.  \\

\begin{figure} [h]  
	\centering
	\includegraphics[width=\linewidth]{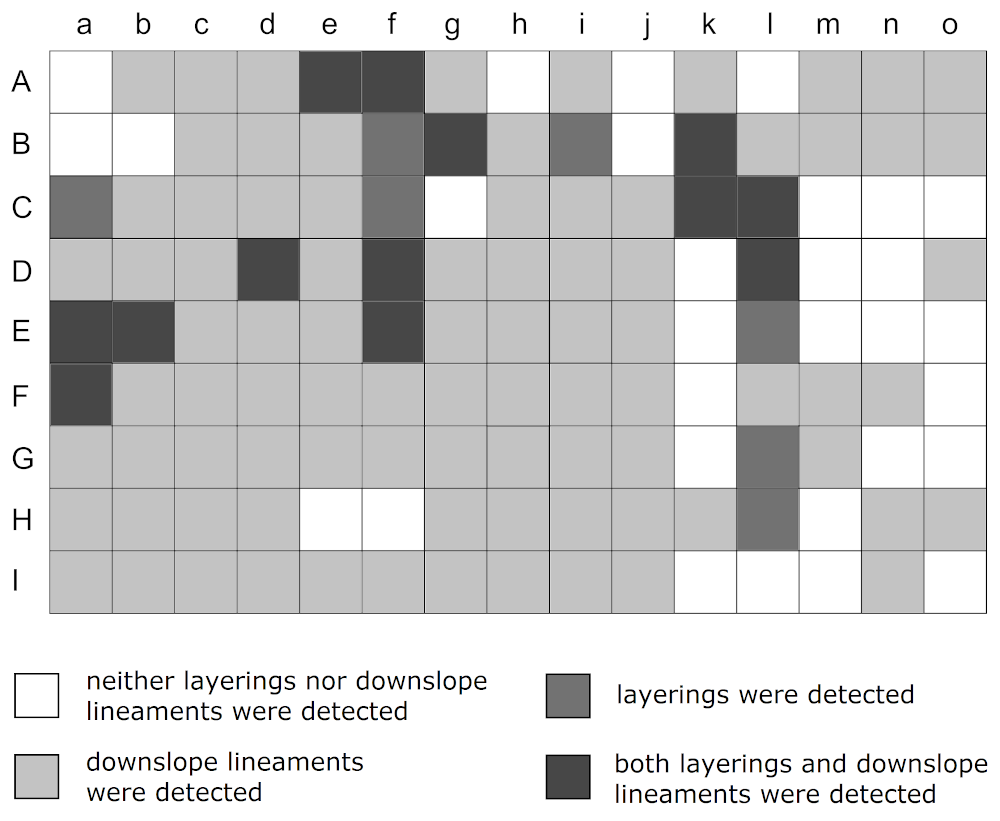}
	\caption[Map of peak detections by lineament type, and where no peaks were detected]
	{Map of peak detections classified by lineament type, also indicating where no peaks were detected.}
	\label{fig:no_peaks}
\end{figure}

\begin{figure} [h]  
	\centering
	\includegraphics[width=\linewidth]{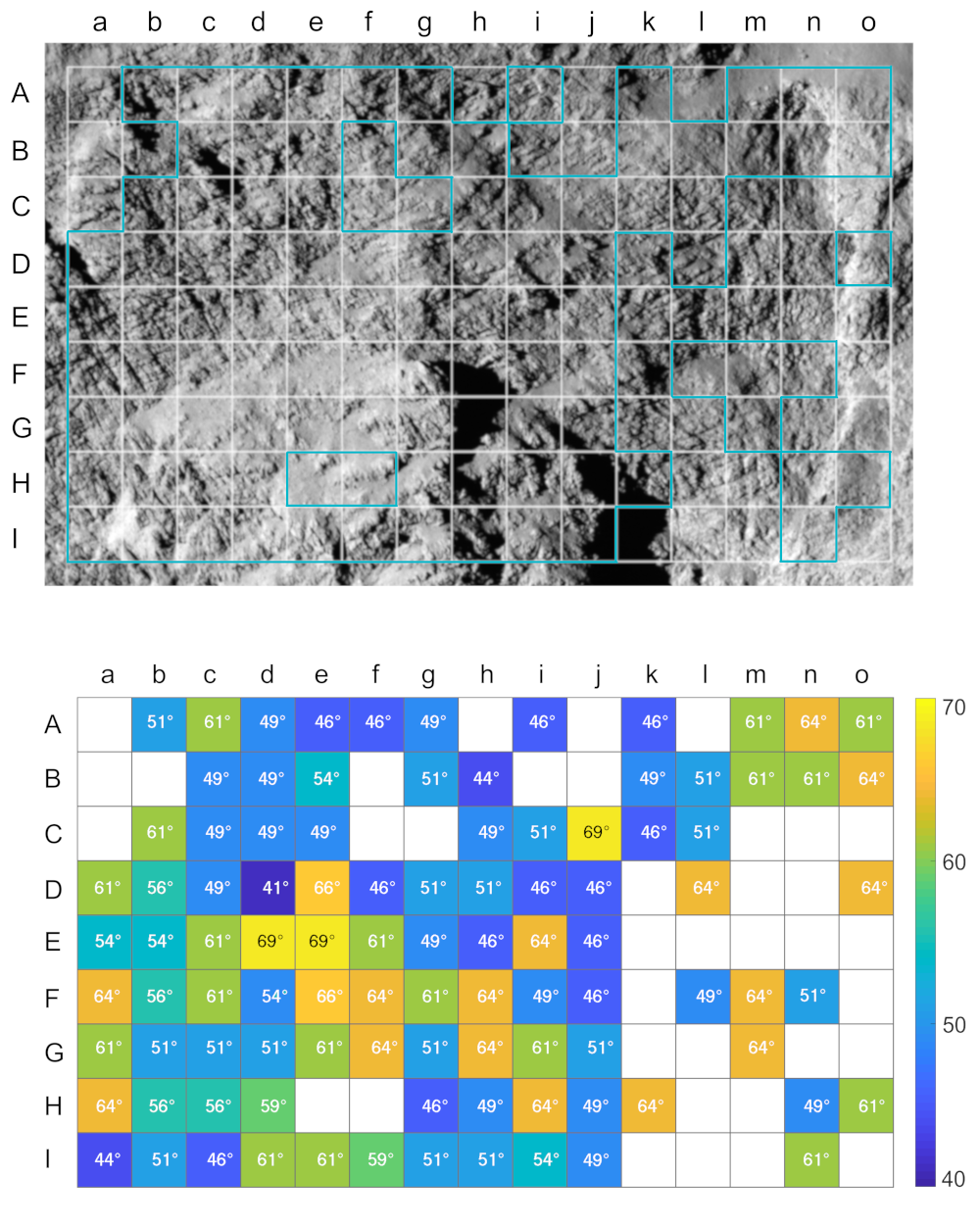}
	\caption[Map of $\theta$ values for the intensity-peaks of downslope lineaments]{Distribution of $\theta$ values (in degrees) for the intensity-peak associated with the \underline{downslope lineaments}. 89 peaks were found, their mean $\theta$ value is 54\textdegree. Input image of the Hathor wall shown on top for comparison.}
	\label{fig:down_theta}
\end{figure}

\begin{figure} [h]  
	\centering
	\includegraphics[width=\linewidth]{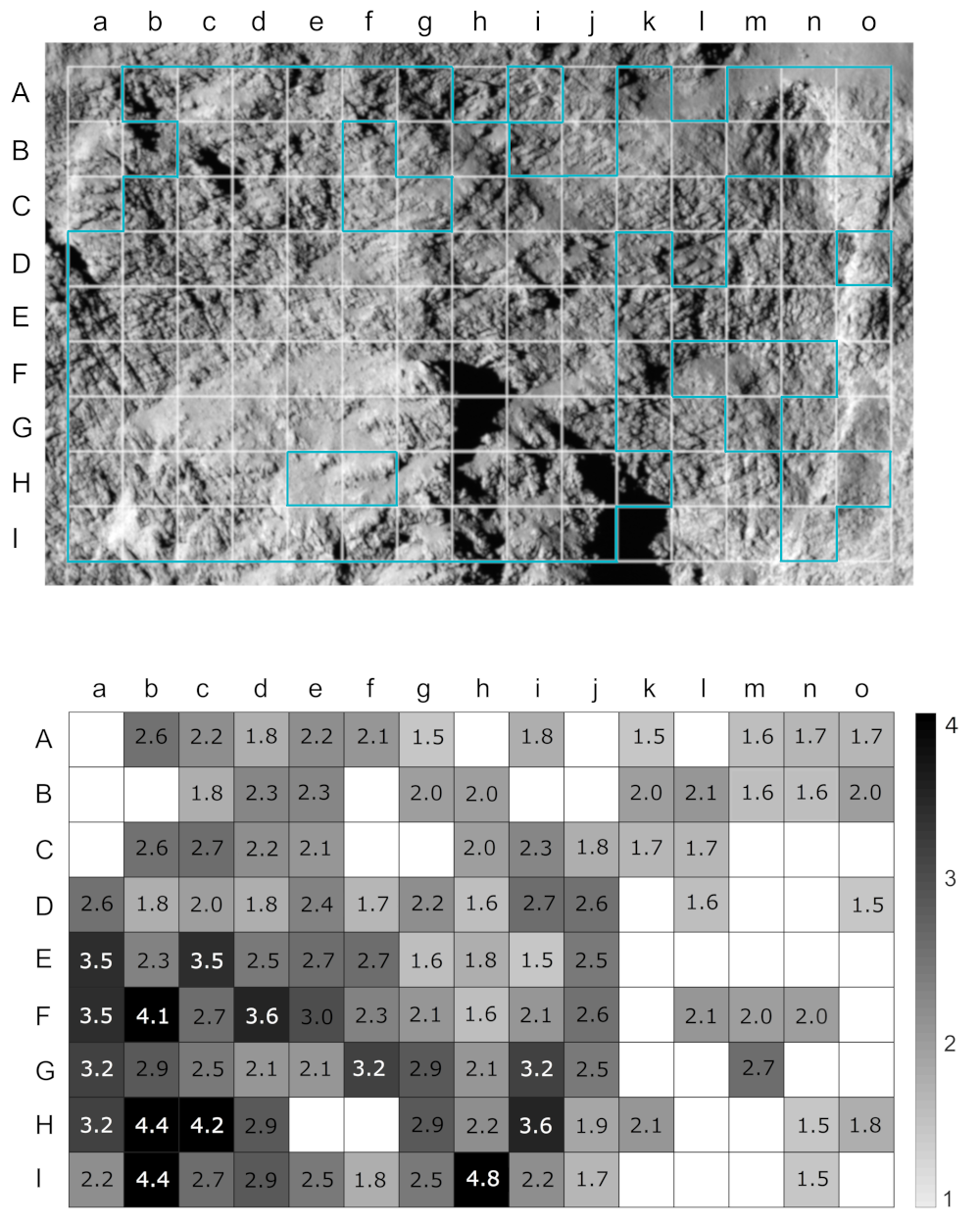}
	\caption[Map of peak-factors for the intensity-peaks of downslope lineaments]{Distribution of peak-factors for the intensity-peak associated with the \underline{downslope lineaments}. Input image of the Hathor wall shown on top for comparison.}
	\label{fig:down_peakfactor}
\end{figure}

\begin{figure} [h]  
	\centering
	\includegraphics[width=\linewidth]{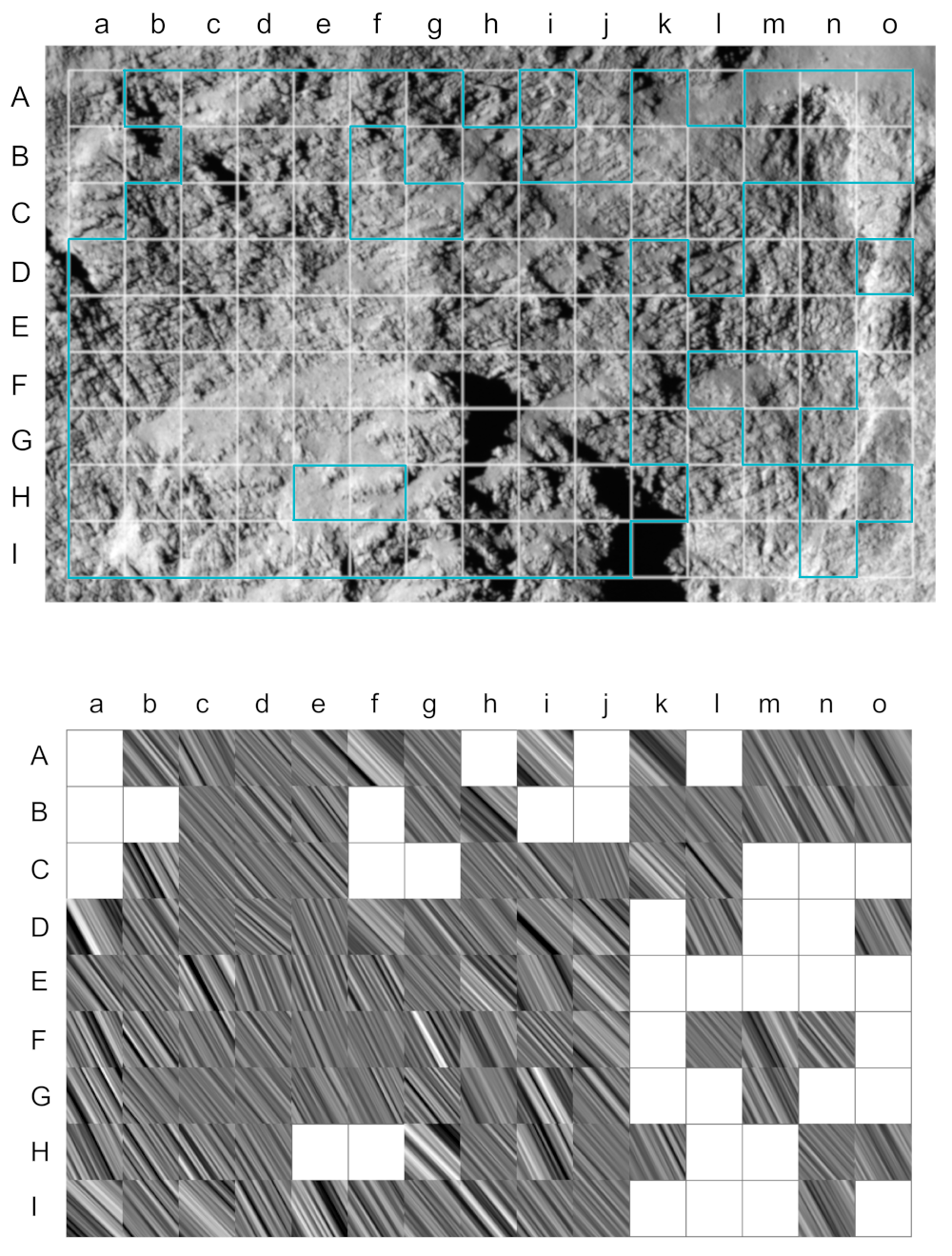}
	\caption[Map of backtransformations along the directions of downslope lineaments]
	{Backtransformation along the direction of the peak associated with \underline{downslope lineaments}. Input image of the Hathor wall shown on top for comparison.}
	\label{fig:down_backtransforms}
\end{figure}

\noindent\textbf{\autoref{fig:no_peaks}} summarises which type of lineament was detected for each tile on the maps. It also indicates where detectable layerings and downslope lineaments overlap, and for which areas no peaks were detected at all. 

\vspace{3cm}

\newpage

\newpage

\newpage

\newpage

\subsubsection*{Dominant wavelengths of the lineaments}

Analysing the spacing of lineaments on the Hathor wall was not the main focus of my study, as Fourier image analysis is not suited to quantify distances in this type of image. Rather, it is a statistical method and the closest it comes to describing the spacing is in finding the dominant wavelength perpendicular to the direction of lineaments. The dominant wavelength is the wavelength of the signal which most strongly contributes to the lineaments of interest in an area. As the lineaments are not uniformly spaced (i.e. periodic), this wavelength is not equal to their spacing. Rather, I expect the spacing to be in the neighbourhood of the two peak values the frame's power spectrum (\autoref{fig:powerspectrum}). 

\begin{figure} [h] 
	\centering
	\includegraphics[width=0.83\linewidth]{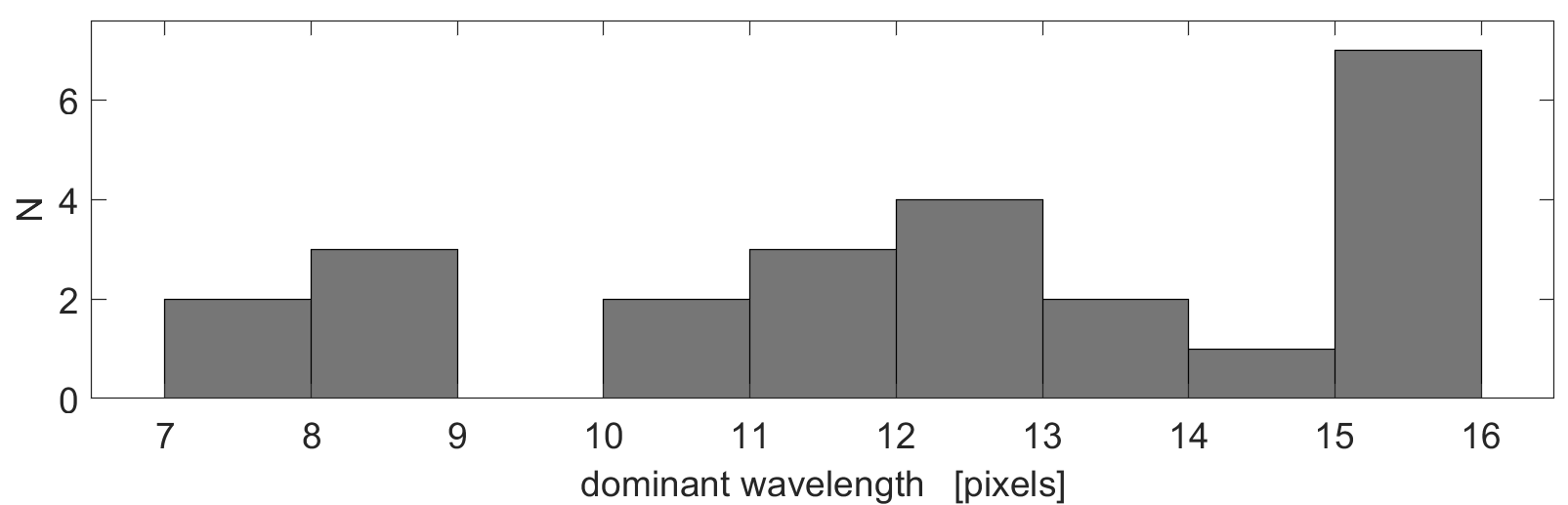}
	\caption[Histogram of wavelengths for layering-associated lineaments]
	{Histogram of wavelengths for layering-associated lineaments} 
	\label{fig:histogram_lambda_lay}
\end{figure}

The dominant wavelengths of signals perpendicular to the direction of the \textbf{layering-related lineaments} range between 7 and 15 pixels, with a mean of about 12 pixels and a standard deviation of 2.7 pixels (\autoref{fig:histogram_lambda_lay} and \ref{fig:lay_lambda}). I therefore expect the spacing of the layerings in the Hathor cliff to be close to 12 pixels, or $\sim$12 metres (cf. \autoref{ch_3_3_2}). 

There is a clustering of tiles with lower wavelength (i.e. higher spatial frequency) in columns a to d, whereas no trends or clusters are discernible within the other columns.

\begin{figure} [h] 
	\centering
	\includegraphics[width=0.83\linewidth]{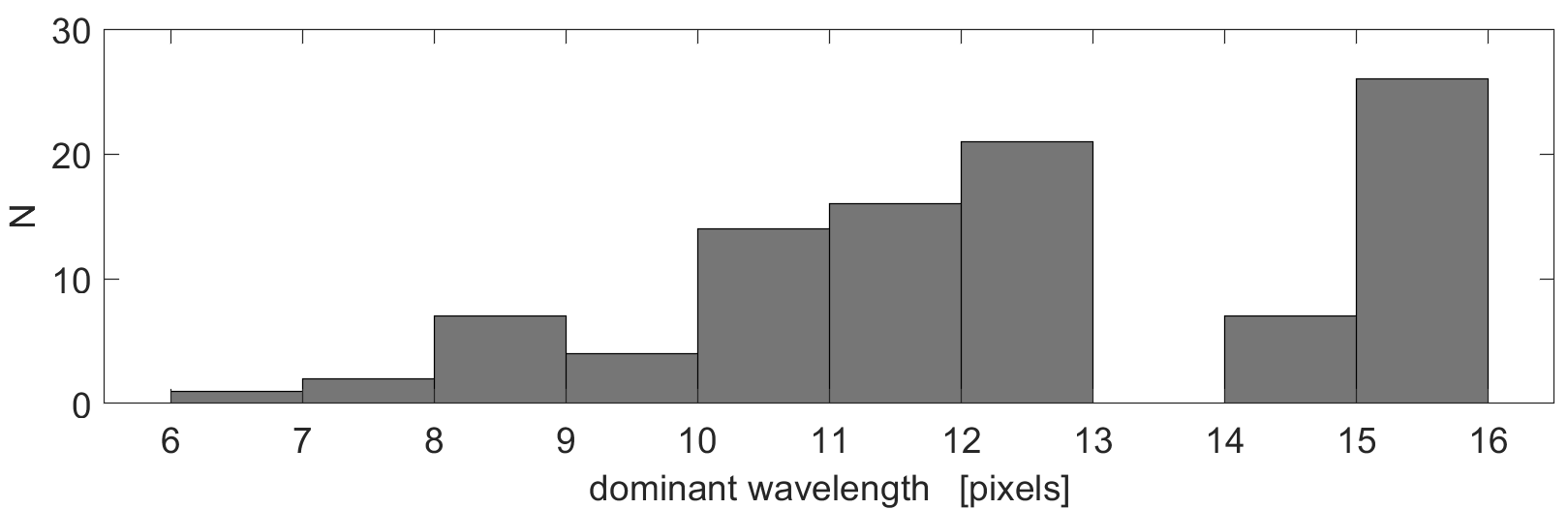}
	\caption[Histogram of wavelengths for downslope lineaments]
	{Histogram of wavelengths for downslope lineaments} 
	\label{fig:histogram_lambda_down}
\end{figure}

\begin{figure} [h]  
	\centering
	\includegraphics[width=\linewidth]{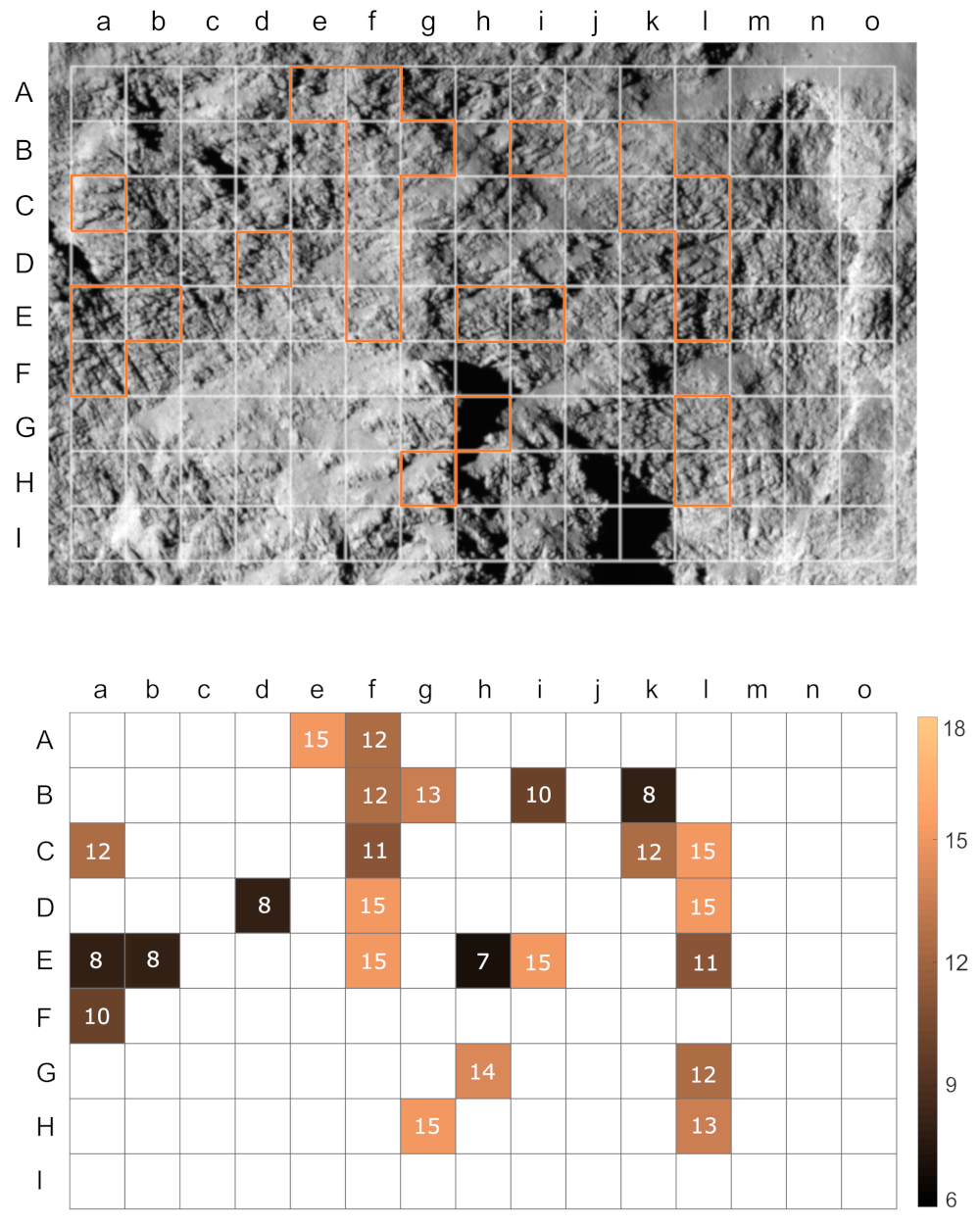}
	\caption[Map of dominant wavelengths found for the layering-associated lineaments]
	{Distribution of dominant wavelengths found for the \underline{layering-related lineaments}, in image pixels. Values range from 7 px to 15 px, the mean is 12.2 px. These wavelengths are statistical image properties and are not indicative of the actual layering thickness.}
	\label{fig:lay_lambda}
\end{figure}

\begin{figure} [h]  
	\centering
	\includegraphics[width=\linewidth]{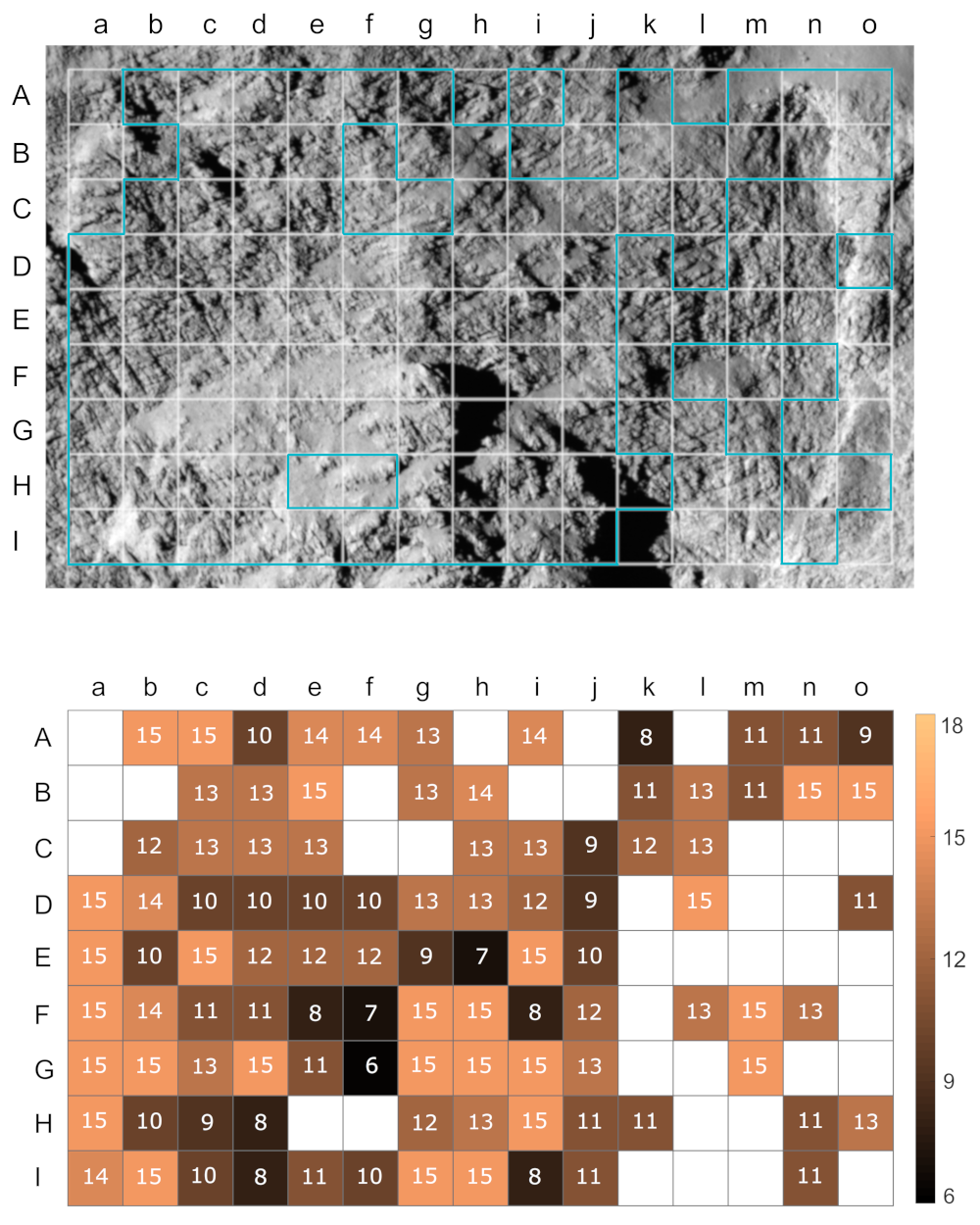}
	\caption[Map of dominant wavelengths found for the downslope lineaments]
	{Distribution of dominant wavelengths found for the \underline{downslope lineaments}, in image pixels. Values range from 6 px to 15 px, the mean is 12.6 px. True values are likely considerably higher, as the frame size was optimised for detecting layerings and these lineaments are separated by greater distances than the layerings.}
	\label{fig:down_lambda}
\end{figure}

The dominant wavelengths of signals perpendicular to the direction of the \textbf{downslope lineaments} range from 6 to 15 pixels, with a mean of about 13 pixels and a standard deviation of 2.3 pixels (\autoref{fig:histogram_lambda_down} and \ref{fig:down_lambda}). However, the true dominant wavelength of the downslope lineaments is likely even higher than 15 pixels, as the algorithm is optimised for detecting layerings. The ideal frame size for detecting layerings is 101 $\times$ 101 pixels, as the layerings are closely spaced. However, a frame size of 201 $\times$ 201 pixels would likely be a more effective frame size to study the spacing of the downslope lineaments, which appears to be separated by greater distances than the layerings.

\newpage

\newpage

\newpage

\subsection{Demonstration of algorithm on different image}

\begin{figure} 
	\includegraphics[width=0.95\textwidth]{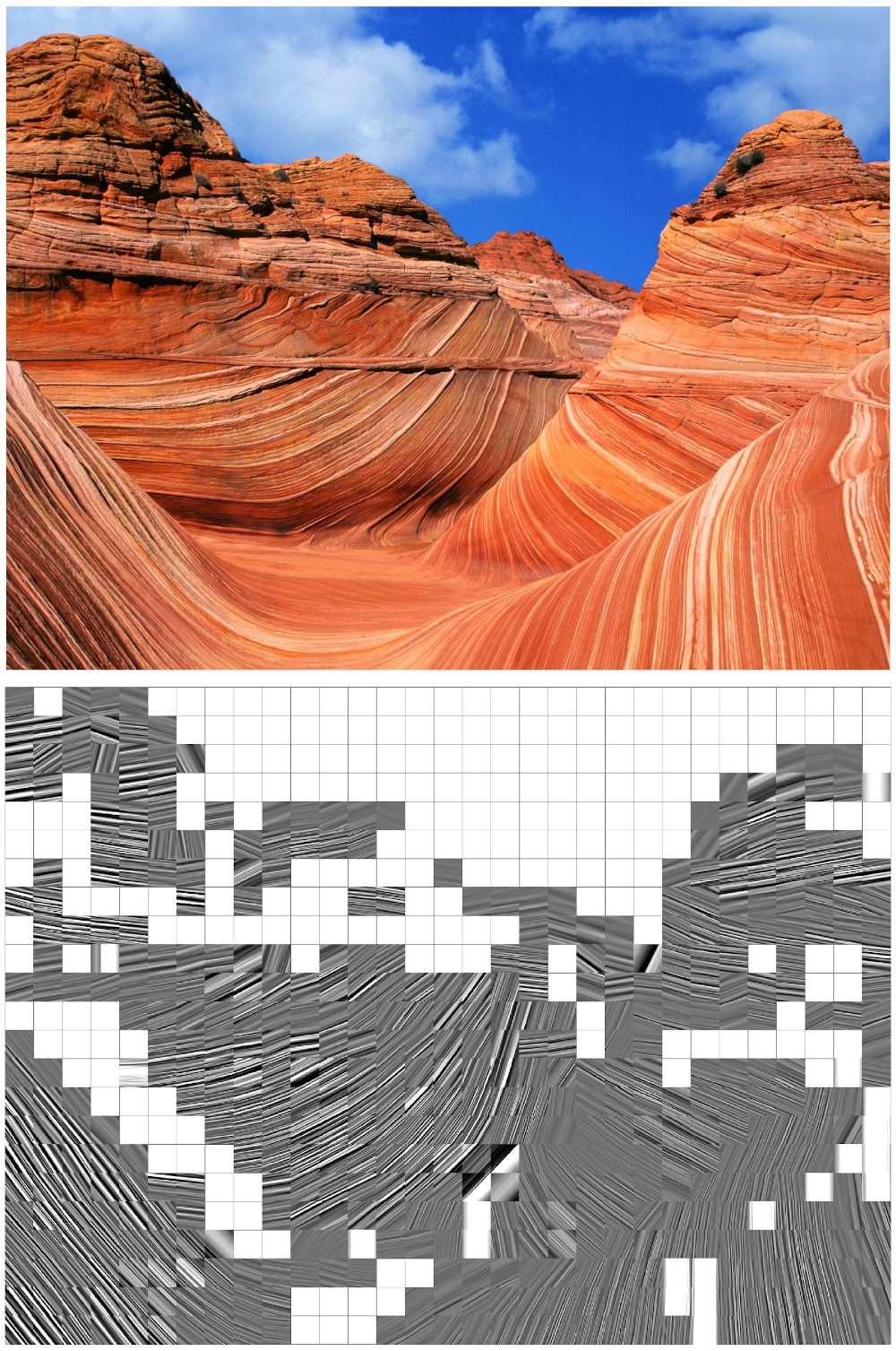}
	\caption[Result of lineament-detection algorithm run on a different photograph]{\textbf{Top:} Photograph of the famous sandstone formation called 'The Wave', located in Arizona (Earth). Image cropped from \citet{lineau_2009_thewave}. \textbf{Bottom:} Map of backtransformed tiles along the direction of the highest peak.}
	\label{fig:thewave}
\end{figure}

\autoref{fig:thewave} is a demonstration of my algorithm on a vastly different example image. The input image is a colour photograph of a sandstone formation in Arizona, USA, called 'The Wave'. The formation has been eroded into cross-bedded, layered eolian sandstone. The layerings differ in colour and hardness due to variations in grain size \citep{freeman_1975_navajo}. 

I ran the \texttt{fullcliffscan.m} algorithm with the same configuration of parameters as I used for the Hathor cliff. As the size of the tiles remains unchanged, the resulting map for this image of size 1201 $\times$ 1601 pixels is more highly resolved than the maps of the 501 $\times$ 801 Hathor image. 

The map of backtransformations along the direction of the highest peak in the Fourier intensity spectrum (\autoref{fig:thewave}, bottom) clearly reproduces the main structure visible within the sandstone. Almost all empty tiles (i.e. no detections) below the skyline are caused by the Fourier energy not clearing the Minimum Peak Height threshold. Lowering the threshold, however, introduces false detections, which is likely due to the fact that the image areas covered by those tiles contain more than one layering orientation. As the algorithm is not designed to handle more than one peak per tile, this leads to empty tiles. The number of such empty tiles could be reduced by either choosing smaller tiles, or better yet using a larger image.

Rightly, no lineaments were detected in the sky.

This, as well as further experiments I ran on a spectrum of different scenes (including images of thin-sections and photographs taken by the Mars rovers) confirms how powerful the algorithm is in identifying layering-related lineaments under close-to-ideal circumstances (i.e. opportune lighting conditions and perspective, and no vegetation or lose sediment shrouding the surface).

\newpage

\section{Conclusions}  \label{ch_3_4}
I studied the lineament structures on the Hathor cliff wall on comet 67P/Churyumov-Gerasimenko, which appear to be surface expressions of the comet's subsurface layerings. Previous works manually mapped layering-associated lineaments on a small part of this cliff \citep{belton_origin_2018} and in other locations on the comet nucleus \citep{ruzicka_mnras_2018}, but manual mapping from images carries inherent limitations. The spatial resolution of manual mapping is generally well below the resolution of the image, the process is time-intensive and bothersome to repeat if a parameter needs to be altered, and it is vulnerable to bias. Sources of bias include inconsistency in method between researchers, hardware quality, and most importantly psychological factors like confirmation bias. 

It was the goal of this study to develop a mapping method for the Hathor cliff that overcomes these limitations, and instead allows an objective characterisation of dominant orientations. When conventional methods of image processes, such as edge-detection algorithms, proved to be inapplicable to images of the cliff due to the bumpy surface texture and the adverse lighting and contrast conditions, I turned to a statistical approach to the issue. I developed an algorithm that splits the image into frames, identifies the directions of the most frequently appearing lineaments in each frame by evaluating its Fourier domain, and then re-assembles the frames into a map that shows where on the image layerings were detected and what their orientation is. This procedure is fully automated within the algorithm, the user merely needs to select suitable values for five 'input parameters' according to the surface- and lighting-conditions of the image. 

When applying the algorithm to a section of an image of the Hathor wall that shows a high spatial concentration of the proposed layerings, three sets of lineaments with different main orientations were detected. One group of linear structures represents the direction of shadows cast by the uneven topography, the second group of structures comprises sub-parallel furrows that are oriented downhill (or parallel to the local gravity vector), and the third group of structures is oriented approximately perpendicular to the local gravity vector and makes up lineaments proposed to be layerings. 

For these layering-associated lineaments, the algorithm's positive detections mostly overlap with those frames where the human brain recognises layerings as well. But the algorithm also shows positive detections for additional frames, meaning that there are areas in the image where layerings are not humanly discernible but where periodic variations in brightness exist within the image data and have an appropriate orientation for detection. In this way, the signal-to-noise detection sensitivity of the algorithm is higher than that of a human researcher. However, the algorithm did not yield a positive detection for every tile that a researcher identified as showing layerings, particularly along the edge of the Hathor cliff and in areas of ambiguous texture. This confirms that appropriate input parameters and thresholds were selected such that the algorithm is sufficiently conservative in its positive detections. I ascribed the 'human false positives' to confirmation bias and the brain's tendency to continue a pattern.

While I was not able to precisely quantify the layerings' thickness, I found their spacing on the small lobe, as exposed by the Hathor cliff, to be on the order of 12 metres. This result, in conjunction with my other findings and observations, leads me to concur with \citet{belton_origin_2018} who propose that the layerings on comet 67P have an average radial thickness of $\sim$14 metres.

In addition to the Hathor wall, I tested the applicability of my algorithm in several different environments, and found that it frequently produced satisfactory results simply using the parameter configuration I derived for the Hathor wall. The success of applying the algorithm to any image hinges on selecting an appropriate frame size with respect to the features of interest, and ideally using orthoimages taken under advantageous lighting conditions. 

A potential practical application of this algorithm lies in in-situ analysis of targets in space. In recent years, CCDs and processors have become exponentially more powerful, while the bandwidth for data transmission back to Earth is limited by physical factors. Processing images directly on board and merely sending results would free up some of this bandwidth for other data. However, the method still requires that a human decides which parameters are appropriate for the environment in which the images are taken. Another potential application might be in assisting geological mapping or digital outcrop analysis from images taken under sub-optimal conditions. Although my algorithm does not result in a geological map, it might be helpful in guiding future researchers in the creation of such maps, should they encounter ambiguous conditions. 

In conclusion, I successfully applied the Fast Fourier Transform to identify layerings on images of comet 67P, constrained their orientation and spacing, and demonstrated the potential of Fourier analysis as a tool for locating linear structures of interest in images where edge detection algorithms fail.

\chapter{Conclusions: Layerings in comet 67P and their formation} \label{ch_4}

The aim of this thesis was to study the layerings in the nucleus of comet 67P/Churyumov-Gerasimenko in order to advance our understanding of the formation of layerings in cometary nuclei in general. When I began my studies, the first high-resolution images of comet 67P's surface were raising questions about the numerous exposures of layered sequences visible on these images:

\begin{enumerate}
    \item Are the layerings a global, intrinsic property of the nucleus, or locally constrained features?
    \item What is their geometry and their geometric relation to each other?
    \item What is their thickness?
    \item What is the composition of the intra-strata material, and what distinguishes it from the boundary planes that separate the layerings? 
\end{enumerate}   

All of these questions aimed at understanding how and when these layerings were formed in the comet. Applying techniques of structural geology, three-dimensional modelling, and statistical image processing, I contribute the following results to this matter: 

1. The layerings are of global extent and appear to be a fundamental property of the nucleus' internal structure. I mapped layering-related, curved lineaments on the nucleus surface and fitted planes through their nodes, to approximate the orientations of the layering-planes represented by these surface lineaments. I found that these orientations are in agreement with a cohesive, regular, concentric inner structure. The cumulative thickness of layerings exposed on cliff faces and within pits, as well as their occurrence at a range of distances from each lobe's centre of gravity, indicate that the layered structure extends to at least several hundred metres below the present-day surface.

2. Strike and dip of the layerings indicate that they are organised in two subsets on the two lobes of the nucleus that are geometrically independent from each other, ruling out the theory that comet 67P was once a single body that assumed its current shape through differential erosion. The layerings' geometry can be approximated by two sets of concentric ellipsoidal shells. The layering exposures on the Big Lobe can be easily fitted with simple ellipsoids, suggesting a regular internal structure, whereas the layerings on the Small Lobe require a more complex shape. Perhaps the internal structure of the Small Lobe was deformed during the collision that connected it to its current partner lobe, or during a previous collision event. Layering-related lineaments can be laterally traced for more than 800\,m on average, and up to 1925\,m in some places. From all available observations, layerings on both lobes show a constant thickness and parallel layer boundaries anywhere they are exposed at the surface.

3. The thickness of individual layerings, as exposed on cliff faces, is less than 20\,m on average and might well be on the order of 12\,m. Constraining their thickness in other locations is challenging, as the nucleus surface is heavily affected by erosion and largely covered in airfall material, both of which obfuscate such minor structures on the surface. For much the same reasons, I did not focus my research on the intra-strata material (4.). 

It is my understanding that two mechanisms of formation are compatible with the aforementioned results. First, the layerings could have been formed during the initial accretion of the nucleus in the early solar system. In this case, the nature of the layerings might be considered as 'sedimentary', and their creation would require periodic variations of either material or conditions during accretion. I hypothesise that these variations might occur if the comet accreted within an inhomogeneous protoplanetary disc, or phases of warming led to sinter processes hardening the surface between phases of accretion.

The second compatible option is the theory proposed by \citet{belton_origin_2018}, in which a primordial and mostly homogeneous rubble-pile nucleus is transformed into a layered body late in its dynamical lifetime. The mechanism that forms the layerings would here be the self-sustaining, dual-mode propagation of amorphous to crystalline water ice phase-change fronts from the surface into the nucleus interior. The authors expect this mechanism to create geologically young layerings of global extent with an approximate radial thickness of $\sim$14\,m. They hypothesise that the direction of propagation of the fronts is controlled by the radial outflow of CO and a coarse layered structure in the primeval material below the front. However, the origin of these primeval layerings remains unclear.

The research on layerings in cometary nuclei is far from exhausted, and will certainly greatly benefit from the entire dataset of level 5 calibrated OSIRIS data expected to become available in the near future.

\bibliography{thesis} 

\appendix
\cleardoublepage
\thispagestyle{empty}
\vspace*{2.2cm}
{\Huge\bf Appendix}

\chapter{Appendix: Supplementary material for chapter 2} \label{app:ch2}

\section{Supplementary figures}

Supplementary material is available online \citep{ruzicka_supplementary_2018}. It includes \textbf{Figures 2.1}, \textbf{2.2} and \textbf{2.3} in full resolution, \textbf{Figure 2.2} as a rotatable \textsc{matlab} figure, and supplemental \textbf{Figure A1}, which is also printed here.

\vspace{1cm}

\begin{figure}[h]
	\includegraphics[width=\textwidth]{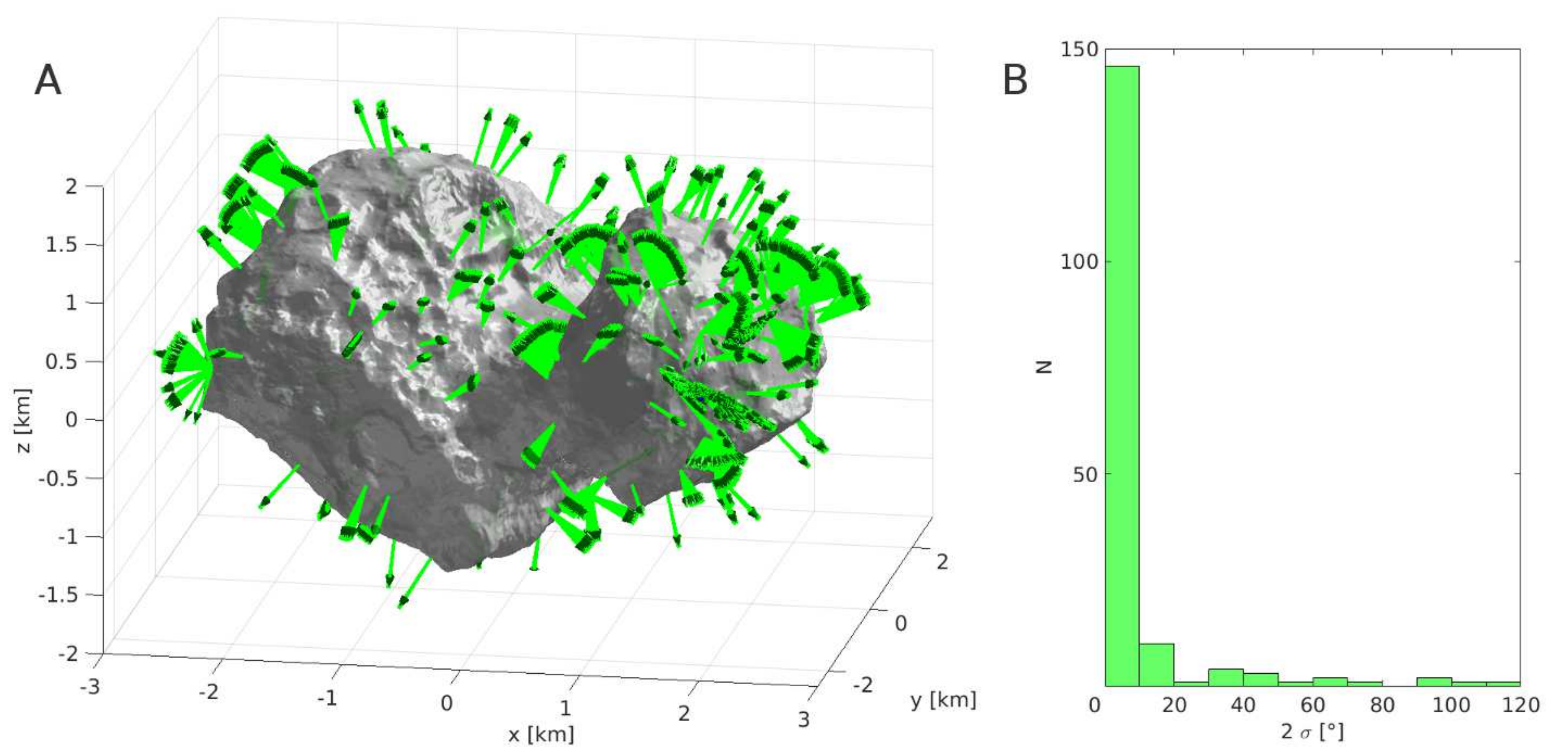}
	\caption[Monte Carlo analysis of plane normals]{Monte Carlo analysis of plane normals, as described in the main text. \textbf{A:} Plot of all Monte Carlo solutions. Each set of green arrows represents the normal vectors of a single linear feature. \textbf{B:} The Monte Carlo solutions of most features show low variability, as seen in this histogram of the standard deviations (2$\sigma$).}
	\label{fig:figure_A1}
\end{figure}

\vspace{1cm}

\section{Supplementary tables}

The additional tables are included in the supplementary file and also printed here: \textbf{Table A1} lists all OSIRIS NAC images (and corresponding CODMAC calibration levels) that were used for mapping, \textbf{Table A2} is a complete table of all feature normals I determined.

\begin{table} [h]
	\centering
	\caption[OSIRIS NAC images used for mapping the lineaments]{OSIRIS NAC images used for mapping the lineaments. The images were retrieved from the ESA Planetary Science Archive and have different calibration levels.}
	\label{tab:tableA1}
	\begin{tabular}{lc}
		\hline
		Image identifier (ID) & CODMAC calibration level\\
		\hline
		N20140808T033734555ID30F22 & 3 \\
		N20140821T204254581ID30F22 & 3 \\
		N20140829T014254551ID30F22 & 3 \\
		N20140902T174322588ID30F22 & 3 \\
		N20140903T034422640ID30F22 & 3 \\
		N20140905T024555555ID30F22 & 3 \\
		N20140905T041055537ID30F22 & 3 \\
		N20140905T053055563ID30F22 & 3 \\
		N20140913T081908366ID40F22 & 4 \\
		N20140913T110900348ID40F22 & 4 \\
		N20140913T130124352ID40F22 & 4 \\
		N20140913T214848344ID40F22 & 4 \\
		N20140915T143919354ID40F22 & 4 \\
		N20140916T174504366ID40F22 & 4 \\
		N20140918T044208376ID40F22 & 4 \\
		N20140919T014248350ID30F22 & 3 \\
		N20140919T093855332ID40F22 & 4 \\
		N20140919T230900386ID40F22 & 4 \\
		N20140920T003052382ID40F22 & 4 \\
		N20140920T183400391ID40F22 & 4 \\
		N20140922T154936334ID40F22 & 4 \\
		N20150214T033352396ID30F22 & 3 \\
		N20150310T142240594ID30F22 & 3 \\
		N20150317T034249413ID30F22 & 3 \\
		N20160102T152325652ID30F22 & 3 \\
		N20160117T002231512ID30F22 & 3 \\
		N20160117T034405521ID30F22 & 3 \\
		N20160117T051405515ID30F22 & 3 \\
		N20160127T110132974ID30F22 & 3 \\
		N20160127T115341927ID30F22 & 3\\
		\hline
	\end{tabular}
\end{table} 

\newpage

\textcolor{white}{placeholder}

\begin{table} [h] 
	\centering
	\caption[Polylines of strata heads (normals)]{Polylines of \textbf{strata heads}: Normal vectors $\boldsymbol{n}$ and their bases $\boldsymbol{nb}$ (reference points). r is the radial distance of each vector base from the gravitational centre of the lobe \citep[Table 3 in][]{jorda_global_2016}. Values are in [km] in the comet-fixed Cheops reference frame.}
	\label{tab:tableA2}
	\begin{tabular}{rrrrrrr}
		\hline
		$\boldsymbol{nb_x}$ & $\boldsymbol{nb_y}$ & $\boldsymbol{nb_z}$ & $\boldsymbol{n_x}$ & $\boldsymbol{n_y}$ & $\boldsymbol{n_z}$ & r\\
		\hline
		\multicolumn{7}{l}{Values for the big lobe:}\\
		0.159 & 1.384 & 0.869 & 0.381 & 0.449 & 0.807 & 1.75\\
		0.141 & 1.307 & 0.915 & 0.183 & 0.692 & 0.698 & 1.71\\
		-1.943 & 0.160 & 1.074 & -0.381 & 0.010 & 0.924 & 1.70\\
		-1.973 & 0.055 & 1.143 & -0.690 & 0.132 & 0.711 & 1.77\\
		-1.936 & -0.021 & 1.253 & -0.224 & -0.001 & 0.974 & 1.82\\
		-1.906 & 0.058 & 1.285 & -0.154 & 0.100 & 0.983 & 1.82\\
		-2.053 & -0.173 & 0.966 & -0.532 & -0.197 & 0.823 & 1.74\\
		-2.036 & -0.126 & 0.919 & -0.767 & 0.064 & 0.637 & 1.69\\
		-0.690 & 0.219 & 1.475 & -0.061 & -0.037 & 0.997 & 1.53\\
		-0.013 & 0.891 & 1.124 & 0.704 & -0.061 & 0.707 & 1.54\\
		0.008 & 0.855 & 1.090 & 0.757 & 0.070 & 0.648 & 1.50\\
		-0.288 & -0.306 & 0.569 & 0.520 & -0.539 & 0.661 & 0.86\\
		-0.162 & -0.106 & 0.648 & 0.426 & -0.601 & 0.675 & 0.90\\
		-1.032 & -0.565 & 1.013 & -0.616 & -0.273 & 0.738 & 1.33\\
		-1.008 & -0.516 & 1.015 & -0.294 & -0.477 & 0.827 & 1.30\\
		-1.199 & -0.257 & 1.282 & 0.019 & -0.535 & 0.844 & 1.49\\
		-1.899 & 0.189 & 1.271 & -0.262 & 0.113 & 0.958 & 1.80\\
		&&&&&&\\
		\multicolumn{7}{l}{Values for the small lobe:}\\
		1.090 & 0.364 & 1.165 & -0.834 & 0.346 & 0.428 & 1.29\\
		\hline
	\end{tabular}
\end{table}

\textcolor{white}{placeholder}
\vspace{6cm}
\textcolor{white}{placeholder}

\begin{table}  
	\centering
	\caption[Polylines of terrace edges (normals)]{Polylines of \textbf{terrace edges}: Normal vectors $\boldsymbol{n}$ and their bases $\boldsymbol{nb}$ (reference points). r is the radial distance of each vector base from the gravitational centre of the lobe \citep[Table 3 in][]{jorda_global_2016}. Values are in [km] in the comet-fixed Cheops reference frame.}
	\label{tab:tableA3}
	\begin{tabular}{rrrrrrr}
		\hline
		$\boldsymbol{nb_x}$ & $\boldsymbol{nb_y}$ & $\boldsymbol{nb_z}$ & $\boldsymbol{n_x}$ & $\boldsymbol{n_y}$ & $\boldsymbol{n_z}$ & r\\
		\hline
		\multicolumn{7}{l}{Values for the big lobe:}\\
		-0.184 & 0.055 & 0.940 & 0.433 & -0.456 & 0.777 & 1.11\\
		-0.703 & 1.098 & -0.408 & 0.808 & -0.500 & 0.310 & 1.02\\ 
		-1.176 & 1.153 & 0.831 & 0.146 & -0.670 & 0.727 & 1.43\\
		0.234 & 0.925 & 0.844 & 0.761 & 0.122 & 0.637 & 1.49\\
		0.383 & 0.823 & 0.646 & 0.567 & 0.305 & 0.764 & 1.43\\
		-0.003 & 1.515 & 0.932 & 0.312 & 0.565 & 0.763 & 1.81\\
		0.218 & 0.899 & 0.860 & 0.733 & 0.298 & 0.610 & 1.48\\
		-1.350 & -0.803 & 0.983 & -0.049 & -0.554 & 0.830 & 1.56\\
		-1.141 & 0.252 & 1.535 & -0.134 &  -0.311 & 0.940 & 1.66\\
		-0.833 & 0.821 & 1.476 & 0.156 & 0.244 & 0.957 & 1.67\\
		-0.389 & 0.815 & 1.403 & 0.054 & -0.035 & 0.997 & 1.62\\
		-1.427 & 0.280 & -0.292 & 0.786 & -0.558 & 0.264 & 0.80\\
		-0.754 & 0.705 & -0.853 & 0.598 & -0.621 & 0.505 & 0.98\\
		-0.610 & 1.151 & -0.583 & 0.643 & -0.679 & 0.353 & 1.14\\
		-1.336 & 0.248 & -0.642 & 0.700 & -0.631 & 0.333 & 0.90\\
		-1.352 & 0.229 & -0.612 &   0.557 & -0.659 & 0.503 & 0.89\\
		-1.366 & 0.252 & -0.461 & 0.413 & -0.577 & 0.703 & 0.81\\
		-1.456 & 0.199 & -0.371 & 0.778 & -0.573 & 0.255 & 0.85\\
		-1.955 & 0.081 & 0.960 & -0.212 & -0.231 & 0.949 & 1.63\\
		-1.986 & 0.067 & 0.871 & -0.506 & 0.164 & 0.846 & 1.61\\
		-2.108 & -0.315 & 0.894 & -0.759 & -0.110 & 0.640 & 1.78\\
		-2.236 & -0.307 & 0.701 & -0.728 & 0.461 & 0.506 & 1.79\\
		-0.523 & 0.886 & 1.429 & 0.185 & 0.069 & 0.980 & 1.66\\
		0.049 & 0.322 & 0.848 & 0.798 & -0.584 & 0.143 & 1.16\\
		-1.538 & -1.378 & 0.292 & -0.300 & -0.934 & 0.192 & 1.79\\
		-1.567 & -1.392 & 0.212 & -0.299 & -0.932 & 0.202 & 1.80\\
		-1.032 & -1.382 & 0.077 & 0.342 & 0.773 & 0.532 & 1.58\\
		-2.431 & -0.948 & -0.228 & -0.878 & -0.389 & 0.278 & 2.08\\
		-2.415 & -0.584 & -0.163 & 0.988 & -0.101 & 0.109 & 1.89\\
		-2.412 & -0.717 & 0.104 & -0.813 & -0.290 & 0.503 & 1.95\\
		-2.384 & -0.451 & 0.262 & -0.906 & -0.252 & 0.338 & 1.84\\
		-0.173	& -0.666 &	0.281 & 0.034 & -1.021 & 0.566 & 1.01\\
		0.049 & 0.322 & 0.848 & 0.448 & 0.029 & 0.919 & 1.16\\
		-0.658 & 1.421 & 0.790 & -0.533 & 1.807 & 1.083 & 1.53\\
		0.913 & 1.145 & 0.048 & 1.320 & 1.207 & 0.332 & 1.88\\
		0.426 & 1.877 & 0.143 & 0.537 & 2.338 & 0.303 & 2.06\\
		-1.024 & 0.043 & -1.195 & -1.341 & 0.269 & -1.508 & 1.20\\
		\multicolumn{7}{c}{...}\\
		\multicolumn{7}{c}{continued on next page}\\
	\end{tabular}	
\end{table}

\textcolor{white}{placeholder}

\begin{table} 
	\centering
	\caption{Table A.3 (continued)}
	\begin{tabular}{rrrrrrr}
		\hline
		$\boldsymbol{nb_x}$ & $\boldsymbol{nb_y}$ & $\boldsymbol{nb_z}$ & $\boldsymbol{n_x}$ & $\boldsymbol{n_y}$ & $\boldsymbol{n_z}$ & r\\
		\hline
		&&&...&&&\\
		\multicolumn{7}{l}{Values for the small lobe:}\\
		
		0.475 & -0.926 & 0.414 & -0.087 & -0.989 & 0.723 & 1.20\\
		0.720 & -0.764 & 0.576 & 0.268 & -0.732 & 0.788 & 0.96\\
		2.143 & 0.531 & 0.619 & 2.546 & 0.595 & 0.908 & 1.18\\
		1.980 & 0.466 & 0.802 & 2.256 & 0.505 & 1.217 & 1.14\\
		2.164 & -0.955 & 0.343 & 2.583 & -1.207 & 0.450 & 0.86\\
		2.235 & -0.522 & 0.648 & 2.719 & -0.594 & 0.752 & 0.85\\
		2.045 & -0.854 & 0.818 & 2.331 & -1.141 & 1.111 & 0.93\\
		2.080 & -1.098 & 0.233 & 2.459 & -1.416 & 0.307 & 0.90\\
		2.094 & -1.043 & 0.294 & 2.510 & -1.280 & 0.440 & 0.87\\
		1.848 & -1.311 & -0.182 & 2.034 & -1.771 & -0.245 & 1.05\\
		1.659 & -1.399 & -0.319 & 1.993 & -1.738 & -0.473 & 1.15\\
		2.091 & 0.795 & 0.047 & 2.222 & 1.269 & -0.044 & 1.32\\
		1.655 & -0.343 & -0.755 & 1.713 & -0.240 & -1.241 & 0.97\\
		1.262 & -0.362 & -0.849 & 1.402 & -0.377 & -1.328 & 1.09\\
		2.073 & 0.790 & 0.163 & 2.201 & 1.258 & 0.041 & 1.30\\
		2.265 & -0.378 & 0.392 & 2.753 & -0.468 & 0.454 & 0.76\\
		2.201 & -0.966 & 0.044 & 2.531 & -1.327 & 0.150 & 0.90\\
		\hline
		\vspace*{4.5in}
\end{tabular}
\end{table}

\section{Supplementary MATLAB code}

In addition to the supplementary material that was made available online, I am including the entirety of MATLAB code that was used for the analysis in \autoref{ch_2} for documentation purposes.

\subsection*{Visualisation of plane normals} 

\begin{lstlisting}
%% Loading a shape model into MATLAB

fname = 'cg-spc-shap5-v1.5-cheops.obj';

% Count the lines in the file:
fid = fopen(fname,'r'); 
n = 0; while ~feof(fid), line = fgetl(fid); n = n+1; end
fclose(fid);

% Import data from the file:
fid = fopen(fname,'r');
f = zeros(n,3); v0 = zeros(n,3);

nv = 0; nf = 0;

while ~feof(fid)
    if ~mod(nv+nf,100000)
        fprintf('%2.1f %%\n',100-(nv+nf)/n*100)
    end
    line = fgetl(fid); 
    if line(1) == 'v'
        nv = nv+1; v0(nv,:) = str2num(line(2:end)); 
    elseif line(1) == 'f'
        nf = nf+1; f(nf,:) = str2num(line(2:end));
    end
end

v0(nv+1:end,:) = []; f(nf+1:end,:) = [];
fclose(fid);
save([fname(1:end-4) '.mat'],'f','v')

% If computational resources are limited, downsample the file (e.g. to 20% of original vertex density):

[redf,redv] = reducepatch(f,v0,0.2);
save([fname(1:end-4) '_red20.mat'],'redf','redv')

\end{lstlisting}
\newpage
\begin{lstlisting}
%% Plotting the plane normals to the PLW measurements

% Display a blank shape-model of the comet:

clear all
load('cg-spc-shap5-v1.5-cheops.mat')    % SHAP5 shape model

figure
shp = patch('Faces',redf,'Vertices',redv,'FaceVertexC',[1 1 1],'FaceC','f','LineS','n','FaceAlpha',1); hold on;
colormap(gray);
lighting gouraud;
hl = light('Style', 'infinite'); set(hl,'Position',[2.5 1 2]);
axis([-3.5 3 -2.5 3 -2 2]); axis equal; axis vis3d; 
rotate3d on; grid on;
xlabel('x [km]'); ylabel('y [km]'); zlabel('z [km]');
view(10,15); zoom(1);

% Define general variables (sl = small lobe, bl = big lobe), all according to Penasa et al. (2017):

% coordinates of the lobe centres
center_sl = [1.06 -0.346 0.01];     center_bl = [-0.473 0.32 -0.167]; 
% ellipsoidal axis ratios
b_sl = 0.76;   c_sl = 0.704;        b_bl = 0.805;   c_bl = 0.544;   
% rotational parameters of axes
ra = 28.1; rb = -11.2; rg = -7.3;    

% rotational matrices  
Rmat_sl=[cosd(ra)*cosd(rb) cosd(ra)*sind(rb)*sind(rg)-cosd(rg)*sind(ra)         
         sind(ra)*sind(rg)+cosd(ra)*cosd(rg)*sind(rb);
         cosd(rb)*sind(ra) cosd(ra)*cosd(rg)+sind(ra)*sind(rb)*sind(rg) 
         cosd(rg)*sind(ra)*sind(rb)-cosd(ra)*sind(rg);
         -sind(rb) cosd(rb)*sind(rg) cosd(rb)*cosd(rg)];

rA = 44.8; rB = 15; rG = 66.3;

Rmat_bl=[cosd(rA)*cosd(rB) cosd(rA)*sind(rB)*sind(rG)-cosd(rG)*sind(rA) 
         sind(rA)*sind(rG)+cosd(rA)*cosd(rG)*sind(rB);
         cosd(rB)*sind(rA) cosd(rA)*cosd(rG)+sind(rA)*sind(rB)*sind(rG) 
         cosd(rG)*sind(rA)*sind(rB)-cosd(rA)*sind(rG);
         -sind(rB) cosd(rB)*sind(rG) cosd(rB)*cosd(rG)];

%% Plot normals from terrace margins (blue)}

% Please note that I had originally called the 'terrace margin' feature by the name 'terrace edge', and the following code still contains the old name (abbreviated as 'E' or 'e').

csv_array = csv2array('all_edges.csv');  % each element of the array holds coordinates for one polyline
EN = zeros(numel(csv_array),3);    % components of plane-normals
n_edge = zeros(numel(csv_array),6);   % matrix containing start-points and direction of EN


for i = 1:numel(csv_array)
    E   = 0.001*csv_array{i}; % one individual polyline, coordinates scaled by 0.001 (shp has different scale!)
    ex  = E(:,1);     ey  = E(:,2);     ez  = E(:,3);    % splits E into x,y,z coordinate vectors (for analysis)
    points = plot3(ex,ey,ez,'.','markersize',8,'color','b'); % re-assembles individual points that make up E (for plotting)
   
    EC    = planefit(ex,ey,ez);    % computes plane equation
    exm   = mean(ex);     eym = mean(ey);    ezm   = exm*EC(1)+eym*EC(2)+EC(3);   % centre of the points
    [x,y] = meshgrid(exm-0.4:0.1:exm+0.4,eym-0.4:0.1:eym+0.4);    
    z = x*EC(1)+y*EC(2)+EC(3);  % grid of the plane in 3D
    
    edgeplane = surf(x,y,z,'FaceAlpha',0,'FaceColor','blue'); 
    
    EN(i,:) = edgeplane.FaceNormals(1,1,:);  % normal to the plane, length = 1}
    
    % the 'arrow3' function requires a start-point and an end-point. The start-point is the centre of the points (exm,...). Inexplicably, some normal-arrows point inward into the shp, their direction needs to be inverted.
    
    % end-point in a regular case:
    exm2 = exm+0.3*EN(i,1); eym2=eym+0.3*EN(i,2); ezm2=ezm+0.3*EN(i,3); 
    % end-point for inverted normals:
    exm3 = exm-0.3*EN(i,1); eym3=eym-0.3*EN(i,2); ezm3=ezm-0.3*EN(i,3);  
    
    if ismember(i,[1:14])    % if feature is located on the Big Lobe  (indices found manually)}
        if pdist2([exm eym ezm],center_bl) < pdist2([exm2 eym2 ezm2],center_bl)    % if normal points inward
            arrow3([exm eym ezm],[exm2 eym2 ezm2],'b1.5',0.5,0.5,'',1); 
            n_edge(i,1:6) = [exm eym ezm EN(i,1) EN(i,2) EN(i,3)]; 
        else
            arrow3([exm eym ezm],[exm3 eym3 ezm3],'b1.5',0.5,0.5,'',1);
            n_edge(i,1:6) = [exm eym ezm -EN(i,1) -EN(i,2) -EN(i,3)]; 
        end
    else    % if feature is located on the Small Lobe
       if pdist2([exm eym ezm],center_sl) < pdist2([exm2 eym2 ezm2],center_sl)
           arrow3([exm eym ezm],[exm2 eym2 ezm2],'b1.5',0.5,0.5,'',1);
           n_edge(i,1:6) = [exm eym ezm EN(i,1) EN(i,2) EN(i,3)];
       else
           arrow3([exm eym ezm],[exm3 eym3 ezm3],'b1.5',0.5,0.5,'',1);
           n_edge(i,1:6) = [exm eym ezm -EN(i,1) -EN(i,2) -EN(i,3)];
       end
    end
end

n_bl_edge = n_edge(1:14,:);  n_sl_edge = n_edge(15,:);   


%% Plot normals from strata heads (pink)

% Please note that I had originally called the 'strata heads' feature by the name 'layerings', and the following code still contains the old name (abbreviated as 'L' or 'l'). Annotations are analogous to the code section above.

csv_array = csv2array('all_layerings.csv'); 
LN = zeros(numel(csv_array),3);   
n_lay = zeros(numel(csv_array),6);   

for i = 1:numel(csv_array)
    L   = 0.001*csv_array{i}; 
    lx  = L(:,1);     ly  = L(:,2);     lz  = L(:,3); 
    points = plot3(lx,ly,lz,'.','markersize',8,'color',[1 0 0.5]); 
   
    LC    = planefit(lx,ly,lz);
    lxm   = mean(lx);     lym = mean(ly);    lzm   = lxm*LC(1)+lym*LC(2)+LC(3); 
    [x,y] = meshgrid(lxm-0.4:0.1:lxm+0.4,lym-0.4:0.1:lym+0.4);    
    z = x*LC(1)+y*LC(2)+LC(3);  
    
    layerplane = surf(x,y,z,'FaceAlpha',0,'FaceColor',[1 0.08 0.4]);
    
    LN(i,:) = layerplane.FaceNormals(1,1,:);
    
    lxm2 = lxm+0.3*LN(i,1); lym2=lym+0.3*LN(i,2); lzm2=lzm+0.3*LN(i,3); 
    lxm3 = lxm-0.3*LN(i,1); lym3=lym-0.3*LN(i,2); lzm3=lzm-0.3*LN(i,3);  
    
    if ismember(i,1:69]) 
        if pdist2([lxm lym lzm],center_bl) < pdist2([lxm2 lym2 lzm2],center_bl)
            arrow3([lxm lym lzm],[lxm2 lym2 lzm2],'_h1.5',0.5,0.5,'',1)
            n_lay(i,1:6) = [lxm lym lzm LN(i,1) LN(i,2) LN(i,3)];
        else
            arrow3([lxm lym lzm],[lxm3 lym3 lzm3],'_h1.5',0.5,0.5,'',1)
            n_lay(i,1:6) = [lxm lym lzm -LN(i,1) -LN(i,2) -LN(i,3)];
        end
    else 
        if pdist2([lxm lym lzm],center_sl) < pdist2([lxm2 lym2 lzm2],center_sl)
            arrow3([lxm lym lzm],[lxm2 lym2 lzm2],'_h1.5',0.5,0.5,'',1)
            n_lay(i,1:6) = [lxm lym lzm LN(i,1) LN(i,2) LN(i,3)];
        else
            arrow3([lxm lym lzm],[lxm3 lym3 lzm3],'_h1.5',0.5,0.5,'',1)
            n_lay(i,1:6) = [lxm lym lzm -LN(i,1) -LN(i,2) -LN(i,3)]; 
        end
    end
end

n_bl_lay = n_lay(1:69,:);   n_sl_lay = n_lay(70:92,:);   

\end{lstlisting}

\newpage

\subsection*{Analysis of plane normals}

\begin{lstlisting}

%% Compare my plane normals to the ellipsoid normals from Penasa et al. (2017) in same location on the Small Lobe (process is analogous for Big Lobe)

% my normal foot-points:            my normal end-points:
xyz_sl_lay  = n_sl_lay(:,1:3);      uvw_sl_lay = n_sl_lay(:,4:6);    
xyz_sl_edge = n_sl_edge(:,1:3);     uvw_sl_edge = n_sl_edge(:,4:6);  

% Compute normal ('gradient') of ellipsoid at location of my foot-points:

for i = 1:size(n_sl_lay,1)    % for the strata heads
    xyz_sl_lay_trans(i,:) = Rmat_sl'*(xyz_sl_lay(i,:)-center_sl)';
    x = xyz_sl_lay_trans(i,1); y = xyz_sl_lay_trans(i,2); 
    z = xyz_sl_lay_trans(i,3);
    t = sqrt((b_sl^2)*(c_sl^2)*(xyz_sl_lay(i,1)^2)+(b_sl^2)
            *(xyz_sl_lay(i,3)^2)+(c_sl^2)*(xyz_sl_lay(i,2)^2));
    gradients_sl_lay(i,1:3) = Rmat_sl*[(b_sl*c_sl*x)/t 
                                       (c_sl*y)/(b_sl*t) 
                                       (b_sl*z)/(c_sl*t)]';
    gradients_sl_lay(i,1:3) = 0.5*gradients_sl_lay(i,:)/norm(gradients_sl_lay(i,:));
end

for i = 1:size(n_sl_edge,1)    % for the terrace margins
    xyz_sl_edge_trans(i,:) = Rmat_sl'*(xyz_sl_edge(i,:)-center_sl)';
    x = xyz_sl_edge_trans(i,1); y = xyz_sl_edge_trans(i,2); 
    z = xyz_sl_edge_trans(i,3);
    t = sqrt((b_sl^2)*(c_sl^2)*(xyz_sl_edge(i,1)^2)+(b_sl^2)
            *(xyz_sl_edge(i,3)^2)+(c_sl^2)*(xyz_sl_edge(i,2)^2));
    gradients_sl_edge(i,1:3) = Rmat_sl*[(b_sl*c_sl*x)/t 
                                        (c_sl*y)/(b_sl*t) 
                                        (b_sl*z)/(c_sl*t)]';
    gradients_sl_edge(i,1:3) = 0.5*gradients_sl_edge(i,:)/norm(gradients_sl_edge(i,:));
end


% Compute angle 'phi' between the two normal vectors:
for i = 1:size(xyz_sl_lay,1)
    u =  uvw_sl_lay(i,1:3) - xyz_sl_lay(i,1:3);
    v = gradients_sl_lay(i,1:3);
    phi_sl_lay(i,1) = acosd(dot(u,v)/dot(norm(u),norm(v)));
end
    
for i = 1:size(xyz_sl_edge,1)
    u =  uvw_sl_edge(i,1:3) - xyz_sl_edge(i,1:3);
    v = gradients_sl_edge(i,1:3);
    phi_sl_edge(i,1) = acosd(dot(u,v)/dot(norm(u),norm(v)));
end

\end{lstlisting}

\chapter{Appendix: Supplementary material for chapter 3}\label{app:ch3}

\section{Supplementary tables}\label{app:ch3_2}

\begin{table}
	\centering
	\caption[Directions $\theta$ of the three highest peaks in the intensity spectrum]{Directions $\theta$ of the three highest peaks found in the intensity spectrum of each frame. Values printed in bold font signify a peak detected as a \underline{layering-related lineament}. Note that layerings were also detected as peak \#4 or \#5 (not included here)\newline}
	\label{tab:table_peaks}
	\begin{tabular}{lrrrclrrr}
		\hline
		Frame & Peak 1 & Peak 2 & Peak 3 &\textcolor{white}{ooooooo}& Frame & Peak 1 & Peak 2 & Peak 3\\
		\hline
		Aa & 78.5  & 103.5 & ---   && Bj & 141.0 & 81.0  & 123.5  \\
		Ab & 51.0  & 73.5  & 141.0 && Bk & 123.5 & 48.5  & \textbf{146.0} \\  
		Ac & 61.0  & 36.0  & ---   && Bl & 81.5  & 51.0  & 121.0\\
        Ad & 81.0  & 113.5 & 48.5  && Bm & 116.0 & 96.0  & 61.0  \\
        Ae & 46.0  & 81.0  & \textbf{161.0} && Bn & 106.0 & 81.0  & 126.0 \\
        Af & \textbf{148.5} & 46.0  & 96.0  && Bo & 96.0  & 63.5  & ---  \\
        Ag & 141.0 & 106.0 & 48.5  && Ca & \textbf{151.0} & 71.0  & ---  \\
        Ah & 141.0 & ---   & ---   && Cb & 61.0  & ---   & ---   \\
        Ai & 128.5 & 46.0  & ---   && Cc & 48.5  & 138.5 & 88.5 \\
        Aj & 38.5  & 123.5 & ---   && Cd & 48.5  & 108.5 & 138.5  \\
        Ak & 123.5 & 46.0  & ---   && Ce & 96.0  & 48.5  & ---   \\
        Al & 128.5 & 88.5  & ---   && Cf & 98.5  & 38.5  & 73.5  \\
        Am & 113.5 & 61.0  & ---   && Cg & 141.0 & 103.5 & ---     \\
        An & 103.5 & 128.5 & 63.5  && Ch & 128.5 & 48.5  & ---     \\
        Ao & 81.0  & 108.5 & 61.0  && Ci & 113.5 & 51.0  &  96.0       \\
        Ba & 38.5  & 81.0  & ---   && Cj & 68.5  & 33.5  &  126.0  \\
        Bb & 81.0  & 38.5  & ---   && Ck & \textbf{148.5} & 46.0  &  71.0  \\
        Bc & 26.0  & 48.5  & ---   && Cl & 71.0  & 126.0 &  51.0    \\
        Bd & 106.0 & 48.5  & ---   && Cm & 113.5 & 71.0  &  ---  \\
        Be & 53.5  & 106.0 & ---   && Cn & 81.0  & 106.0 &  ---   \\
        Bf & 38.5  & 98.5  & 71.0  && Co & 103.5 & 123.5 &  ---\\
        Bg & 106.0 & 51.0  & 128.5 && Da & 61.0  & 138.5 &  ---   \\
        Bh & 126.0 & 43.5  & ---   && Db & 128.5 & 56.0  &  88.5 \\
        Bi & 138.5 & 116.0 & \textbf{156.0} &&  Dc & 48.5  & ---   &  ---   \\
        \multicolumn{9}{c}{...table continued on next page...}\\
	\end{tabular}
\end{table}

\vspace{-0.3cm}
\begin{table}
    \caption[Table B.1 (continued)]{Table B.1 (continued)}
	\centering
	\begin{tabular}{lrrrclrrr}
		\hline
		Frame & Peak 1 & Peak 2 & Peak 3 &\textcolor{white}{ooooooo}& Frame & Peak 1 & Peak 2 & Peak 3\\
		\hline
		Dd & 23.5  & 41.0  & 78.5  && Gc & 51.0  & 138.5 & 103.5 \\
        De & 66.0  & 83.5  & ---   && Gd & 103.5 & 51.0  & 83.5  \\
        Df & \textbf{153.5} & 66.0 & 46.0 && Ge & 61.0  & 96.0  & 126.0   \\
        Dg & 51.0  & 141.0 & ---   && Gf & 63.5  & 96.0  & 128.5\\
        Dh & 136.0 & 51.0  & ---   && Gg & 51.0  & ---   & ---\\
        Di & 46.0  & 128.5 & 71.0  && Gh & 63.5& \textbf{171.0} &---  \\
        Dj & 46.0  & ---   & ---   && Gi & 61.0  & 36.0  & ---  \\
        Dk & 141.0 & 38.5  & 96.0  && Gj & 51.0  & 103.5 & --- \\
        Dl & \textbf{161.0} & 63.5  & --- && Gk & 71.0  & 106.0 & 36.0  \\
        Dm & 73.5  & 126.0 & 96.0  && Gl & 36.0  & \textbf{158.5} & 88.5  \\
        Dn & 113.5 & 78.5  & ---   && Gm & 96.0  & 63.5  & 118.5\\
		Do & 113.5 & 81.0  & 63.5    && Gn & 81.0  & 106.0 & ---      \\
        Ea & 53.5  & \textbf{161.0} & 71.0 && Go & 91.0  & 108.5 & 73.5 \\
        Eb & 53.5  & 136.0 & \textbf{161.0}  && Ha & 63.5  & ---   & ---   \\
        Ec & 61.0  & ---   & ---   && Hb & 56.0  & 106.0 & ---     \\
        Ed & 51.0  & 68.5  & ---   && Hc & 56.0  & 103.5 & ---   \\
		Ee & 68.5  & ---   & ---   && Hd & 58.5  & 101.0 & 76.0  \\
        Ef & 61.0  & 91.0  & 153.5 && He & 106.0 & 71.0  & 126.0  \\
        Eg & 141.0 & 71.0  & 91.0  && Hf & 73.5  & 96.0  & ---\\
        Eh & 138.5 & 46.0  & \textbf{156.0} && Hg & 46.0 & \textbf{153.5}&---\\
        Ei & 46.0  & 81.0  & 161.0 && Hh & 106.0 & 48.5  & 73.5  \\
		Ej & 46.0  & ---   & ---   && Hi & 63.5  & 88.5  & ---  \\
        Ek & 138.5 & 106.0 & 71.0  && Hj & 106.0 & 66.0  & 48.5 \\
        El & 96.0  & 151.0 & ---   && Hk & 91.0  & 63.5  & ---\\
        Em & 96.0  & ---   & ---   && Hl & 71.0  & 88.5  & \textbf{156.0} \\
        En & 106.0 & 83.5  & 128.5 && Hm & 81.0  & ---   & ---  \\   
        Eo & 96.0  & ---   & ---   && Hn & 106.0 & 83.5  & 48.5 \\  
        Fa & 63.5  & \textbf{161.0} & ---  && Ho & 106.0 & 81.0  & 61.0  \\
        Fb & 56.0  & ---   & ---   && Ia & 81.0  & 46.5  & 121.0 \\
        Fc & 61.0  & ---   & ---   && Ib & 51.0  & 71.0  & --- \\
        Fd & 53.5  & ---   & ---   && Ic & 71.0  & 46.0  & ---   \\
        Fe & 66.0  & ---   & ---   && Id & 61.0  & 93.5  & ---  \\
        Ff & 63.5  & 88.5  & ---   && Ie & 96.0  & 61.0  & 78.5 \\
        Fg & 61.0  & 38.5  & ---   && If & 96.0  & 128.5 & 58.5 \\
        Fh & 141.0 & 38.5  & 63.5  && Ig & 51.0  & 103.5 & ---  \\
        Fi & 48.5  & 81.0  & ---   && Ih & 51.0  & 106.0 & ---   \\
        Fj & 46.0  & 73.5  & ---   && Ii & 98.5  & 53.5  & 71.0  \\
        Fk & 71.0  & 36.0  & ---   && Ij & 81.0  & 48.5  & --- \\
        Fl & 81.0  & 48.5  & 116.0 && Ik & 81.0  & 133.5 & 103.5 \\
        Fm & 96.0  & 63.5  & ---   && Il & 88.5  & 38.5  & --- \\
        Fn & 73.5  & 96.0  & 51.0  && Im & 88.5  & 106.0 & --- \\
        Fo & 106.0 & ---   & ---   && In & 106.0 & 128.5 & 81.0  \\
        Ga & 61.0  & ---   & ---   && Io & 113.5 & 138.5 & 81.0 \\
        Gb & 51.0  & 96.0  & ---    \\
        \hline
	\end{tabular}
\end{table}

\newpage

\section{Supplementary MATLAB code} \label{app:ch3_3}

\subsection*{The function 'fftdir.m'}

\begin{lstlisting}[literate={°}{\textdegree}1]

function [peak_max_x,peak_max_y,peak_lay_x,peak_lay_y,peak_down_x,...
    backtrnsf_down,backtrnsf_lay,peak_factor_max,peak_factor_lay,...
    wavelength_max,wavelength_lay,wavelength_down] ... 
    = fftdir(image,step,disp,name)
% FFTDIR(image,step,disp,name) calculates the cumulative FFT2-
% intensities in radial sectors of an [image], to determine the 
% prevalent direction(s) of linear features in the image. The 
% sectors are defined by radius and angular width. 
%   
%   image       image to perform this function on. can be unedited.
%
%   step        angular step/width of the radial sector in degrees 
%               (default: 5)
%
%   disp        toggles display of function outputs:
%               0   outputs are passed to variables, no other display
%               1   (option removed) text-display of outputs
%               2   results are shown as composite figure
%
%   name        name of the input-image (e.g. number of frame)

%% set up image and parameters

if ~isequal(size(image,1),size(image,2))
    disp('attention: image is not square, image was cropped for use!')
    if size(image,1) > size(image,2)
        image = image(1:size(image,2),1:size(image,2));
    elseif size(image,1) < size(image,2)
        image = image(1:size(image,1),1:size(image,1));
    end
end

if size(image,3) > 1
    image = rgb2gray(image);
end

w = size(image,2);  % image width
c = round(w/2);     % location of central pixel is (c,c)


window = mat2gray(fspecial('Gaussian',w,40));
% standard FFT, used for further computations:
imageFT = fftshift(fft2(image.*window));    
% squared FFT, used for peaks detection:
imageFT2 = fftshift(fft2(image.*window)).^2; 
    imageFT_b = imageFT2;
    imageFT_b(c,c) = 0;   % blot out centre pixel (re: brightness)
    imageFTlog_b = log(1+abs(imageFT_b));
imageFTlog = log(1+abs(imageFT2));

%define parameters and defaults:

if ~exist('name','var')
    name = inputname(1); % reads image identifier into variable
end

if ~exist('step','var')
    step = 5;
end

if ~exist('disp','var')
    disp = 2;
end

rmin = 6;                              % minimal radius
rmax = round(0.4 * w);                 % maximal radius
nsec = 360/step;                       % number of sectors
sectors = [0:(step/2):360 (step/2)]';  % vector of overlapping sector
                                       % boundaries, ends on 'step/2'
                                       % because last sector wraps 
                                       % across zero
lbi_lay = (145+step/2)/(step/2);       % boundary indices of layering
                                       % -associated angles
ubi_lay = (180+step)/(step/2); 
sectors_lay = sectors(lbi_lay:ubi_lay);% subset of layering-
                                       % associated sectors


%% find direction of FFT-intensity-peaks

% mask that maps pixel-coordinates to [theta,radius]:
t = zeros(w); r = zeros(w);
for i = 1:w
    for j = 1:w
        [th,ra] = cart2pol(-(i-c),j-c);
        if rad2deg(th) >= 0
            t(i,j) = rad2deg(th);
        else
            t(i,j) = rad2deg(th) + 360;
        end
        r(i,j) = ra;
    end
end












% setting up sectormatrices
sectormatrices = cell(2*nsec,1); sectorpixelsum = zeros(2*nsec,1);
for k = 1:(2*nsec)-1
    m = zeros(w);
    for j = 1:w
        for i = 1:w
            if (t(i,j) >= sectors(k)) && (t(i,j) < sectors(k+2)) ...
                    && (r(i,j) > rmin) && (r(i,j) <= rmax)
               m(i,j) = 1;
            end
        end
    end
    sectormatrices{k} = m;
    sectorpixelsum(k) = sum(sum(sectormatrices{k}));
end
for k = 2*nsec      % special case: crossing zero angle
    m = zeros(w);
    for j = 1:w
        for i = 1:w
            if (t(i,j) >= sectors(k)) && (t(i,j) <= 360) ...
                    && (r(i,j) > rmin) && (r(i,j) <= rmax)
               m(i,j) = 1;
            end
            if (t(i,j) >= 0) && (t(i,j) < sectors(k+2)) ...
                    && (r(i,j) > rmin) && (r(i,j) <= rmax)
               m(i,j) = 1;
            end
        end
    end
    sectormatrices{k} = m;
    sectorpixelsum(k) = sum(sum(sectormatrices{k}));
end        


% sectorsum (cumulative FFT-intensities per radial sector):
sectorsum = zeros(2*nsec,1); sps = zeros(2*nsec,1);
for k = 1:(2*nsec)
    sectorsum(k) = sum(sum(sectormatrices{k}.*abs(imageFT2)));
    % and then normalize by amount of pixels in sector:
    sps(k) = sectorpixelsum(k)/max(sectorpixelsum); % scaling factor
    sectorsum(k) = sectorsum(k) / sps(k); 
end

% subset of sectors where layering might be found
sectorsum_subset = sectorsum(145*2/step:180*2/step);    


% Peak-finding within sectorsum:
[pks,locs] = findpeaks(sectorsum(1:nsec+1),sectors(1:nsec+1)+1,...
            'SortStr','descend','MinPeakDistance',15,...
            'MinPeakWidth',5,'MinPeakHeight',1.5*mean(sectorsum));




% largest peak (overall):
if ~isempty(locs)
    peak_max_x = locs(1);
    peak_max_y = pks(1);
    peak_factor_max = peak_max_y/mean(sectorsum); 
else
    peak_max_x = 0;
    peak_max_y = 0;
    peak_factor_max = 0;
end

% second-largest peak:
if length(locs) >= 2
    peak_max2_x = locs(2);
    peak_max2_y = pks(2);
end

% third-largest peak:
if length(locs) >= 3
    peak_max3_x = locs(3);
    peak_max3_y = pks(3);
end

% layering-associated peak:
peak_lay_index = find(locs>=145,1,'last'); 
% 'last' selects last index within layering-angles, if there is more than one peak. Be aware that 'last' does not mean highest peak.

if ~isempty(peak_lay_index)
    peak_lay_x = locs(peak_lay_index);
    peak_lay_y = pks(peak_lay_index);
    dir_lin = peak_lay_x - 90;
    peak_factor_lay = peak_lay_y/mean(sectorsum); 
else
    peak_lay_x = 0;
    peak_lay_y = 0;
    peak_factor_lay = 0;
end


% downslope lineaments peak: 
peak_down_index = find(locs>=40 & locs<=70,1,'last');

if ~isempty(peak_down_index)
    peak_down_x = locs(peak_down_index);
    peak_down_y = pks(peak_down_index);
    dir_down = peak_down_x - 90;
    peak_factor_down = peak_down_y/mean(sectorsum); 
else
    peak_down_x = 0;
    peak_down_y = 0;
    peak_factor_down = 0;
end


% backtransform in direction of largest peak:
if ~isempty(locs) % (only perform if any peak at all is found)
    imageFT_rueck_max = imageFT;
    keep = [peak_max_x-(step/2) peak_max_x+(step/2)]; 
    for j = 1:w
        for i = 1:w
            if ((t(i,j) >= keep(1)) && (t(i,j) <= keep(2))) || ...
                ((t(i,j) >= keep(1)+180) && (t(i,j) <= keep(2)+180))
                %do nothing
            else
                imageFT_rueck_max(i,j) = 0;
            end
        end
    end
    backtrnsf_max = ifft2(ifftshift(imageFT_rueck_max));
else
    backtrnsf_max = ones(w); % create placeholder image
    backtrnsf_max(:,[1 end]) = 0; backtrnsf_max([1 end],:) = 0;
end

% backtransform in direction of layering-associated peak:
if ~isempty(peak_lay_index) % (only perform if peak is found)
    imageFT_rueck_lay = imageFT;
    keep = [peak_lay_x-(step/2) peak_lay_x+(step/2)]; 
    for j = 1:w
        for i = 1:w
            if ((t(i,j) >= keep(1)) && (t(i,j) <= keep(2))) || ...
                ((t(i,j) >= keep(1)+180) && (t(i,j) <= keep(2)+180))
                %do nothing
            else
                imageFT_rueck_lay(i,j) = 0;
            end
        end
    end
    backtrnsf_lay = ifft2(ifftshift(imageFT_rueck_lay));
else
    backtrnsf_lay = ones(w); % create placeholder image
    backtrnsf_lay(:,[1 end]) = 0; backtrnsf_lay([1 end],:) = 0; 
end

% backtransform in direction of downslope peak:
if ~isempty(peak_down_index)
    imageFT_rueck_down = imageFT;
    keep = [peak_down_x-(step/2) peak_down_x+(step/2)]; 
    for j = 1:w
        for i = 1:w
            if ((t(i,j) >= keep(1)) && (t(i,j) <= keep(2))) || ...
                ((t(i,j) >= keep(1)+180) && (t(i,j) <= keep(2)+180))
                %do nothing
            else
                imageFT_rueck_down(i,j) = 0;
            end
        end
    end
    backtrnsf_down = ifft2(ifftshift(imageFT_rueck_down));
else
    backtrnsf_down = ones(w); % create placeholder image
    backtrnsf_down(:,[1 end]) = 0; backtrnsf_down([1 end],:) = 0;
end


%% find FFT power spectrum in direction of peaks

% power spectrum in direction of largest peak:
if ~isempty(locs)
    count = 1;
    for j = 1:w         
        for i = 1:w     
            if (t(i,j) >= peak_max_x-(step/2)) ...
            && (t(i,j) <= peak_max_x+(step/2)) ...
            && (r(i,j) >= rmin) && (r(i,j) <= rmax)
                power_max(count,1) = r(i,j);                 
                power_max(count,2) = w/power_max(count,1);   % lambda 
                power_max(count,3) = abs(imageFT2(i,j));     % power
                power_max(count,4) = t(i,j);                 % theta
                count = count + 1;
            end
        end
    end
    power_max = sortrows(power_max,2);
    [Mpm,Ipm] = max(power_max(:,3)); 
    wavelength_max = power_max(Ipm,2);
else
    wavelength_max = 0;
end

% power spectrum in direction of second-largest peak:
if length(locs) >=2
    count = 1;
    for j = 1:w         
        for i = 1:w     
            if (t(i,j) >= peak_max2_x-(step/2)) ...
            && (t(i,j) <= peak_max2_x+(step/2)) ...
            && (r(i,j) >= rmin) && (r(i,j) <= rmax)
                power_max2(count,1) = r(i,j);                 
                power_max2(count,2) = w/power_max2(count,1);  % lambda 
                power_max2(count,3) = abs(imageFT2(i,j));     % power
                power_max2(count,4) = t(i,j);                 % theta
                count = count + 1;
            end
        end
    end
    power_max2 = sortrows(power_max2,2);
    [Mpm2,Ipm2] = max(power_max2(:,3)); 
    wavelength_max2 = power_max2(Ipm,2);
else
    wavelength_max2 = 0;
end


% power spectrum in direction of third-largest peak:
if length(locs) >=3
    count = 1;
    for j = 1:w         
        for i = 1:w     
            if (t(i,j) >= peak_max3_x-(step/2)) ...
            && (t(i,j) <= peak_max3_x+(step/2)) ...
            && (r(i,j) >= rmin) && (r(i,j) <= rmax)
                power_max3(count,1) = r(i,j);                 
                power_max3(count,2) = w/power_max3(count,1);  % lambda 
                power_max3(count,3) = abs(imageFT2(i,j));     % power
                power_max3(count,4) = t(i,j);                 % theta
                count = count + 1;
            end
        end
    end
    power_max3 = sortrows(power_max3,2);
    [Mpm3,Ipm3] = max(power_max3(:,3)); 
    wavelength_max3 = power_max3(Ipm3,2);
else
    wavelength_max3 = 0;
end

% power spectrum in direction of layering-associated peak:
if ~isempty(peak_lay_index)
    count = 1;
    for j = 1:w         
        for i = 1:w     
            if (t(i,j) >= peak_lay_x-(step/2)) ...
            && (t(i,j) <= peak_lay_x+(step/2)) ...
            && (r(i,j) >= rmin) && (r(i,j) <= rmax)
                power_lay(count,1) = r(i,j);                 
                power_lay(count,2) = w/power_lay(count,1);   % lambda 
                power_lay(count,3) = abs(imageFT2(i,j));     % power
                power_lay(count,4) = t(i,j);                 % theta
                count = count + 1;
            end
        end
    end
    power_lay = sortrows(power_lay,2);
    [Mpl,Ipl] = max(power_lay(:,3)); 
    wavelength_lay = power_lay(Ipl,2);
else
    wavelength_lay = 0;
end

% power spectrum in direction of layering-associated peak:
if ~isempty(peak_down_index)
    count = 1;
    for j = 1:w         
        for i = 1:w     
            if (t(i,j) >= peak_down_x-(step/2)) ...
            && (t(i,j) <= peak_down_x+(step/2)) ...
            && (r(i,j) >= rmin) && (r(i,j) <= rmax)
                power_down(count,1) = r(i,j);                 
                power_down(count,2) = w/power_down(count,1);   % lambda 
                power_down(count,3) = abs(imageFT2(i,j));     % power
                power_down(count,4) = t(i,j);                 % theta
                count = count + 1;
            end
        end
    end
    power_down = sortrows(power_down,2);
    [Mpd,Ipd] = max(power_down(:,3)); 
    wavelength_down = power_down(Ipd,2);
else
    wavelength_down = 0;
end
    
%% visual output

% display figures
if disp == 2
    multiplot = figure; set(gcf,'color','w');
else
    multiplot = figure('visible','off'); 
end
% 1
subplot(2,5,1); imshow(image,[]); title(['frame ' name ]); hold on
    txt1 = ['size of area: ' num2str(w) ' x ' num2str(w) ' px'];
    text(0.08*w,0.96*w,txt1,'Color',[1. 1. .99],'Fontsize',8)
% 2
subplot(2,5,2); imshow(image.*window);title('image after windowing');
% 3
subplot(2,5,3); imshow(abs(imageFTlog_b),[]); 
    title('inspected area in the FFT2'); hold on;
    for e = 1:360
        plot(c + sind(e)*rmax, c + cosd(e)*rmax,'.',...
        'Markersize',(w/rmax),'Color',[1. 1. .99]); hold on
        plot(c + sind(e)*rmin, c + cosd(e)*rmin,'.',...
        'Markersize',(w/rmax),'Color',[1. 1. .99]); hold on
    end
    txt1 = {{['sector width: ' num2str(step) 'deg']},...
    'sector overlap: 50%',...
     {['min. radius: ' num2str(rmin) ' px, max. radius: ' ...
     num2str(rmax) ' px.']}};
    text([0.08 0.08 0.08]*w,[0.84 0.90 0.96]*w,txt1,...
    'Color',[1 1 .99],'Fontsize',8);
    hold off
% 4
subplot(2,5,4); 
    title(['image backtransformed from' newline 'max. peak']);
    if ~isempty(locs)
        imshow(backtrnsf_max,[]); 
    else
        imshow(backtrnsf_max,[]); hold on;
        text(c-20,c,['no peaks' newline 'fit the criteria'],...
        'Color',[0 0 0],'Fontsize',12);
    end

% 5
subplot(2,5,5);
    title(['image backtransformed from' newline 'layering peak']);
    if ~isempty(peak_lay_index)
        imshow(backtrnsf_lay,[]); 
    else
        imshow(backtrnsf_lay,[]); hold on;
        text(c-20,c,['no peaks' newline 'fit the criteria'],...
        'Color',[0 0 0],'Fontsize',12);
    end


% 6 (bottom row begins here)
subplot(2,2,3); ax1 = gca;
    stairs(sectors(1:nsec+1),sectorsum(1:nsec+1),...
    'Color','k','parent',ax1); hold on; 
    avg = plot(xlim,[mean(sectorsum) mean(sectorsum)],'-b'); hold on; 
    if ~isempty(locs)
        peaks = plot(locs,pks+max(sectorsum)/40,'v','MarkerSize',6,...
        'Color',[0 .7 .7],'MarkerFaceColor',[0 .7 .7]); hold on;
         text(locs+2.5,pks+max(sectorsum)/40,num2str((1:numel(pks))'));
        legend([avg peaks],...
          {['average' newline 'intensity of' newline 'all sectors'],...
          ['{}' newline 'peaks' newline '(sorted by' newline ... 
          'descending' newline 'height)']},...
          'Location','northeastoutside');  
    else
        legend(avg,{['average' newline 'intensity of' newline ...
        'all sectors']}, 'Location','northeastoutside'); 
    end
    title({['cumulative FFT-intensities per sector', ...
    newline '{}' newline '{}']});
    ax1 = gca; 
    set(ax1,'Position',[0.1300 0.1100 0.2685 0.3412],...
        'XTick',[0:30:180],'XTicklabel',[0:30:180]);
    xlabel(ax1,'{\theta} : direction in the Fourier space [°]'); 
    ylabel(ax1,'\Sigma of FFT intensities')
    axis(ax1,[-1 183 0 inf]); %up to 28500
    ax2 = axes('Position',[0.1300 0.1100 0.2685 0.3412]);
    axis(ax2,[-91 93 0 inf]);
    set(ax2,'Color','none','XAxislocation','top','YAxislocation',...
        'right','XTick',[-90:30:90],'XTicklabel',[-90:30:90],...
        'YTick',[],'YTicklabel',[]);
    xlabel(ax2,'{\alpha} : direction in the image [°]');
        










% 7
subplot(2,2,4); % several cases need to be accounted for to plot right:
% if one peak exists:
if ~isempty(locs)
    plot(power_max(:,2),power_max(:,3),'-','Color',[.6 0 0],...
    'Marker','d','MarkerFaceColor',[.6 0 0],'Markersize',4); hold on;
    if isequal(peak_lay_index,1) % if peak1 = layering
       legend({['Peak 1: max. at ' num2str(wavelength_max,3) ...
        ' px, (' num2str(thickness_max,3) ' m)' newline ...
        '(this is the layering)']},'Location','northwest');
    else
       legend({['Peak 1: max. at ' num2str(wavelength_max,3) ...
        ' px, (' num2str(thickness_max,3) ' m)' newline ...
        'no layering detected']},'Location','northwest');
    end
else
    text(.4,.5,['no peaks' newline 'fit the criteria'],...
    'Color',[0 0 0],'Fontsize',12);
end

% if two peaks exists:
if length(locs) >=2
    plot(power_max2(:,2),power_max2(:,3),'-','Color',[1 0 0],...
    'Marker','s','MarkerFaceColor',[1 0 0],'Markersize',4);hold on;
    % set layering marker:
    if isequal(peak_lay_index,1)
       legend({['Peak 1: max. at ' num2str(wavelength_max,3) ...
                ' px, (' num2str(thickness_max,3) ' m)' newline ... 
                '(this is the layering)'] ...
                ['Peak 2: max. at ' num2str(wavelength_max2,3) ...
                ' px, (' num2str(thickness_max2,3) ' m)']},'Location','northwest');
    elseif isequal(peak_lay_index,2)
       legend({['Peak 1: max. at ' num2str(wavelength_max,3) ...
                ' px, (' num2str(thickness_max,3) ' m)'] ...
                ['Peak 2: max. at ' num2str(wavelength_max2,3) ...
                ' px, (' num2str(thickness_max2,3) ' m)' newline ... 
                '(this is the layering)']},'Location','northwest');    
    else
       legend({['Peak 1: max. at ' num2str(wavelength_max,3) ...
                ' px, (' num2str(thickness_max,3) ' m)'] ... 
                ['Peak 2: max. at ' num2str(wavelength_max2,3) ...
                ' px, (' num2str(thickness_max2,3) ' m)' newline ...
                'no layering detected']},'Location','northwest');
    end
end

% if three peaks exists:
if length(locs) >=3
    plot(power_max3(:,2),power_max3(:,3),'-','Color',[1 .5 .2],...
    'Marker','o','MarkerFaceColor',[1 .5 .2],'Markersize',4); hold on;
    % set layering marker:
    if isequal(peak_lay_index,1)
       legend({['Peak 1: max. at ' num2str(wavelength_max,3) ...
                ' px, (' num2str(thickness_max,3) ' m)' newline ... 
                '(this is the layering)'] ...
                ['Peak 2: max. at ' num2str(wavelength_max2,3) ...
                ' px, (' num2str(thickness_max2,3) ' m)'] ...
                ['Peak 3: max. at ' num2str(wavelength_max3,3) ...
                ' px, (' num2str(thickness_max3,3) ' m)']},'Location','northwest');
    elseif isequal(peak_lay_index,2)
       legend({['Peak 1: max. at ' num2str(wavelength_max,3) ...
                ' px, (' num2str(thickness_max,3) ' m)'] ...
                ['Peak 2: max. at ' num2str(wavelength_max2,3) ...
                ' px, (' num2str(thickness_max2,3) ' m)' newline ... 
                '(this is the layering)'] ...
                ['Peak 3: max. at ' num2str(wavelength_max3,3) ...
                ' px, (' num2str(thickness_max3,3) ' m)']},'Location','northwest');    
    elseif isequal(peak_lay_index,3)
       legend({['Peak 1: max. at ' num2str(wavelength_max,3) ...
                ' px, (' num2str(thickness_max,3) ' m)'] ...
                ['Peak 2: max. at ' num2str(wavelength_max2,3) ...
                ' px, (' num2str(thickness_max2,3) ' m)'] ...
                ['Peak 3: max. at ' num2str(wavelength_max3,3) ...
                ' px, (' num2str(thickness_max3,3) ' m)' newline ...
                '(this is the layering)']},'Location','northwest');
    else
       legend({['Peak 1: max. at ' num2str(wavelength_max,3) ...
                ' px, (' num2str(thickness_max,3) ' m)'] ... 
                ['Peak 2: max. at ' num2str(wavelength_max2,3) ...
                ' px, (' num2str(thickness_max2,3) ' m)'] ...
                ['Peak 3: max. at ' num2str(wavelength_max3,3) ...
                ' px, (' num2str(thickness_max3,3) ' m)' newline ...
                'no layering detected']},'Location','northwest');
    end
end
title('FFT power spectrum along the direction of...');
xlabel('wavelength of the Fourier mode [in px]');
ylabel('relative power of the Fourier mode')

        
% save multiplot figure to file:
filename = sprintf(['multiplot for frame ',num2str(name)]);
saveas(multiplot,[filename, '.png']);

end % end-of-function       
\end{lstlisting}

\newpage

\subsection*{The script 'fullcliffscan.m'}

\begin{lstlisting}

%% This script splits a big picture into frames, and runs fftdir.m on all of them, producing vectors that contain the output variables for all frames

clear

%% set up parameters:

bigpic = imread('CROP.tiff');   % input file, ideally size *01x*01 px
bigpic = im2double(bigpic);     % converts file from uint8 to double
w = 101;                        % frame size, ideally *01x*01 px
sz = w-1;                       
c = round(w/2);

step = 5;                       % angular step width
rmin = 6;                       % minimum radius
rmax = round(0.4 * w);          % maximal radius
nsec = 360/step;                % number of sectors

%% set up frames:

nframes_h = round((size(bigpic,2)-1)/sz) ; % horizontal frames,     
                                           % non-overlapping
nframes_v = round((size(bigpic,1)-1)/sz) ; % vertical frames, 
                                           % non-overlapping
nframes=(2*nframes_h-1) * (2*nframes_v-1); % number of frames, 
                                           % overlapping
f = 1;
for v = 1:0.5:nframes_v     % the '0.5' produces the overlap
    for h = 1:0.5:nframes_h
        updown = ((v-1)*sz)+1:(v*sz)+1;     % vertical extent of frame
        leftright = ((h-1)*sz)+1:(h*sz)+1;  % horizontal extent
        frames{f} = bigpic(updown,leftright);
        f = f+1;
    end
end

%% apply fftdir.m to frames

% pre-define matrices of output-parameters for speedy execution

peak_max_x = zeros(nframes,1); peak_max_y = zeros(nframes,1); 
peak_down_x = zeros(nframes,1);
peak_lay_x = zeros(nframes,1); peak_lay_y = zeros(nframes,1);
backtrnsf_d = cell(nframes,1); backtrnsf_l = cell(nframes,1);
peak_factor_max = zeros(nframes,1); peak_factor_lay = zeros(nframes,1);
peak_factor_down = zeros(nframes,1);
wavelength_max = zeros(nframes,1); wavelength_lay = zeros(nframes,1);
wavelength_down = zeros(nframes,1);
dir_lay = zeros(nframes,1); dir_max = zeros(nframes,1); 
dir_down = zeros(nframes,1);

for f = 1:nframes
    [peak_max_x(f),peak_max_y(f),peak_lay_x(f),peak_lay_y(f),...
    peak_down_x(f),backtrnsf_down,backtrnsf_lay,peak_factor_max(f),...
    peak_factor_lay(f),peak_factor_down(f),wavelength_max(f),...
    wavelength_lay(f),wavelength_down(f)] = fftdir(frames{f},step,0,f); 

    backtrnsf_d{f} = backtrnsf_down(25:75,25:75); % adjust to sz
    backtrnsf_l{f} = backtrnsf_lay(25:75,25:75);  % adjust to sz
    if peak_max_x(f) > 0    % if any peaks were found...
        dir_max(f) = peak_max_x(f) - 90; % converts theta to alpha
    end
    if peak_lay_x(f) > 0    % if layering-peaks were found...
        dir_lay(f) = peak_lay_x(f) - 90;
    end
    if peak_down_x(f) > 0   % if downslope-peaks were found...
        dir_down(f) = peak_down_x(f) - 90;
    end
end

% optionally: save output variable to a .mat file for later
save('dir_lay.mat','dir_lay');



%% sample output map (various maps can be created, here an example for a map tiled with backtransforms:)

map1 = ones(size(bigpic,1),size(bigpic,2));
 
f = 1;
for v = 1:0.5:nframes_v
    for h = 1:0.5:nframes_h
    updown = ((v-1)*sz)+25:((v-1)*sz)+75;
    leftright = ((h-1)*sz)+25:((h-1)*sz)+75;
    map1(updown,leftright) = backtrnsf_d{f};
    f = f+1;
    end
end
 
figure; imshow(map1,[-0.08 0.1]);   % display range adjusted for 
                                    % better visibility
title('map of backtransformations along direction of downslope lineaments')
% optionally: (interesting but misleads the eye!)
figure; imshow(bigpic + 3*map1);
title('overlay of Hathor wall image and map1');



%% sample map showing single red lines:

map2 = ones(size(bigpic,1),size(bigpic,2));    % white background

figure; imshow(map2); hold on

% plot black grid:
for xx = 25:50:825
    plot([xx xx],[25 475],'-k','LineWidth',1.5); hold on
end
for yy = 25:50:475
    plot([25 775],[yy yy],'-k','LineWidth',1.3); hold on
end

% plot red lines: 
f = 1; 
for v = 1:0.5:nframes_v
    for h = 1:0.5:nframes_h
        updown = ((v-1)*sz)+25:((v-1)*sz)+75;
        leftright = ((h-1)*sz)+25:((h-1)*sz)+75;
        no_lay = backtrnsf_l{f};  
        if ~isequal(no_lay(2,2),1)
            plot([mean(leftright)-22*sind(dir_lay(f)) 
                  mean(leftright)+22*sind(dir_lay(f))],...
                 [mean(updown)+22*cosd(dir_lay(f)) 
                  mean(updown)-22*cosd(dir_lay(f))],...
                 '-r','LineWidth',2); hold on;
        end
        f = f+1;
    end
end

\end{lstlisting}

\newpage

\section{State Vector Calculations} \label{app:ch3_4}

The following are two outputs of the NASA SPICE Geometry Calculator found at https://wgc.jpl.nasa.gov:8443/webgeocalc/ (Version 2.1.0 (4386 N66 11-FEB-2019), accessed 01 Sept 2019).\\

\noindent \textbf{State Vector Calculation: The Rosetta spacecraft as seen from comet 67P}\\

\noindent Calculation Inputs\\

\begin{tabular}{ll}
Calculation type     &= State Vector\\
Target               &= ROSETTA\\
Observer             &= 67P/CHURYUMOV-GERASIMENKO (1969 R1)\\
Reference frame      &= 67P/C-G\_CK\\
Light propagation    &= No correction\\
Time system          &= UTC\\
Time format          &= Calendar date and time\\
Input time           &= 2014-08-28T12:44:03.538\\
State representation &= Rectangular\\
\end{tabular}

\vspace{0.4cm}
\noindent State Vector Results\\

\begin{tabular}{lll}
UTC calendar date    &= 2014-08-28 12:44:03.538000 UTC\\
Distance (km)        &= 56.75364967\\
Speed (km/s)         &= 0.00579590\\
X (km)               &= -40.78008583\\
Y (km)               &= 3.02932856\\
Z (km)               &= 39.35459972 \\
dX/dt (km/s)         &= 0.00070295\\
dY/dt (km/s)         &= 0.00575160\\
dZ/dt (km/s)         &= 0.00013231\\
Time at Target       &= 2014-08-28 12:44:03.538000 UTC\\
Light Time (s)       &= 0.00018931\\
\end{tabular}{}

\vspace{0.4cm}
\noindent Kernels Used\\

\noindent\path{pds/wgc/mk/ground_stations_v0007.tm}\\
\path{pds/wgc/mk/solar_system_v0028.tm}\\
\path{pds/wgc/mk/latest_lsk_v0004.tm}\\
\path{pds/data/ro_rl-\e_m_a_c-spice-6-v1.0/rossp_1000/EXTRAS/MK/ROS_V06.TM}\\

\newpage

\noindent \textbf{State Vector Calculation: The Sun as seen from comet 67P}\\

\noindent Calculation Inputs\\

\begin{tabular}{ll}
Calculation type     &= State Vector\\
Target               &= SUN\\
Observer             &= 67P/CHURYUMOV-GERASIMENKO (1969 R1)\\
Reference frame      &= 67P/C-G\_CK\\
Light propagation    &= No correction\\
Time system          &= UTC\\
Time format          &= Calendar date and time\\
Input time           &= 2014-08-28T12:44:03.538\\
State representation &= Rectangular\\
\end{tabular}

\vspace{0.4cm}
\noindent State Vector Results\\

\begin{tabular}{lll}
UTC calendar date    &= 2014-08-28 12:44:03.538000 UTC\\
Distance (km)        &= 519124813.89904130\\
Speed (km/s)         &= 53145.40696210\\
X (km)               &= -162404434.65031720\\
Y (km)               &= 341089248.32254330\\
Z (km)               &= 356052659.99622524 \\
dX/dt (km/s)         &= 47985.18250844\\
dY/dt (km/s)         &= 22844.17652012\\
dZ/dt (km/s)         &= -11.82865241\\
Time at Target       &= 2014-08-28 12:44:03.538000 UTC\\
Light Time (s)       &= 1731.61398843\\
\end{tabular}{}

\vspace{0.4cm}
\noindent Kernels Used\\

\noindent\path{pds/wgc/mk/ground_stations_v0007.tm}\\
\path{pds/wgc/mk/solar_system_v0028.tm}\\
\path{pds/wgc/mk/latest_lsk_v0004.tm}\\
\path{pds/data/ro_rl-\e_m_a_c-spice-6-v1.0/rossp_1000/EXTRAS/MK/ROS_V06.TM}\\

\newpage

\section{Graphical display of fftdir.m results for all tiles} \label{app:ch3_5}

On the following pages, 135 figures show a graphical display of the output given by \texttt{fftdir.m} for the analysed part of the Hathor cliff. The frame designation is given in the top-left corner of each figure, and corresponds to the frame coordinates used in all parameter-maps in \autoref{ch_3_3_2}. The reader is reminded that only the central section of each frame (first image on the following pages) makes up the corresponding tile in the output-maps (cf. also \autoref{fig:frames_composite}).

\begin{figure} [h]
	\centering 
	\includegraphics[width=0.96\linewidth]{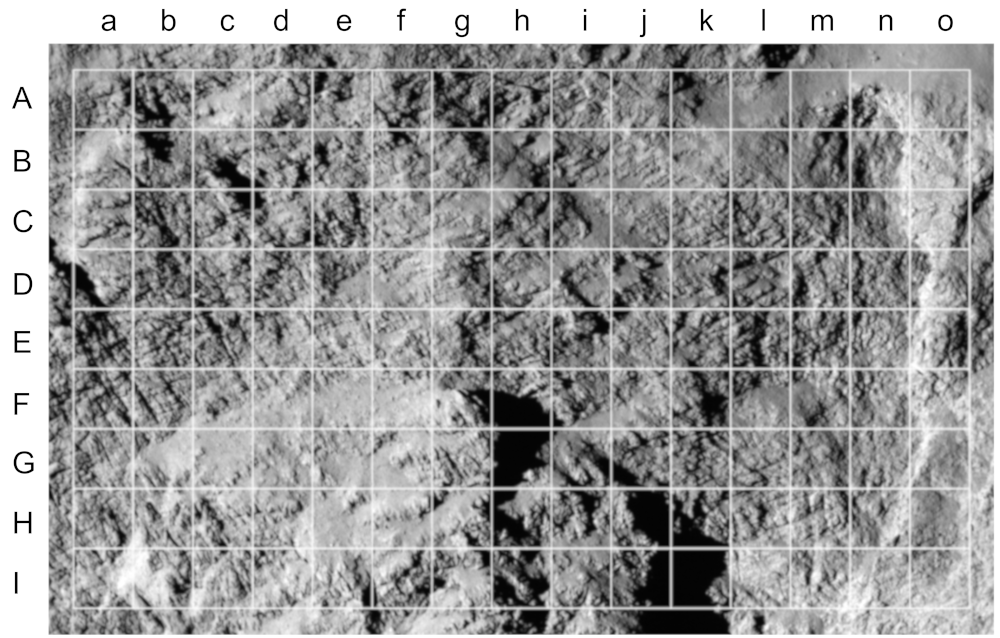}
	\caption{Input image, cropped from an OSIRIS NAC image, overlaid with coordinate grid where each square signifies a frame.}
	\label{fig:crop_raster_cs}
\end{figure}

For ease of reference, the input image and its coordinate grid are shown again in \autoref{fig:crop_raster_cs}.\\

\textcolor{red}{Note: In this version of my thesis (published online e.g. at the arXiv and my website), only the first plate of Figure B.2 is included, in order to reduce the size of this document. Please refer to the full version published under ISBN 978-3-947208-20-3, also available through the online catalogue of the SUB library G\"ottingen.}

\vspace{1cm}

\begin{figure}

\caption[Graphical display of fftdir.m (135 figures)]{Graphical display of fftdir.m (135 figures on the following pages)\textcolor{white}{ooooooo}}
\end{figure}

\newpage

\centering\frame{
\includegraphics[angle=90,scale=0.56,trim={5.2cm 0 4.2cm 0},clip]{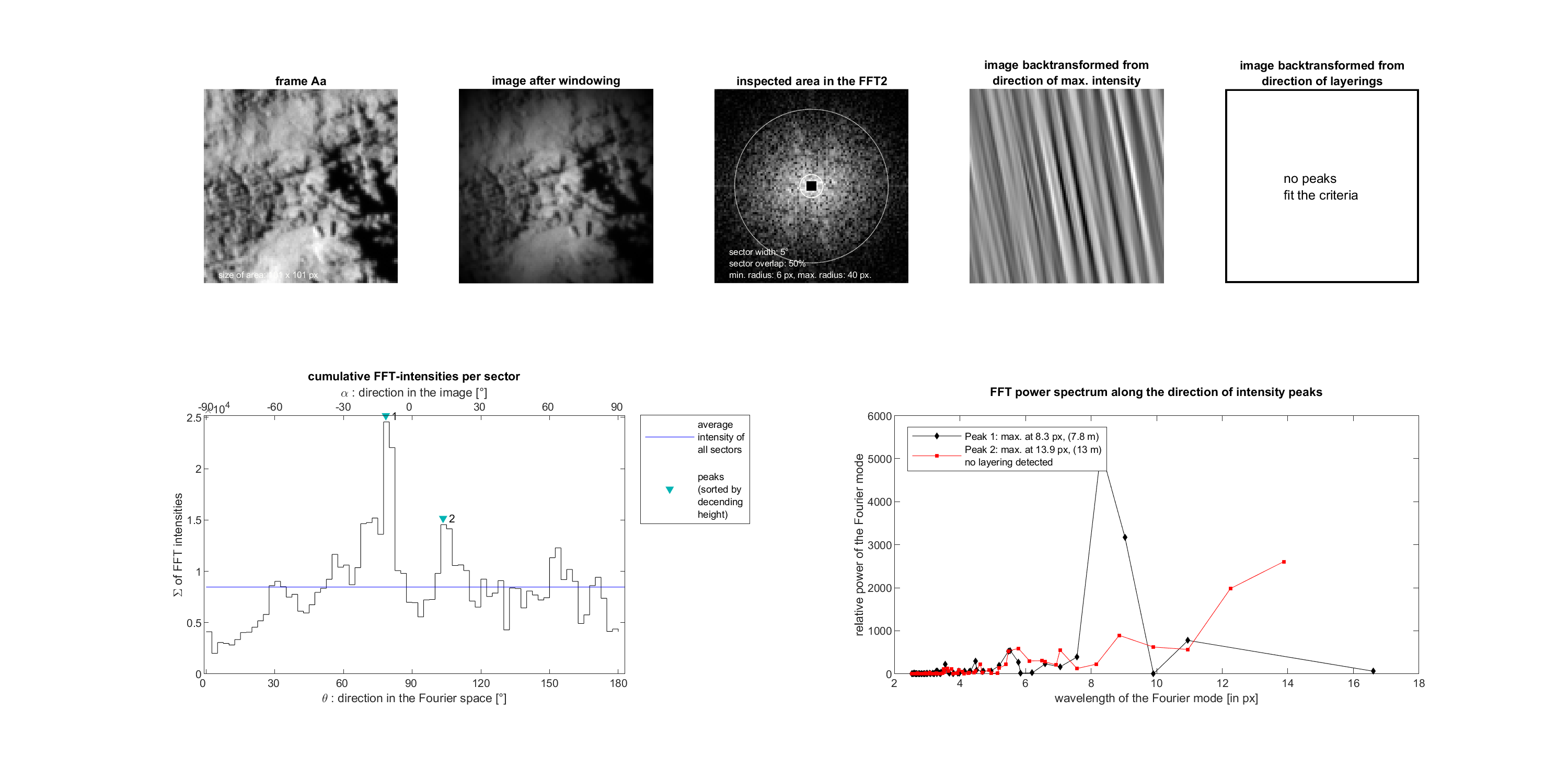}}

\raggedright 

\chapter*{Publications\markboth{Publications}{Publications}}
\addcontentsline{toc}{chapter}{Publications}

{\bf\large Refereed publications}\\
\vspace{0.5cm}
Ruzicka, B.-K., Penasa, L., Boehnhardt, H., Pack, A., Dolives, B., Souvannavong, F., and Remetean, E.: Analysis of layering-related linear features on comet 67P/Churyumov-Gerasimenko. Monthly Notices of the Royal Astronomical Society, 2018, Volume 482, Issue 4, p. 5007--5011.

\vspace{3cm}

{\bf\large Conference proceedings}
\begin{itemize}
\item Ruzicka, B.-K., Boehnhardt, H., Pack, A.: Layerings in cometary nuclei. European Planetary Science Congress 2017, Riga. EPSC Abstracts Vol. 11, EPSC2017-482
\item Ruzicka, B.-K., Boehnhardt, H., Pack, A.: Layerings in cometary nuclei. Rocks \& Stars Conference 2017, G\"ottingen 
\item Ruzicka, B.-K., Boehnhardt, H., Penasa, L., Pack, A.: Layering-Related Linear Features on Comet 67P. European Planetary Science Congress 2018, Berlin. EPSC Abstracts Vol. 12, EPSC2018-664
\item Ruzicka, B.-K., Schr\"oter, M., Boehnhardt, H., and Pack, A.: Fourier-based analysis of lineament structures on the Hathor cliff of comet 67P. EPSC Abstracts Vol. 13, EPSC-DPS2019-1328, European Planetary Science Congress-Division of Planetary Science Joint Meeting 2019

\end{itemize}

\chapter*{Acknowledgements\markboth{Acknowledgements}{Acknowledgements}} \label{ch_ackn}
\addcontentsline{toc}{chapter}{Acknowledgements}

This work was supported by the International Max-Planck Research School (IMPRS) for Solar System Science at the University of G\"ottingen, in the department for Planets and Comets under the direction of Ulrich Christensen, who was also a member of my advisory committee.

\vspace{1cm}

In the words of astronaut Peggy Whitson, it takes a village to make a space mission successful, and my personal 'space mission' of completing this doctoral thesis certainly is built on the support of a large number of people. 

First and foremost I would like to thank the members of my thesis advisory committee for their guidance, instruction, and countless inspiring discussions. I would like to thank my supervisor Hermann Boehnhardt for teaching me to look at comets and the solar system through the eyes of an astrophysicist, and I hope that supervising my doctoral thesis gave him a glimpse of a geologist's perspective. He also opened up phenomenal opportunities by connecting me to researchers at CNES and the University of Padua, and I am very grateful for that. I would also like to thank Andreas Pack for his contagious enthusiasm for cometary geology, and for guiding me back on track whenever I scienced myself into a corner.

I am immensely grateful to Luca Penasa for inexhaustibly and patiently mentoring me through the process of writing (and re-writing, and re-writing) my first scientific manuscript. Bis his example, Luca also unknowingly challenged me to broaden the range of my computer- and programming skills, which gave me the confidence to tackle scientific problems I would never have dared to approach otherwise. 

Matthias Schr\"oter was instrumental in the conception of the second half of this thesis, and I would like to thank him for his unwavering encouragement and tutorship regarding \textsc{MATLAB} in general, and Fourier analysis in particular. He believed in my coding abilities before I did, and it made a world of difference.

I gratefully acknowledge assistance from the OSIRIS team: Holger Sierks for providing image data sets, and Carsten G\"uttler, Cecilia Tubiana, and Jakob Deller for patiently helping me to make sense of the data.

I was very glad to collaborate with Emile Remetean and Beno\^it Dolives, who went above and beyond to customise the \textsc{PLW} software for me and to make me feel welcome at CNES. Thank you so much to Mattheo Massironi, for hosting me in Padua and introducing me to the wonderful people in your lab.

I sincerely would like to thank Frank Preusker for enlightening me about shape models in ways that textbooks never could, and for generously sending me converted versions of his shape model until we found a file format I could wrangle into my processing pipeline. Similarly, I need to thank Jean-Baptiste Vincent for providing me with his Shape Viewer software and including functionalities to help me figure out geometries on the nucleus, and also for identifying specific OSIRIS images for me on short notice. 

I am grateful to Sonja Schuh for helping me navigate the bureaucracy, always having an open ear, and doing such an admirable job at coordinating the IMPRS.

I also owe gratitude to Daniel Maase and Bernhardt Bandow, for having my back through all of the tech trouble I ran into. 

This thesis greatly benefited from the mental support I received from Alina Lira Lorca at the International Writing Centre at the Universit\"at G\"ottingen. Her guidance with regards to time management and structuring this thesis were invaluable and greatly helped my composure during its creation.

Lastly, but definitely most importantly, I need to thank my personal 'ground segment', or as I should more correctly call it, my personal growth system. Anke Friedrich: Your mentorship and leadership through the years made me the scientist and human that I am today. You taught me to love the Earth and encouraged me when I started to love the skies. Urs Mall: Your support and friendship kept me going, and I will miss our discussions dearly. Sarah: You kept me grounded and wildly increased the amount of grammar in my manuscripts and beauty in my life. Katherine, James, Deni: Thank you for checking my math and for helping overcome writer's block again and again. Sebastian: Thank you for not letting me give up. Renard: Oh, where do I even start. Your friendship means the world to me, and so much of the completion of this thesis is on you. Tasya: You are my North Star. Out of everything this PhD endeavour has given me, you are the greatest and most unexpected gift of all. Mom and dad: Thank you with all my heart for providing the best possible launch pad anyone could wish for. I am proud to be your daughter. 

\end{document}